\documentclass{aa}  
\usepackage{dcolumn}
\usepackage{aalongtable}
\usepackage{longtable}
\usepackage{graphicx}
\usepackage{txfonts}
\begin{document}

\newcommand{\swift}{{\it Swift}}
\newcommand{\he}{{HE\,1136-2304}}
\newcommand{\plm}{$\pm$}
\newcommand{\rb}[1]{\raisebox{1.5ex}[-1.5ex]{#1}}
\newcommand{\Ha}{H$\alpha$}
\newcommand{\Hb}{H$\beta$}
\newcommand{\Hg}{H$\gamma$}
\newcommand{\Lya}{Ly$\alpha$}
\newcommand{\sVl}[3]{#1\,{\sc #2}]\,$\lambda{#3}$}
\newcommand{\Nl}[3]{#1\,{\sc #2}\,$\lambda{#3}$}
\makeatletter
 \def\hlinewd#1{%
   \noalign{\ifnum0=`}\fi\hrule \@height #1 \futurelet
    \reserved@a\@xhline}
\makeatother
\newcommand{\htopline}{\hlinewd{.8pt}}
\newcommand{\hmidline}{\hlinewd{.2pt}}
\newcommand{\hbotline}{\htopline}
\newcommand{\mcc}[1]{\multicolumn{1}{c}{#1}}
\newcommand{\mcl}[1]{\multicolumn{1}{l}{#1}}
\newcommand{\mcr}[1]{\multicolumn{1}{r}{#1}}
\newcommand{\kms}{km\,s$^{-1}$}
\newcolumntype{d}{D{.}{.}{-1}}


    \title{Long-term optical, UV, and X-ray
continuum variations in the changing-look AGN
 HE\,1136-2304
}

   \author{M. Zetzl \inst{1},
           W. Kollatschny \inst{1}, 
           M. W. Ochmann \inst{1},
           D. Grupe \inst{2},
           M. Haas \inst{3},
           M. Ramolla \inst{3},
           D. Chelouche \inst{4}, 
           S. Kaspi \inst{5}, 
           N. Schartel \inst{6}
          }

   \institute{Institut f\"ur Astrophysik, Universit\"at G\"ottingen,
              Friedrich-Hund Platz 1, D-37077 G\"ottingen, Germany\\
              \email{wkollat@astro.physik.uni-goettingen.de}
         \and
          Department of Earth and Space Sciences, Morehead State University, Morehead, KY 40351, USA
         \and
          Astronomisches Institut, Ruhr-Universit\"at Bochum,
               Universit\"atsstrasse 150, 44801 Bochum, Germany
         \and
   Physics Department and the Haifa Research Center for Theoretical Physics and
   Astrophysics, University of Haifa, Haifa 3498838, Israel
         \and  
          School of Physics \& Astronomy and the Wise Observatory,
   The Raymond and Beverly Sackler Faculty of Exact Sciences
   Tel-Aviv University, Tel-Aviv 69978, Israel
         \and
         XMM-Newton Science Operations Centre, ESA, Villafranca del Casuntilo, Apartado 78, 
              28691 Villanueva de la Ca{\~nada}, Spain\\ 
}

   \date{Received 20 December 2017; Accepted 6 April 2018}
   \authorrunning{Zetzl et al.}
   \titlerunning{HE\,1136-2304 variability}

 
  \abstract
   {}
   {A strong outburst in the X-ray continuum
and a change of its Seyfert spectral type
was detected in HE\,1136-2304 in  2014.
The spectral type changed
 from nearly Seyfert  2  type (1.95) to Seyfert  1.5 type
 in comparison to previous observations taken ten to twenty years before.
In a subsequent variability campaign 
we wanted to investigate whether this
outburst was a single event
 or whether the variability pattern following
the outburst was similar to those seen in other variable Seyfert galaxies.}
   {In addition to a SALT spectral variability campaign,
we carried out optical continuum as well as X-ray and UV (\swift{}) monitoring 
studies from 2014 to 2017.
}  
   {HE\,1136-2304 strongly
varied on  timescales of days to months from 2014 to 2017.
 No systematic trends were found in the variability behavior
  following the outburst in 2014. A general decrease in flux
would have been expected for a tidal 
disruption event. This could not be confirmed.
More likely the flux variations are connected
to irregular fluctuations in the accretion
rate. The strongest variability amplitudes have been found in the X-ray regime:
HE\,1136-2304 varied by a factor of eight
during 2015.
The amplitudes of the continuum variability (from the UV to the optical)
 systematically decreased
with wavelength following a power law
$F_{\rm var}=a\cdot\lambda^{-c}$  with c = 0.84.
There is a trend that the B-band
continuum shows a delay of three light days with respect to the
variable X-ray flux. The Seyfert type 1.5 did not change despite
the strong continuum variations
for the period between 2014 and 2017.
        } 
{}
\keywords {Galaxies: active --
                Galaxies: Seyfert  --
                Galaxies: nuclei  --
                Galaxies: individual: HE\,1136-2304 --   
                (Galaxies:) quasars: emission lines 
               }

   \maketitle
%

\section{Introduction}

It is generally known that Seyfert 1 galaxies are variable in the optical
and in X-rays on timescales of days to decades. Several
active galactic nuclei (AGN) have shown variations in the X-ray
continuum by a factor of more than 20  (e.g., Grupe et al. 2001, 2010).
AGN that show 
extreme X-ray flux variations in combination with
X-ray spectral variations, i.e.,   when a
Compton-thick AGN becomes Compton-thin and vice versa,  were designated as
changing-look AGN (e.g., Guainazzi\citealt{guainazzi02}). 
By analogy, optical changing-look AGN exhibit transitions from type 1 to type 2 and vice
versa. In this case, the optical spectral classification can change due
to a variation in the intrinsic  nuclear power/accretion power, a variation
in reddening, or  a combination  of the two.
Typical transition timescales are months to years. 

To date about a dozen Seyfert galaxies are known to have changed their
 optical
spectral type, for example,   NGC\,3515 (Collin-Souffrin et al.\citealt{souffrin73}),
 NGC\,4151 (Penston \& Perez\citealt{penston84}), Fairall\,9 
 (Kollatschny et al.\citealt{kollatschny85}), NGC\,2617 (Shappee
 et al.\citealt{shappee14}), Mrk\,590 (Denney at al.\citealt{denney14}) 
and references therein. Further recent findings are based on spectral
variations detected by means of the Sloan Digital Sky Survey (SDSS)
(e.g., Komossa et al.\citealt{komossa08}, LaMassa et al.\citealt{lamassa15},
Runnoe et al.\citealt{runnoe16}, MacLeod et al.\citealt{macleod16}). 
In most of these recent findings only a few optical spectra
of the individual SDSS galaxies 
have been secured to prove their changing-look character.

HE\,1136-2304 ($\alpha_{2000}$ = 11h 38m 51.1s,
$\delta$ = $-23^{\circ}$ 21$^{'}$ 36$^{''}$)
has been detected as a variable X-ray source by the XMM-Newton slew
survey in 2014 (Parker et al.\citealt{parker16}).
The 0.2--2 keV flux increased by a factor of about 30 in comparison
to the ROSAT all-sky survey in 1990. However, no clear evidence of
X-ray absorption variability has been seen.
HE\,1136-2304  changed its optical
spectral classification from 1994 (Seyfert 2/1.95) to 2014 (Seyfert 1.5)
and can be considered  an optical changing-look AGN. 

We decided to study the variability behavior of HE\,1136-2304 
subsequent to its X-ray outburst in 2014 in detail.
We carried out 
 optical photometric and spectroscopic variability follow-up studies 
in combination with \swift{} UV and X-ray photometric observations
 to investigate
the variability behavior  of this changing-look galaxy on timescales of weeks
to years. 
The outburst in HE\,1136-2304 could have been caused by a 
tidal disruption event, by a less drastic variation in the intrinsic nuclear power/accretion power, or by significant
variation in the absorption.
Although detailed and long-term optical variability
studies exist for many AGN, for example 
NGC\,5548 (Peterson et al.\citealt{peterson02}, Pei et al.\citealt{pei17}),
NGC\,7603 (Kollatschny et al.\citealt{kollatschny00}), and
3C\,390.3  (Shapovalova et al.\citealt{shapovalova10}),  no detailed
follow-up studies have been reported for 
the changing-look-type AGN mentioned above.

This is the first paper in a series. We will discuss
the spectral variations seen in 2015 and the broad-line region structure
in HE\,1136-2304 in a second paper in detail.

Throughout this paper, we assume that H$_0$~=~70~km s$^{-1}$ Mpc$^{-1}$ 
with a $\Lambda$CDM cosmology with $\Omega_{\Lambda}$=0.73 and $\Omega_{\rm M}$=0.27. With a redshift of z=0.0271 this results in a luminosity distance of $D_{\rm L}$ = 118 Mpc using the Cosmology Calculator developed by Wright \cite{wright06}.

\section{Observations and data reduction}
\subsection{{\bf Optical spectroscopy with the SALT telescope}}
\begin{table}
\tabcolsep+6mm
\caption{Log of spectroscopic observations of HE1136 with SALT.
Listed are the Julian date, the UT date, and the exposure time.}
 
\centering
\begin{tabular}{ccr}
\hline 
\noalign{\smallskip}
Julian Date &  & Exp. time \\
2\,400\,000+&  \rb{UT Date}       &  [s]   \\
\hline 
56846.248       &       2014-07-07      &      1200    \\
57016.559       &       2014-12-25      &       985     \\
57070.399       &       2015-02-16      &       985     \\
57082.362       &       2015-02-28      &       985     \\
57088.594       &       2015-03-07      &       985     \\
57100.539       &       2015-03-19      &       985     \\
57112.285       &       2015-03-30      &       985     \\
57121.256       &       2015-04-08      &       985     \\
57131.243       &       2015-04-18      &      1230    \\
57167.359       &       2015-05-24      &      1144     \\
57171.364       &       2015-05-28      &      1144      \\
57182.330       &       2015-06-08      &      1144       \\
57187.319       &       2015-06-13      &      1144      \\
57192.308       &       2015-06-18      &      1144       \\
57196.295       &       2015-06-22      &      1144      \\
57201.271       &       2015-06-27      &      1144       \\
57206.265       &       2015-07-02      &      1144      \\
57217.227       &       2015-07-13      &      1144       \\
57399.510       &       2016-01-12      &      1200    \\
57519.391       &       2016-05-10      &      1196     \\
57540.351       &       2016-05-31      &      1196     \\
57889.381       &       2017-05-15      &      1218      \\ 
\hline 
\vspace{-.7cm}
\end{tabular}
\label{saltlog}
\end{table}

We took a first optical spectrum of the Seyfert nucleus in HE\,1136-2304
with the 10~m Southern African Large Telescope (SALT)
nearly simultaneously with X-ray observations by XMM-Newton
on 2014 July 07, just after the X-ray flaring
 (Parker et al.\citealt{parker16}).
To study the subsequent variability behavior, we took additional optical
spectra  at 17 epochs
with the SALT telescope between  2014 December 25 and 2015 July 13.
To examine the long-term trend, 
four additional spectra were taken: three spectra in 2016 between January 12 and May 31 and one spectrum in 2017 on May 15. The log of our spectroscopic observations
with SALT is given in Table~\ref{saltlog}. The Julian dates in all tables 
mark the beginning of the observations.
We acquired 16 spectra between 2015 February and
2015 July with a mean interval of 9 days.
The spacing of our observations was not regular. The shortest time interval
between two subsequent observations was  four days.

All spectroscopic observations were taken under identical instrumental
conditions with the Robert Stobie Spectrograph
attached to the SALT telescope using the PG0900 grating.
The slit width was fixed to
2\arcsec\hspace*{-1ex}.\hspace*{0.3ex}0 projected onto the
sky at an optimized position angle to minimize differential refraction.
Furthermore, all observations were taken at the same air mass
thanks to the particular design feature of the SALT.
All spectra were taken with exposure times of 
10 to 20 minutes (see Table~\ref{saltlog}).
Typical seeing full width at half maximum (FWHM) values 
were 1 to 2 arcsec.

We covered the wavelength range from
4355 to 7230~\AA\  at a spectral resolution of 6.5~\AA\ .
The observed wavelength range corresponds to 
a wavelength range from 4240 to 7040~\AA\ in the rest frame of the galaxy. 
 There are two gaps in the spectrum caused by the gaps between the three CCDs:
one between the blue and the central CCD chip, and one between the
central and red CCD chip covering the wavelength ranges
5206--5263~\AA\  and 6254--6309~\AA\ (5069--5124~\AA\ and 6089--6142~\AA\
in the rest frame){.}
All spectra were wavelength corrected to the rest frame of the galaxy
(z=0.0271). 

In addition to the galaxy spectra, we also observed necessary flat-field and
Xe arc frames, as well as spectrophotometric standard stars for flux calibration
 (EG274, LTT3218, LTT7379).
The spatial
resolution per binned pixel was 0\arcsec\hspace*{-1ex}.\hspace*{0.3ex}2534
for our SALT spectra.
We extracted eight columns from our object spectrum 
corresponding to 2\arcsec\hspace*{-1ex}.\hspace*{0.3ex}03.
The reduction of the spectra (bias subtraction, cosmic ray correction,
flat-field correction, 2D wavelength calibration, night sky subtraction, and
flux calibration) was done in a homogeneous way with IRAF reduction
packages (e.g., Kollatschny et al.\citealt{kollatschny01}).  
We obtained typical S/N values of 40 in the continua of
 the galaxy spectra.

Great care was taken to ensure high-quality intensity and wavelength
calibrations to keep the intrinsic measurement errors very low, as described in
Kollatschny et al.\cite{kollatschny01,kollatschny03,kollatschny10}.
Our AGN spectra and  our
 calibration star spectra were not always taken
 under photometric conditions. 
Therefore,
all spectra were calibrated to the same absolute
[\ion{O}{iii}]\,$\lambda$5007 flux of
$1.75 \times 10^{-13} \rm erg\,s^{-1}\,cm^{-2}$ 
(Reimers et al.\citealt{reimers96}).
The flux of the narrow emission line [\ion{O}{iii}]\,$\lambda$5007
is considered to be constant on timescales of many years.
A relative flux  accuracy on the order of 1\% was achieved for most of
 our spectra.

\subsection{Optical, UV, and X-ray observations with \swift{}}

\begin{sidewaystable*}
\centering
\tabcolsep1.mm
\caption{\swift{} monitoring: V, B, U, UVOT W1, M2, and W2 observed flux 
densities in units of 10$^{-15}$\,erg\,s$^{-1}$\,cm$^{-2}$\,\AA$^{-1}$ (columns 2 to 7)
and magnitudes in the Vega system (columns 8 to 13).
}
\begin{tabular}{lcccccc|cccccccc}
\hline 
\noalign{\smallskip}
JD-2400000& V & B & U &UVW1  & UVM2 & UVW2  & {V} & {B} & {U} & {UV W1} & {UV M2} & {UVW2}   \\
\noalign{\smallskip}
(1) & (2) & (3) & (4) & (5) & (6) & (7) & (8) & (9) & (10) & (11) & (12) & (13)  \\  
\noalign{\smallskip}
\hline 
\noalign{\smallskip}
55350.7604    &             &                 &                 &                 & 1.12 $\pm$ 0.04 &                 &                    &                &        &               & 16.23\plm0.06 &                  \\     
55420.4688    &             &                 &                 & 1.00 $\pm$ 0.06 &              &                 &                   &                &        & 16.27\plm0.06 &               &                  \\          
55424.4167    &             &                 &                 & 0.93 $\pm$ 0.05 &              &                 &                   &                &        & 16.35\plm0.06 &               &                  \\          
\noalign{\smallskip}
\hline                                                                                                                
\noalign{\smallskip}
56833.6090    & 2.64 $\pm$ 0.11 & 2.35 $\pm$ 0.09 & 2.49 $\pm$ 0.12 & 1.92 $\pm$ 0.13 & 1.84 $\pm$ 0.09 & 1.83 $\pm$ 0.10 & 15.38\plm0.05 &  15.96\plm0.06 & 15.20\plm0.06 & 15.56\plm0.07 & 15.69\plm0.08 & 15.89\plm0.07    \\    
56840.3437    &             &                 &                 &                     &                  & 1.77 $\pm$ 0.07 &                   &             &               &               &               & 15.78\plm0.06    \\ 
56844.1424    & 2.46 $\pm$ 0.13 & 2.20 $\pm$ 0.10 & 2.35 $\pm$ 0.13 & 1.81 $\pm$ 0.14 & 1.70 $\pm$ 0.10 & 1.55 $\pm$ 0.09 & 15.35\plm0.06     &  16.04\plm0.06 & 15.26\plm0.06 & 15.63\plm0.08 & 15.77\plm0.09 & 16.08\plm0.07    \\       
56847.2729    &             &                 &                 &                 & 1.56 $\pm$ 0.06 &                  &                   &                &        &               & 15.86\plm0.07 &                  \\           
56850.9375    &             &                 &                 &                 & 1.61 $\pm$ 0.06 &                  &                   &                &        &               & 15.83\plm0.06 &                  \\       
56854.8646    &             &                 &                 &                 & 1.77 $\pm$ 0.07 &                  &                   &                &        &               & 15.72\plm0.07 &                  \\       
56861.9479    &             &                 &                 &                 & 2.14 $\pm$ 0.07 &                  &                   &                &        &               & 15.51\plm0.07 &                  \\       
56862.6007    &             &                 &                 &                 & 2.21 $\pm$ 0.08 &                  &                   &                &        &               & 15.49\plm0.06 &                  \\        
56866.9965    &             &                 &                 &                 & 2.11 $\pm$ 0.07 &                  &                   &                &        &               & 15.54\plm0.06 &                  \\       
56871.0729    &             &                 &                 &                 & 2.22 $\pm$ 0.08 &                  &                   &                &        &               & 15.48\plm0.06 &                  \\        
56874.1007    &             &                 &                 &                 & 2.06 $\pm$ 0.07 &                  &                   &                &        &               & 15.56\plm0.06 &                  \\        
56878.5347    &             &                 &                 &                 & 1.88 $\pm$ 0.08 &                  &                   &                &        &               & 15.66\plm0.07 &                  \\        
56882.7340    &             &                 &                 &                 & 1.67 $\pm$ 0.06 &                  &                   &                &        &               & 15.79\plm0.06 &                  \\      
\noalign{\smallskip}
\hline                                                                                                                
\noalign{\smallskip}
57085.3507    &             &                 & 2.02 $\pm$ 0.08 &                 &                  &                 &                   &                & 15.43\plm0.05 &        &               &                  \\           
57087.5208    &             &                 &                 & 1.73 $\pm$ 0.10 &              &                 &                   &                &        & 15.67\plm0.03 &               &                  \\        
\noalign{\smallskip}
\hline                                                                                                                
\noalign{\smallskip}
57132.3125    & 2.30 $\pm$ 0.11 & 1.94 $\pm$ 0.08 & 1.77 $\pm$ 0.10 & 1.44 $\pm$ 0.11 & 1.34 $\pm$ 0.08 & 1.34 $\pm$ 0.09 & 15.42\plm 0.06    &  16.17\plm0.06 & 15.57\plm0.06 & 15.87\plm0.08 & 16.03\plm0.09 & 16.24\plm0.08    \\
57138.7500    & 2.12 $\pm$ 0.09 & 1.69 $\pm$ 0.07 & 1.60 $\pm$ 0.08 & 1.13 $\pm$ 0.08 & 1.09 $\pm$ 0.06 & 1.03 $\pm$ 0.06 & 15.50\plm 0.05    &  16.32\plm0.06 & 15.68\plm0.06 & 16.13\plm0.08 & 16.25\plm0.08 & 16.52\plm0.07    \\
57145.8368    & 2.20 $\pm$ 0.11 & 1.75 $\pm$ 0.08 & 1.60 $\pm$ 0.09 & 1.20 $\pm$ 0.10 & 1.05 $\pm$ 0.07 & 1.10 $\pm$ 0.07 & 15.46\plm 0.06    &  16.28\plm0.06 & 15.68\plm0.06 & 16.07\plm0.08 & 16.29\plm0.09 & 16.45\plm0.08    \\  
57152.6875    & 2.13 $\pm$ 0.09 & 1.65 $\pm$ 0.06 & 1.32 $\pm$ 0.07 & 0.98 $\pm$ 0.07 & 1.01 $\pm$ 0.11 & 0.87 $\pm$ 0.05 & 15.50\plm 0.05    &  16.34\plm0.05 & 15.89\plm0.06 & 16.30\plm0.08 & 16.33\plm0.14 & 16.70\plm0.07    \\
57160.9028    & 2.12 $\pm$ 0.14 & 1.59 $\pm$ 0.09 & 1.45 $\pm$ 0.10 & 1.16 $\pm$ 0.12 & 0.88 $\pm$ 0.10 & 0.92 $\pm$ 0.08 & 15.51\plm 0.08    &  16.39\plm0.08 & 15.79\plm0.08 & 16.11\plm0.10 & 16.48\plm0.14 & 16.65\plm0.10    \\
57167.0833    & 2.09 $\pm$ 0.09 & 1.76 $\pm$ 0.07 & 1.57 $\pm$ 0.08 & 1.27 $\pm$ 0.09 & 1.02 $\pm$ 0.06 & 1.07 $\pm$ 0.06 & 15.53\plm 0.05    &  16.28\plm0.05 & 15.70\plm0.06 & 16.01\plm0.07 & 16.33\plm0.08 & 16.48\plm0.07    \\
57173.8160    & 2.05 $\pm$ 0.10 & 1.74 $\pm$ 0.08 & 1.39 $\pm$ 0.08 & 0.97 $\pm$ 0.08 & 0.90 $\pm$ 0.06 & 0.91 $\pm$ 0.06 & 15.56\plm 0.06    &  16.29\plm0.06 & 15.83\plm0.07 & 16.30\plm0.09 & 16.57\plm0.10 & 16.66\plm0.08    \\
57181.0000    & 1.97 $\pm$ 0.08 & 1.53 $\pm$ 0.06 & 1.19 $\pm$ 0.06 & 0.87 $\pm$ 0.06 & 0.67 $\pm$ 0.04 & 0.71 $\pm$ 0.04 & 15.59\plm 0.05    &  16.43\plm0.05 & 16.00\plm0.06 & 16.42\plm0.08 & 16.78\plm0.09 & 16.88\plm0.07    \\
57187.8056    & 2.01 $\pm$ 0.09 & 1.43 $\pm$ 0.06 & 1.09 $\pm$ 0.07 & 0.71 $\pm$ 0.06 & 0.60 $\pm$ 0.04 & 0.67 $\pm$ 0.04 & 15.56\plm 0.05    &  16.50\plm0.06 & 16.09\plm0.07 & 16.63\plm0.09 & 16.89\plm0.10 & 16.98\plm0.08    \\
57194.9236    & 1.98 $\pm$ 0.09 & 1.47 $\pm$ 0.07 & 1.05 $\pm$ 0.07 & 0.77 $\pm$ 0.06 & 0.60 $\pm$ 0.04 & 0.66 $\pm$ 0.04 & 15.58\plm 0.05    &  16.47\plm0.06 & 16.14\plm0.07 & 16.56\plm0.09 & 16.90\plm0.10 & 17.00\plm0.08    \\
57201.8785    & 1.95 $\pm$ 0.10 & 1.49 $\pm$ 0.07 & 1.18 $\pm$ 0.07 & 0.83 $\pm$ 0.07 & 0.77 $\pm$ 0.05 & 0.75 $\pm$ 0.05 & 15.60\plm 0.06    &  16.46\plm0.06 & 16.01\plm0.07 & 16.48\plm0.09 & 16.63\plm0.10 & 16.86\plm0.08    \\
57209.3986    & 2.06 $\pm$ 0.09 & 1.66 $\pm$ 0.06 & 1.32 $\pm$ 0.07 & 0.91 $\pm$ 0.07 & 0.79 $\pm$ 0.04 & 0.86 $\pm$ 0.05 & 15.54\plm 0.05    &  16.34\plm0.06 & 15.89\plm0.06 & 16.38\plm0.08 & 16.60\plm0.08 & 16.71\plm0.07    \\
57218.2188    & 1.88 $\pm$ 0.09 & 1.59 $\pm$ 0.07 & 1.34 $\pm$ 0.07 & 0.96 $\pm$ 0.07 & 0.78 $\pm$ 0.05 & 0.73 $\pm$ 0.05 & 15.64\plm 0.05    &  16.48\plm0.06 & 15.87\plm0.06 & 16.31\plm0.08 & 16.62\plm0.09 & 16.90\plm0.08    \\
57222.6285    & 1.98 $\pm$ 0.09 & 1.51 $\pm$ 0.06 & 1.21 $\pm$ 0.07 & 0.94 $\pm$ 0.07 & 0.82 $\pm$ 0.05 & 0.79 $\pm$ 0.05 & 15.58\plm 0.05    &  16.44\plm0.06 & 15.98\plm0.06 & 16.33\plm0.08 & 16.56\plm0.09 & 16.81\plm0.07    \\
57229.9167    & 2.03 $\pm$ 0.09 & 1.56 $\pm$ 0.07 & 1.25 $\pm$ 0.07 & 0.89 $\pm$ 0.07 & 0.69 $\pm$ 0.04 & 0.77 $\pm$ 0.05 & 15.55\plm 0.05    &  16.40\plm0.06 & 15.95\plm0.06 & 16.39\plm0.08 & 16.65\plm0.09 & 16.73\plm0.08    \\
57236.9687    & 2.09 $\pm$ 0.09 & 1.58 $\pm$ 0.07 & 1.28 $\pm$ 0.07 & 0.90 $\pm$ 0.07 & 0.86 $\pm$ 0.05 & 0.87 $\pm$ 0.06 & 15.52\plm 0.05    &  16.39\plm0.06 & 15.92\plm0.06 & 16.39\plm0.08 & 16.51\plm0.09 & 16.70\plm0.07    \\
\noalign{\smallskip}
\hline                                                                                                                
\noalign{\smallskip}
57337.5763    & 2.13 $\pm$ 0.10 & 1.95 $\pm$ 0.08 & 1.79 $\pm$ 0.09 & 1.44 $\pm$ 0.10 & 1.23 $\pm$ 0.07 & 1.29 $\pm$ 0.08 & 15.51\plm 0.05    &  16.17\plm0.06 & 15.56\plm0.06 & 15.88\plm0.07 & 16.11\plm0.08 & 16.28\plm0.07    \\  
\noalign{\smallskip}
\hline                                                                                                                
\noalign{\smallskip}
57421.3507    & 2.34 $\pm$ 0.10 & 2.00 $\pm$ 0.07 & 1.88 $\pm$ 0.09 & 1.41 $\pm$ 0.09 & 1.23 $\pm$ 0.07 & 1.24 $\pm$ 0.07 & 15.40\plm 0.05    &  16.13\plm0.05 & 15.50\plm0.06 & 15.90\plm0.07 & 16.12\plm0.08 & 16.32\plm0.07    \\
\noalign{\smallskip}
\hline 
\noalign{\smallskip}
\label{swiftmag}
\end{tabular}
\end{sidewaystable*}

\begin{table*}
\caption
{\swift{} monitoring: Julian date, UT date, XRT 0.3--10 keV count rates (CR)
and hardness ratios (HR$^1$), X-ray photon index $\Gamma$, the observed 
0.3--10 keV X-ray flux in units of $10^{-11}$ erg s$^{-1}$ cm$^{-2}$,
 reduced $\chi^2$ of the simple power-law model fit (pl),
X-ray photon index $\Gamma$ for an intrinsic absorber, and
 reduced $\chi^2$ for an intrinsic absorbed power-law model fit (zwa * pl)
of \he.}
\tabcolsep+1mm
\begin{tabular*}{\textwidth}{@{\extracolsep{\fill} } lcclcccccc}
\hline 
\noalign{\smallskip}
Julian Date &  \\
2\,400\,000+ & \rb{UT Date} &\rb{CR} & \,\quad\rb{HR} &    \rb{$\Gamma_{\rm pl}$} & \rb{XRT flux} & \rb{$(\chi^2/\nu)_{\rm pl}$}&\rb{$\Gamma_{\rm zwa * pl}$} & \rb{$N_{\rm H, intr}^2$}  & \rb{$(\chi^2/\nu)_{\rm zwa * pl}$}    \\
\hline 
55350.7604 &2010-06-02&  0.35\plm0.01 & 0.57\plm0.02     & 1.55\plm0.08 & 1.40 $\pm$ 0.06& 38.2/42     &  1.79\plm0.16 & 1.04\plm0.58 & 28.8/41  \\     
55420.4688 &2010-08-11&  0.24\plm0.01 & 0.44\plm0.06     & 1.54\plm0.16 & 1.19 $\pm$ 0.07& 10.1/14     &       ---     & 0$^3$        &  ---     \\       
55424.4167 &2010-08-15&  0.17\plm0.01 & 0.63\plm0.03     & 1.41\plm0.12 & 1.00 $\pm$ 0.05& 21.1/24     &  1.58\plm0.24 & 0.84\plm0.84 & 19.0/23  \\     
56833.6090 &2014-06-24&  0.46\plm0.02 & 0.53\plm0.04     & 1.58\plm0.10 & 1.87 $\pm$ 0.10& 48.7/26     &  1.94\plm0.23 & 1.31\plm0.73 & 39.3/25  \\     
56840.3437 &2014-07-01&  0.47\plm0.03 & 0.55\plm0.04     & 1.49\plm0.13 & 2.30 $\pm$ 0.16& 27.4/18     &  1.84\plm0.27 & 1.56\plm1.09 & 21.5/17  \\     
56844.1424 &2014-07-05&  0.46\plm0.03 & 0.61\plm0.03     & 1.37\plm0.12 & 2.25 $\pm$ 0.14& 20.7/18     &  1.68\plm0.26 & 1.46\plm1.05 & 15.2/17  \\      
56847.2729 &2014-07-08&  0.38\plm0.02 & 0.51\plm0.05     & 1.61\plm0.16 & 1.70 $\pm$ 0.15& 5.8/13      &       ---     & 0$^3$        & ---      \\     
56850.9375 &2014-07-11&  0.45\plm0.03 & 0.64\plm0.04     & 1.51\plm0.16 & 2.12 $\pm$ 0.24& 20.0/12     &  2.17\plm0.40 & 3.40\plm1.85 & 10.0/11  \\      
56854.8646 &2014-07-15&  0.35\plm0.02 & 0.44\plm0.03     & 1.44\plm0.20 & 1.65 $\pm$ 0.15& 12.1/9      &       ---     &    0$^3$     & ---      \\     
56861.9479 &2014-07-22&  0.61\plm0.03 & 0.57\plm0.03     & 1.43\plm0.11 & 2.64 $\pm$ 0.16& 30.4/23     &  1.63\plm0.22 & 0.89\plm0.81 & 27.0/22  \\     
56862.6007 &2014-07-23&  0.43\plm0.03 & 0.52\plm0.05     & 1.78\plm0.24 & 1.94 $\pm$ 0.16& 6.5/8       &       ---     &     0$^3$    & ---      \\     
56866.9965 &2014-07-27&  0.71\plm0.04 & 0.55\plm0.05     & 1.53\plm0.11 & 2.96 $\pm$ 0.17& 36.7/24     &  2.04\plm0.24 & 1.95\plm0.83 & 19.1/23  \\     
56871.0729 &2014-08-01&  0.70\plm0.04 & 0.40\plm0.05$^4$ & 1.61\plm0.11 & 3.02 $\pm$ 0.20& 37.8/27     &  2.06\plm0.25 & 1.45\plm0.70 & 24.0/26  \\   
56874.1007 &2014-08-04&  0.36\plm0.04 & 0.40\plm0.05$^4$ & 1.56\plm0.11 & 2.61 $\pm$ 0.14& 30.1/26     &  1.97\plm0.23 & 1.34\plm0.65 & 16.72/25 \\   
56878.5347 &2014-08-08&  0.39\plm0.03 & 0.54\plm0.07$^4$ & 1.48\plm0.17 & 2.09 $\pm$ 0.25& 14.8/14     &  1.84\plm0.34 & 1.51\plm1.25 & 10.5/13  \\    
56882.7340 &2014-08-12&  0.37\plm0.04 & 0.62\plm0.10$^4$ & 1.27\plm0.23 & 3.53 $\pm$ 0.50& 11.2/15     &      ---      &    0$^3$     & ---      \\      
57085.3507 &2015-03-03&  0.49\plm0.03 & 0.51\plm0.04     & 1.51\plm0.15 & 2.35 $\pm$ 0.20& 19.2/17     &      ---      &    0$^3$     & ---      \\      
57087.5208 &2015-03-05&  0.74\plm0.04 & 0.53\plm0.04     & 1.46\plm0.16 & 3.75 $\pm$ 0.29& 15.4/13     &  1.90\plm0.34 & 2.12\plm1.48 & 9.6/12   \\      
57132.3125 &2015-04-19&  0.28\plm0.02 & 0.52\plm0.05     & 1.67\plm0.19 & 1.18 $\pm$ 0.10& 13.9/11     &  1.80\plm0.28 & 0.68\plm0.64 & 13.5/10  \\      
57138.7500 &2015-04-25&  0.17\plm0.01 & 0.56\plm0.04     & 1.64\plm0.17 & 0.78 $\pm$ 0.07& 13.7/11     &  1.87\plm0.35 & 1.45\plm1.45 & 12.2/13  \\      
57145.8368 &2015-05-02&  0.32\plm0.02 & 0.53\plm0.05     & 1.44\plm0.17 & 1.56 $\pm$ 0.14& 15.2/14     &  1.48\plm0.18 & 0.21\plm0.21 & 15.2/13  \\      
57152.6875 &2015-05-09&  0.10\plm0.01 & 0.55\plm0.06     & 1.45\plm0.25 & 0.48 $\pm$ 0.10& 1.0/6       &      ---      &    0$^3$     & ---      \\       
57160.9028 &2015-05-17&  0.30\plm0.03 & 0.57\plm0.06     & 1.29\plm0.19 & 1.71 $\pm$ 0.15& 93.6/130$^5$&      ---      &    0$^3$     & ---      \\
57167.0833 &2015-05-24&  0.42\plm0.02 & 0.57\plm0.03     & 1.47\plm0.08 & 2.07 $\pm$ 0.10& 54.4/36     &  1.85\plm0.17 & 1.46\plm0.57 & 33.3/35  \\       
57173.8160 &2015-05-30&  0.24\plm0.01 & 0.60\plm0.05     & 1.34\plm0.17 & 1.33 $\pm$ 0.12& 7.1/11      &  1.80\plm0.39 & 2.38\plm1.80 & 2.1/10   \\        
57181.0000 &2015-06-07&  0.16\plm0.01 & 0.65\plm0.04     & 1.51\plm0.13 & 0.77 $\pm$ 0.05& 26.1/16     &  2.12\plm0.29 & 3.61\plm1.56 & 8.9/15   \\       
57187.8056 &2015-06-13&  0.09\plm0.01 & 0.56\plm0.06     & 1.59\plm0.20 & 0.48 $\pm$ 0.05& 118/119$^5$ &  2.06\plm0.38 & 1.89\plm1.28 & 112/118$^5$ \\  
57194.9236 &2015-06-21&  0.13\plm0.01 & 0.52\plm0.06     & 1.36\plm0.18 & 0.61 $\pm$ 0.08& 137/142$^5$ &      ---      &    0$^3$     & ---      \\  
57201.8785 &2015-06-27&  0.15\plm0.01 & 0.59\plm0.05     & 1.52\plm0.25 & 0.75 $\pm$ 0.11& 8.5/7       &      ---      &    0$^3$     & ---      \\   
57209.3986 &2015-07-05&  0.16\plm0.01 & 0.56\plm0.05     & 1.54\plm0.19 & 0.76 $\pm$ 0.07& 12.5/14     &      ---      &    0$^3$     & ---      \\   
57218.2188 &2015-07-14&  0.21\plm0.01 & 0.64\plm0.04     & 1.57\plm0.15 & 1.02 $\pm$ 0.10& 13.9/16     &  1.89\plm0.30 & 1.95\plm1.58 & 9.7/15   \\   
57222.6285 &2015-07-18&  0.15\plm0.01 & 0.67\plm0.05     & 1.33\plm0.19 & 0.83 $\pm$ 0.12& 14.9/10     &  1.80\plm0.44 & 3.41\plm2.88 & 11.0/9   \\   
57229.9167 &2015-07-25&  0.16\plm0.01 & 0.65\plm0.05     & 1.21\plm0.15 & 0.86 $\pm$ 0.10& 16.9/11     &  1.57\plm0.31 & 2.44\plm1.88 & 12.0/10  \\   
57236.9687 &2015-08-01&  0.15\plm0.01 & 0.60\plm0.05     & 1.52\plm0.18 & 0.69 $\pm$ 0.05& 16.5/10     &      ---      &    0$^3$     & ---      \\   
57337.5763 &2015-11-10&  0.37\plm0.02 & 0.57\plm0.04     & 1.52\plm0.11 & 1.72 $\pm$ 0.09& 25.1/27     &  1.74\plm0.20 & 1.07\plm0.83 & 20.4/26   \\   
57421.3507 &2016-02-02&  0.25\plm0.01 & 0.29\plm0.03     & 1.47\plm0.12 & 1.33 $\pm$ 0.08& 21.1/18     &  1.73\plm0.21 & 1.22\plm0.85 & 15.3/17   \\    
\hline 
\end{tabular*}
$^1$ The hardness ratio is defined as HR = $\frac{hard-soft}{hard+soft}$ where {\it soft} and {\it hard} are the background corrected counts in the 0.3--1.0 keV and 1.0--10.0 keV bands, respectively 

$^2$ The intrinsic $N_{\rm H, intr}$ is given in units of $10^{21}$ cm$^{-2}$

$^3$ No additional absorber required. The fit is consistent with  Galactic absorption alone. 

$^4$ These observations were performed in windowed timing mode. 

$^5$ Fit using Cash Statistics (Cash\citealt{cash79})
\label{swiftdata} 
\end{table*}

After the discovery of the X-ray flaring in June 2014 (Parker et al.\citealt{parker16}), we started monitoring \he\ 
with \swift\ 
(Gehrels et al.\citealt{gehrels04}) in X-rays and the UV/optical. All \swift\ 
observing dates and exposure times are listed in Table\,\ref{swiftlog}. 
In this paper, we focus on the \swift\ observations between 2014 June 06 and 2016 February 02. 
However, \he\ had been observed previously by \swift\ during three epochs in
2010. For comparison purposes,
we list these observations in all the tables relevant to \swift\ data 
(see Tables\, \ref{swiftmag}, \ref{swiftdata}, and \ref{swiftlog}).

 Most X-ray observations with the \swift\ X-ray Telescope (XRT, Burrows et al.\citealt{burrows05}) were performed in photon counting mode (pc-mode, Hill et al.\citealt{hill04}). However, the four observations in 2014 August  were performed in windowed timing mode.
For the pc-mode data, 
source counts were collected in a circular region with a radius of 30 pixels (equivalent to 70$^{''}$) and background counts in a nearby source-free  circular region with a radius of 90 pixels (equal to 210${''}$). The windowed timing source and background spectra were selected in boxes with a width of 40 pixels each. 
Spectra were extracted with the FTOOL {\it XSELECT}. An auxiliary response file (ARF) was created for each observation using  {\it xrtmkarf}. We applied the \swift\ XRT response file {\it swxpc0to12s6\_20130101v014.rmf} and {\it swxwt0to2s6\_20131212v015.rmf} for the pc and WT data, respectively. 
Most spectra were rebinned to have at least 20 counts per bin using {\it
  grppha}. For some spectra the number of counts was too low to allow $\chi^2$
statistics. These data were fitted by Cash statistics (Cash\citealt{cash79}).
The spectral analysis was performed in {\it XSPEC}
(Arnaud\citealt{arnaud96}).

We fitted the X-ray spectra first with a simple power-law model with the
 absorption parameter fixed to the Galactic value. In addition, we fitted a
 power-law model with redshifted intrinsic absorption (zwa) to the data with
 the redshift fixed to the redshift of HE~1136-2304.
For some spectra we found some evidence of a low intrinsic absorption
 on the order of $1\times 10^{21}$ cm$^{-2}$; however,  in most cases the absorption
column density of the absorber was consistent with the Galactic value and
the fits did not require any additional absorber. 
Finally, all spectra were fitted with a single power-law
 model with Galactic absorption ($N_{\rm H,gal} = 3.3\times 10^{20}$ cm$^{-2}$;
Kalberla et al. \citealt{kalberla05}). As indicated in Table\,\ref{swiftdata},
we also fitted the data with a redshifted intrinsic absorber model.

Count rates, hardness ratios, and the best fit values obtained are listed
in Table\,\ref{swiftdata}. 
The hardness ratio is defined as 
HR = counts(0.3--1.0 keV)/counts(1.0--10.0 keV).
In order to determine a background corrected hardness ratio, we
applied the program  {\it BEHR} by Park et al.\cite{park06}. 

During most observations, the \swift\ UV-Optical Telescope (UVOT, Roming et
al. \citealt{roming05}) observed in all six photometric filters UVW2 (1928
\AA{}), UVM2 (2246 \AA{}),  UVW1 (2600 \AA{}), u (3465\AA{}), b (4392 \AA{}),
and v (5468 \AA{}). Before analyzing the data, all snapshots in one segment were combined with the UVOT tool {\it uvotimsum}.
The flux densities and magnitudes in each filter were determined by the tool {\it uvotsource} using the count rate conversion and 
calibration, as described in  Poole et al.\citealt{poole08} and  Breeveld et al.\citealt{breeveld10}. 
Source counts were extracted in a circle with a radius of $5^{''}$ and
background counts in a nearby source-free region with a radius of
20$^{''}$. The UVOT fluxes listed in Table \ref{swiftmag} are not corrected for Galactic reddening. The reddening value in the direction of \he{} is
E$_\text{B-V}$ = 0.03666,  deduced from
 the Schlafly \& Finkbeiner\cite{schlafly11} re-calibration of the 
Schlegel et al.\cite{schlegel98} infrared-based dust map. Applying equation 2 in Roming et al.\cite{roming09}, 
who used the standard reddening correction curves by Cardelli et al.\cite{cardelli89}, we calculated the following 
 magnitude corrections:
v$_\text{corr}$ = 0.110, b$_\text{corr}$ = 0.143, u$_\text{corr}$ = 0.180,
UVW1$_\text{corr}$ = 0.226, UVM2$_\text{corr}$ = 0.324, UVW2$_\text{corr}$ = 0.270.
For all \swift\, UVOT magnitudes used in this publication we adopted the Vega magnitude system.

\subsection{Optical photometry with the MONET  North and South
 telescopes}

Additional optical B-, V-, and R-band photometric data were collected  with
the 1.2m MONET/North telescope between 2014 November 17 and 2015 February 04
and with the twin MONET/South telescope between 2016 April 25 and June 21.
Table~\ref{monetlog} lists the Julian dates of the MONET observations.
The MONET/North and South  
telescopes are located at McDonald Observatory in  Texas, USA,  and 
Sutherland, South Africa, respectively. 
The data were obtained with the MONET browser-based remote-observing 
interface. The photometric data were taken with a SBIG STF-8300M CCD camera 
at MONET/North and with a Spectral Instruments 1100 CCD camera at MONET/South.
Typical exposure times were 60\,s and 120\,s. The photometry was performed 
relative to three comparison stars approximately 2.0 arcmin west of 
HE\,1136-2304. 

\subsection{Optical photometry with the Bochum telescopes at Cerro Armazones}

Between 2015 April and 2016 July HE1136$-$2304 was monitored in the B and V bands and in a  narrowband
filter NB$_{670}$ covering the redshifted H$\alpha$ line 
using the 40 cm Bochum Monitoring Telescope (BMT)
of the Universit\"atssternwarte Bochum near 
Cerro Armazones, Chile (Ramolla et al.\citealt{ramolla13}). 
Per night and broadband filter 15 dithered 60 s
exposures with a size of 41.2$\arcsec$ $\times$ 27$\arcsec$ were obtained;
for the narrowband filter NB$_{670}$ we took 25 exposures. 

We performed additional B-band monitoring using the 25 cm 
Berlin Exoplanet Search Telescope-II 
(BEST-II\footnote{https://www.astro.rub.de/Astrophysik/BESTII.html})
with 1.7$^{\circ}$  $\times$ 1.7$^{\circ}$ FoV
(Kabath et al.\citealt{kabath09}), 
also located at the Universit\"atssternwarte Bochum.
Per night 15 dithered 60 s exposures were obtained.
We performed standard data reduction including corrections
for bias, dark current, flatfield, astrometry, and astrometric
distortion before combining the dithered images, separated by telescope,
night, and filter. As in Pozo Nu\~nez et al.\cite{pozonunez15}  
a 7$\farcs$5 diameter aperture was used to extract the
photometry and to create flux-normalized light curves relative
to 15 nonvariable stars located in the same images, within 10$\arcmin$
around HE1136$-$2304, and of similar brightness to HE1136$-$2304. Absolute
calibration was performed using standard reference stars from
Landolt\cite{landolt09} 
observed on the same nights as the AGN. We
also corrected for atmospheric (Patat et al.\citealt{patat11}) 
and Galactic
foreground extinction (Schlafly \& Finkbeiner\citealt{schlafly11}).

A list of the photometric
observations is given in
Table~\ref{bochumlog}.

\section{Results}
\begin{figure*}[h]
\centering
 \includegraphics[width=16.9cm,angle=0]{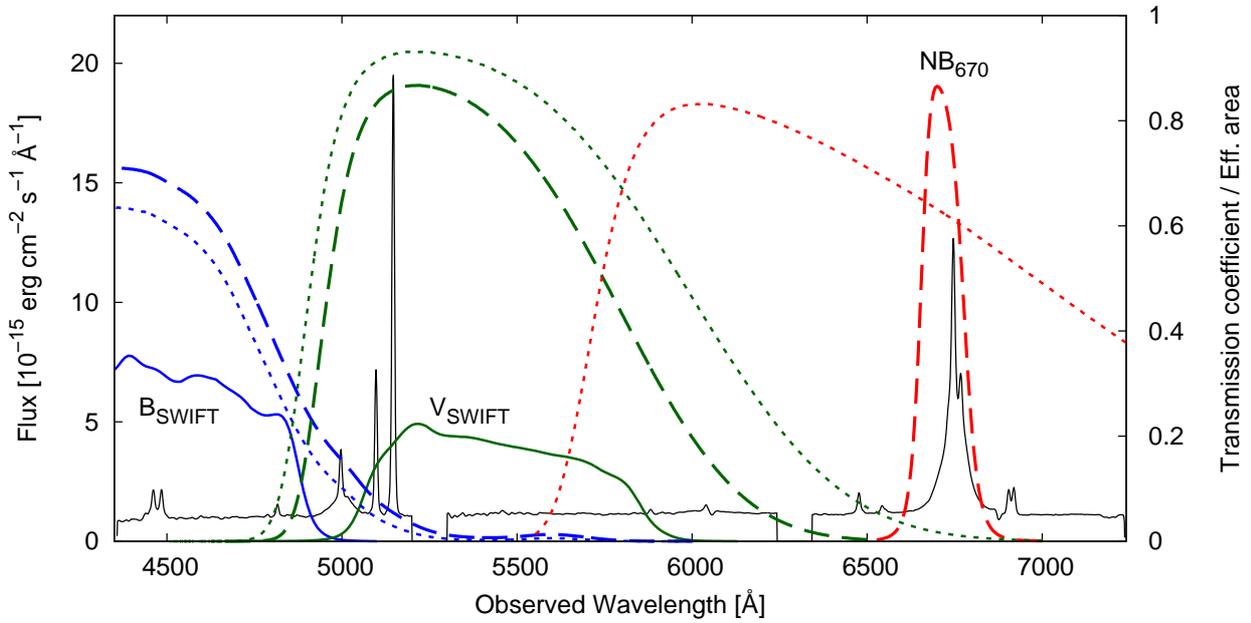}
      \caption{Mean spectrum of HE\,1136-2304 based on the
     observations performed with the SALT telescope together with the
      effective areas of the \swift{}
      B and V bandpass curves (in units of cm$^2$ divided by 100)
      (solid lines);
      the Bochum B, V, and NB$_\text{670}$ filter curves
      (dashed lines); and the MONET B, V, and R
      filter curves (dotted lines).
                          }
       \vspace*{-3mm} 
         \label{filter_format.eps}
   \end{figure*}

Figure~\ref{filter_format.eps}
shows the mean spectrum of HE 1136-2304 based
on our SALT variability campaign in 2015 together with the
\swift{} B- and V-band filter curves; the Bochum B, V, and NB$_{670}$ 
filter curves; and the MONET B, V, and R filter curves. The
B-band filter curves are shown in blue, the V-band filter curves
are given in green, and the R-band filter curves in red.
The fluxes in the filter bands are contaminated by both constant
and variable emission line contributions.
\subsection{Optical, UV, and X-ray continuum variations}
\begin{figure*}
    \centering
    \includegraphics[width=8.5cm,angle=-90]{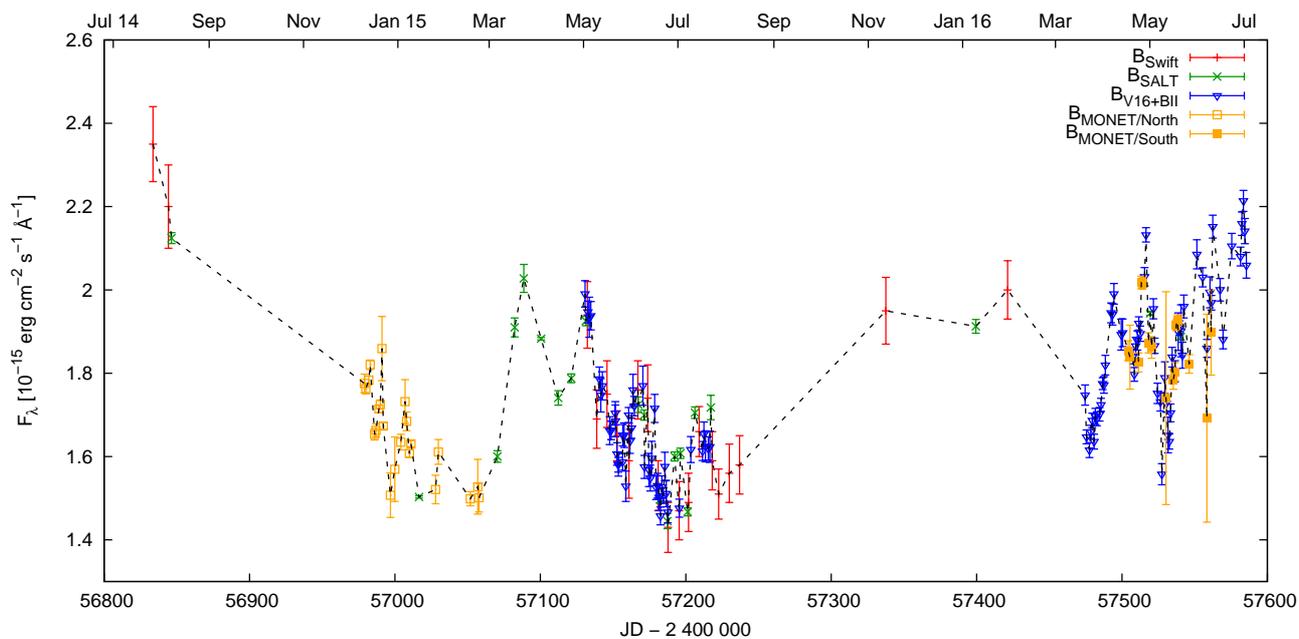}
      \caption{Combined B-band continuum light curve (\swift{}, SALT, MONET,
          VYSOS-16, BEST II) calibrated with respect to the \swift{} 
             data from 2014 to  2016. The time stamps at the top indicate the
 first day of the month. 
              }
       \vspace*{-3mm} 
         \label{LC_B_all_ochm.ps}
   \end{figure*}

\begin{figure*}
\centering
 \includegraphics[width=8.5cm,angle=-90]{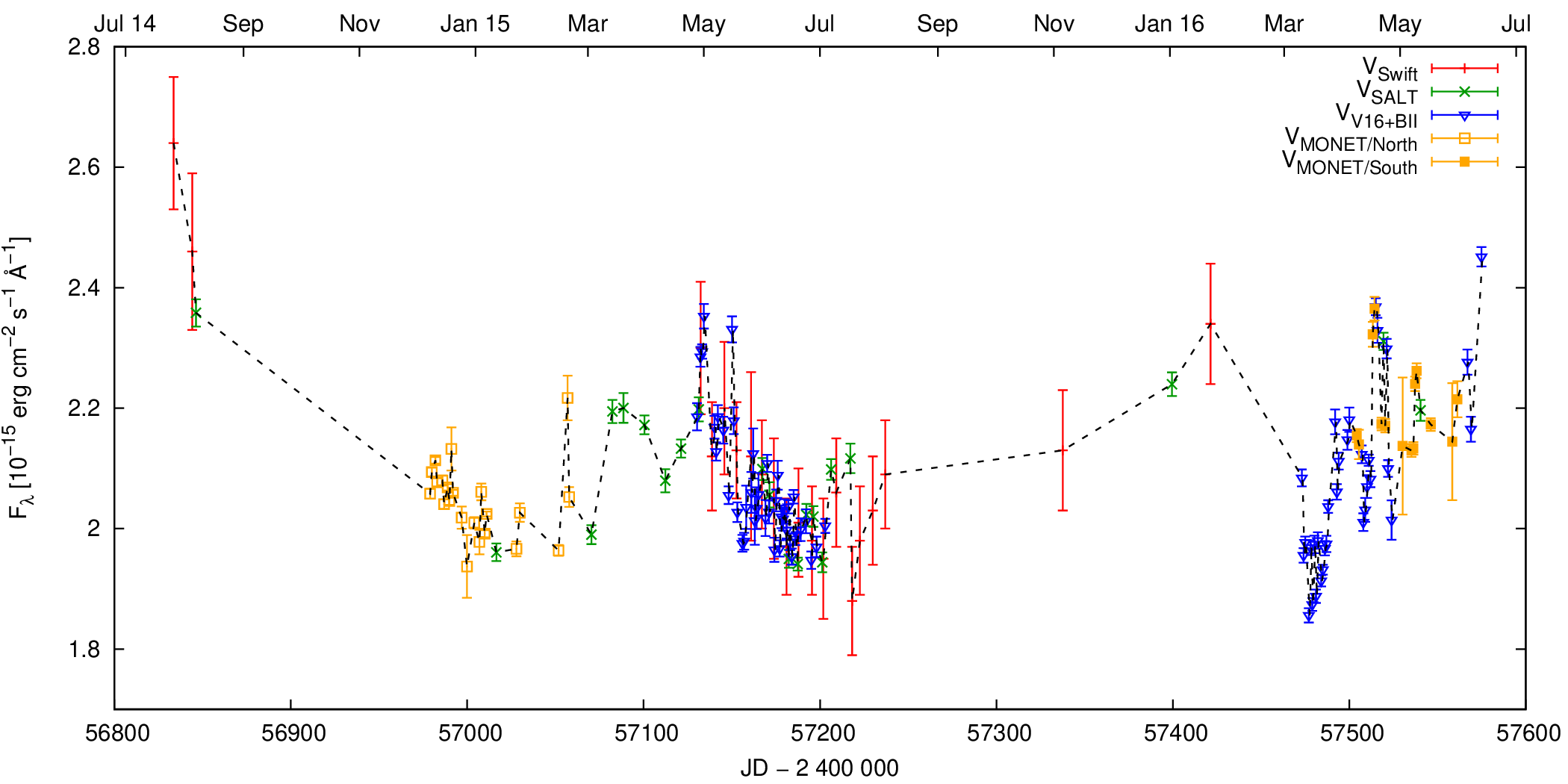}
      \caption{Combined V-band continuum light curve (\swift{}, SALT, MONET,
          VYSOS-16, BEST II) calibrated with respect to the \swift{} data from
           2014 to 2016. The time stamps at the top indicate
the
 first day of the month.  
                       }
       \vspace*{-3mm} 
         \label{LC_V_all_ochm.ps}
   \end{figure*}

First we created optical B- and V-band light curves based on the absolute
calibration of the \swift{} data. Then we generated B and V light curves  
by measuring the continuum flux in the SALT spectra at 4570 \AA{} and 5360
\AA{} in the rest frame.
Afterwards we created light curves based on the B- and V-band intensities
taken with the MONET telescopes. Additionally, we created light curves based
on the B and V photometry observed at Cerro Armazones. We intercalibrated
all these light curves to the B- and V-band \swift{} data
(Table\, \ref{optcontfluxlog}).
The fluxes in these light curves are
not corrected for Galactic absorption.
We applied a multiplicative scale factor and an additional flux adjustment
component to put the light curves on the same scale and to correct
for differences in the host galaxy contribution.
These differences are caused by different aperture sizes
and by the different instruments attached to our telescopes.
Figures~\ref{LC_B_all_ochm.ps} and \ref{LC_V_all_ochm.ps}
show the combined B- and V-band continuum light curves of HE\,1136-2304  
from 2014 to 2016.
Overall, there is a good agreement
between the light curves from the different telescopes.

\begin{figure*}
\centering
\includegraphics[width=13.9cm,angle=0]{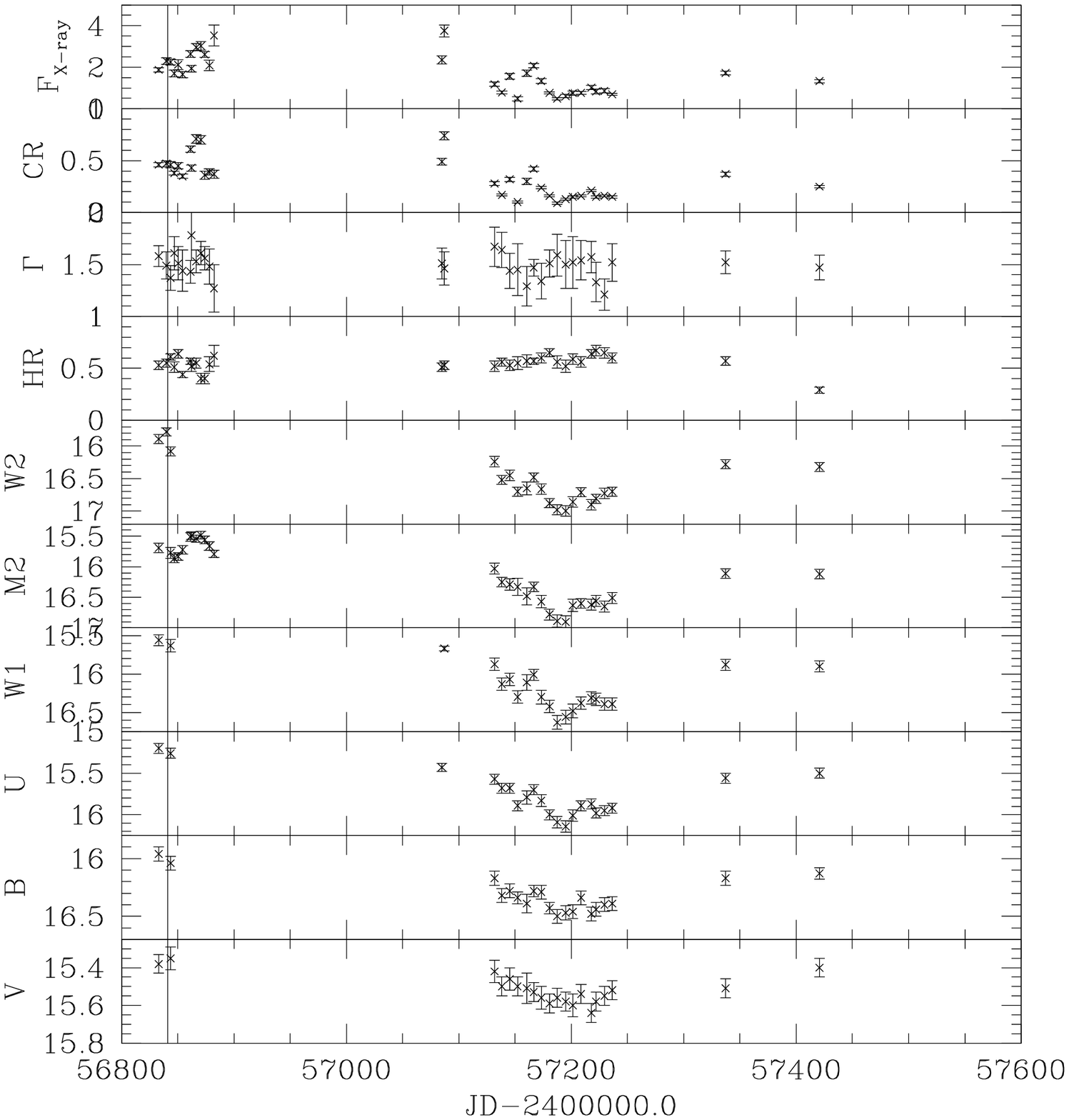}
\caption{\swift{} X-ray light curve from 2014 to 2016. 
The solid line at JD 2456841 marks the time of the XMM observation discussed in 
Parker et al. \cite{parker16}. 
The observed 0.3--10 keV X-ray flux is given in units of  $10^{-11}$ ergs s$^{-1}$ cm$^{-2}$. The X-ray count rates (CR), the X-ray photon
index $\Gamma$ of a simple power-law model, and the hardness ratios (HR) are also shown. 
The UVOT  W2, M2, W1, U, B, and V magnitudes are given in the Vega system.
       }
\label{lc_swift_2014_2016}
\end{figure*}

Optical  and 0.3--10 keV \swift{} light curves are shown in
 Figure~\ref{lc_swift_2014_2016} along with
the X-ray photon index values 
and the hardness ratios (see Sect. 2.2). All measurements are listed in 
Table\,\ref{swiftdata}.  The X-ray 0.3--10 keV flux and count rate are
clearly variable by a factor of about 5. 
 We also checked whether there is any significant variability in the hardness ratio 
and photon index $\Gamma$.
The 2015 data may suggest  hardening. However, testing whether there is any
correlation between the count rate and the hardness ratio and $\Gamma$,
we only found a
weak trend with a Spearman rank order correlation coefficient of -0.30 between
the count rate and hardness ratio,
but with a probability of 6\% that this result is just random. This random
result is confirmed when checking the correlation between the count rate and
$\Gamma,$ which results in a Spearman rank order correlation coefficient of
$r_{\rm s} = 0.06$ with a probability P = 0.74 of a random result. 
We therefore
conclude that there is no obvious connection with the X-ray flux/count rate
variability and the variability of the hardness ratio. The distribution of
the hardness ratios is almost Gaussian.

%
\begin{longtable}{lllllll}
\caption{Julian date, UT date, and B- and V-band fluxes taken with the
 MONET North/South (MN/MS) and  VYSOS 16 (V16) telescopes, and with the \swift{} satellite (SW),
 as well as corresponding continuum fluxes taken with SALT (S).}\\
  \hline 
Julian Date  & UT Date& Cont. Flux& Julian Date  & UT Date& Cont. Flux& Tel. \\
2\,400\,000+&         &   B band, 4570\,\AA& 2\,400\,000+&         &  V band, 5360\,\AA &   \\
 \hline 
  \endfirsthead
\caption{continued.}\\
  \hline 
 Julian Date  & UT Date& Cont. Flux& Julian Date  & UT Date& Cont. Flux& Tel. \\
2\,400\,000+&         &  4570\,\AA, B band& 2\,400\,000+&         &  5360\,\AA,
V band&   \\
 \hline 
\endhead
\hline
\multicolumn{3}{l}{Continuum flux in units of  10$^{-15}$\,erg\,s$^{-1}$\,cm$^{-2}$\,\AA$^{-1}$.}\\
\endfoot
49066.500       &       1993-03-20      &       1.480   $\pm$   0.089   &       49066.500       &       1993-03-20      &       2.046   $\pm$   0.061   &       ESO     \\
52410.920       &       2002-05-16      &       2.051   $\pm$   0.089   &       52410.920       &       2002-05-16      &       2.130   $\pm$   0.031   &       6dF     \\
56833.609       &       2014-06-25      &       2.350   $\pm$   0.090   &       56833.609       &       2014-06-25      &       2.640   $\pm$   0.110   &       SW      \\
56844.142       &       2014-07-05      &       2.200   $\pm$   0.101   &       56844.142       &       2014-07-05      &       2.460   $\pm$   0.130   &       SW      \\
56846.248       &       2014-07-07      &       2.124   $\pm$   0.013   &       56846.248       &       2014-07-07      &       2.358   $\pm$   0.024   &       S         \\
56979.010       &       2014-11-17      &       1.776   $\pm$   0.023   &       56979.011       &       2014-11-17      &       2.058   $\pm$   0.009   &       MN      \\
56979.968       &       2014-11-18      &       1.763   $\pm$   0.013   &       56979.969       &       2014-11-18      &       2.094   $\pm$   0.008   &       MN      \\
56981.980       &       2014-11-20      &       1.783   $\pm$   0.011   &       56981.981       &       2014-11-20      &       2.113   $\pm$   0.006   &       MN      \\
56982.971       &       2014-11-21      &       1.820   $\pm$   0.011   &       56982.972       &       2014-11-21      &       2.077   $\pm$   0.007   &       MN      \\
56985.947       &       2014-11-24      &       1.653   $\pm$   0.014   &       56985.948       &       2014-11-24      &       2.080   $\pm$   0.008   &       MN      \\
56986.936       &       2014-11-25      &       1.663   $\pm$   0.016   &       56986.937       &       2014-11-25      &       2.041   $\pm$   0.008   &       MN      \\
56988.945       &       2014-11-27      &       1.714   $\pm$   0.009   &       56988.946       &       2014-11-27      &       2.069   $\pm$   0.005   &       MN      \\
56989.936       &       2014-11-28      &       1.725   $\pm$   0.011   &       56989.937       &       2014-11-28      &       2.046   $\pm$   0.006   &       MN      \\
56991.034       &       2014-11-29      &       1.859   $\pm$   0.078   &       56991.035       &       2014-11-29      &       2.132   $\pm$   0.036   &       MN      \\
56991.987       &       2014-11-30      &       1.673   $\pm$   0.009   &       56991.988       &       2014-11-30      &       2.060   $\pm$   0.005   &       MN      \\
56996.924       &       2014-12-05      &       1.507   $\pm$   0.054   &       56996.925       &       2014-12-05      &       2.018   $\pm$   0.019   &       MN      \\
56999.890       &       2014-12-08      &       1.570   $\pm$   0.078   &       56999.892       &       2014-12-08      &       1.937   $\pm$   0.052   &       MN      \\
57004.007       &       2014-12-12      &       1.635   $\pm$   0.019   &       57004.008       &       2014-12-12      &       2.010   $\pm$   0.010   &       MN      \\
57006.928       &       2014-12-15      &       1.732   $\pm$   0.054   &       57006.929       &       2014-12-15      &       1.978   $\pm$   0.021   &       MN      \\
57008.023       &       2014-12-16      &       1.685   $\pm$   0.023   &       57008.025       &       2014-12-16      &       2.061   $\pm$   0.014   &       MN      \\
57009.928       &       2014-12-18      &       1.607   $\pm$   0.009   &       57009.929       &       2014-12-18      &       1.992   $\pm$   0.006   &       MN      \\
57011.018       &       2014-12-19      &       1.631   $\pm$   0.008   &       57011.019       &       2014-12-19      &       2.024   $\pm$   0.005   &       MN      \\
57016.559       &       2014-12-25      &       1.503   $\pm$   0.004   &       57016.559       &       2014-12-25      &       1.961   $\pm$   0.015   &       S         \\
57027.909       &       2015-01-05      &       1.521   $\pm$   0.034   &       57027.910       &       2015-01-05      &       1.966   $\pm$   0.013   &       MN      \\
57029.911       &       2015-01-07      &       1.611   $\pm$   0.030   &       57029.912       &       2015-01-07      &       2.027   $\pm$   0.016   &       MN      \\
57051.844       &       2015-01-29      &       1.499   $\pm$   0.018   &       57051.845       &       2015-01-29      &       1.964   $\pm$   0.009   &       MN      \\
57056.946       &       2015-02-03      &       1.528   $\pm$   0.066   &       57056.947       &       2015-02-03      &       2.217   $\pm$   0.038   &       MN      \\
57057.827       &       2015-02-04      &       1.501   $\pm$   0.035   &       57057.828       &       2015-02-04      &       2.052   $\pm$   0.017   &       MN      \\
57070.399       &       2015-02-16      &       1.600   $\pm$   0.014   &       57070.399       &       2015-02-16      &       1.990   $\pm$   0.016   &       S         \\
57082.362       &       2015-02-28      &       1.910   $\pm$   0.023   &       57082.362       &       2015-02-28      &       2.195   $\pm$   0.020   &       S         \\
57088.594       &       2015-03-07      &       2.028   $\pm$   0.034   &       57088.594       &       2015-03-07      &       2.200   $\pm$   0.025   &       S         \\
57100.539       &       2015-03-19      &       1.883   $\pm$   0.004   &       57100.539       &       2015-03-19      &       2.172   $\pm$   0.016   &       S         \\
57112.285       &       2015-03-30      &       1.741   $\pm$   0.018   &       57112.285       &       2015-03-30      &       2.080   $\pm$   0.020   &       S         \\
57121.256       &       2015-04-08      &       1.788   $\pm$   0.011   &       57121.256       &       2015-04-08      &       2.133   $\pm$   0.015   &       S         \\
57130.736       &       2015-04-18      &       1.990   $\pm$   0.032   &       57130.755       &       2015-04-18      &       2.186   $\pm$   0.024   &       V16     \\
57131.243       &       2015-04-18      &       1.925   $\pm$   0.011   &       57131.243       &       2015-04-18      &       2.198   $\pm$   0.021   &       S         \\
57132.313       &       2015-04-19      &       1.940   $\pm$   0.081   &       57132.313       &       2015-04-19      &       2.300   $\pm$   0.110   &       SW      \\
57132.712       &       2015-04-20      &       1.947   $\pm$   0.026   &       57132.731       &       2015-04-20      &       2.285   $\pm$   0.017   &       V16     \\
57133.659       &       2015-04-21      &       1.934   $\pm$   0.048   &       57133.677       &       2015-04-21      &       2.294   $\pm$   0.013   &       V16     \\
57134.681       &       2015-04-22      &       1.938   $\pm$   0.034   &       57134.700       &       2015-04-22      &       2.353   $\pm$   0.021   &       V16     \\
57138.750       &       2015-04-26      &       1.690   $\pm$   0.070   &       57138.750       &       2015-04-26      &       2.120   $\pm$   0.090   &       SW      \\
57140.653       &       2015-04-28      &       1.786   $\pm$   0.029   &       57140.672       &       2015-04-28      &       2.168   $\pm$   0.019   &       V16     \\
57141.604       &       2015-04-29      &       1.747   $\pm$   0.040   &       57141.622       &       2015-04-29      &       2.128   $\pm$   0.015   &       V16     \\
57142.602       &       2015-04-30      &       1.770   $\pm$   0.034   &       57142.620       &       2015-04-30      &       2.186   $\pm$   0.020   &       V16     \\
        &               &                               &       57145.670       &       2015-05-03      &       2.163   $\pm$   0.023   &       V16     \\
57145.837       &       2015-05-03      &       1.750   $\pm$   0.081   &       57145.837       &       2015-05-03      &       2.200   $\pm$   0.110   &       SW      \\
57147.651       &       2015-05-05      &       1.662   $\pm$   0.034   &               &               &                               &       V16     \\
57148.656       &       2015-05-06      &       1.663   $\pm$   0.023   &       57148.675       &       2015-05-06      &       2.056   $\pm$   0.015   &       V16     \\
        &               &                               &       57150.644       &       2015-05-08      &       2.331   $\pm$   0.022   &       V16     \\
57151.581       &       2015-05-09      &       1.704   $\pm$   0.029   &       57151.600       &       2015-05-09      &       2.179   $\pm$   0.023   &       V16     \\
57151.663       &       2015-05-09      &       1.686   $\pm$   0.039   &               &               &                               &       BII     \\
57152.688       &       2015-05-10      &       1.650   $\pm$   0.061   &       57152.688       &       2015-05-10      &       2.130   $\pm$   0.080   &       SW      \\
57152.727       &       2015-05-10      &       1.606   $\pm$   0.027   &               &               &                               &       BII     \\
57153.585       &       2015-05-11      &       1.580   $\pm$   0.018   &       57153.603       &       2015-05-11      &       2.027   $\pm$   0.016   &       V16     \\
57153.636       &       2015-05-11      &       1.582   $\pm$   0.029   &               &               &                               &       BII     \\
57156.555       &       2015-05-14      &       1.587   $\pm$   0.020   &       57156.574       &       2015-05-14      &       1.976   $\pm$   0.014   &       V16     \\
57156.667       &       2015-05-14      &       1.651   $\pm$   0.026   &               &               &                               &       BII     \\
57157.663       &       2015-05-15      &       1.651   $\pm$   0.030   &       57157.556       &       2015-05-15      &       1.979   $\pm$   0.014   &       BII, V16     \\
57158.735       &       2015-05-16      &       1.529   $\pm$   0.037   &       57158.653       &       2015-05-16      &       2.035   $\pm$   0.036   &       BII, V16     \\
57160.681       &       2015-05-18      &       1.700   $\pm$   0.018   &               &               &                               &       BII     \\
57160.903       &       2015-05-18      &       1.590   $\pm$   0.090   &       57160.903       &       2015-05-18      &       2.120   $\pm$   0.140   &       SW      \\
57161.637       &       2015-05-19      &       1.637   $\pm$   0.029   &       57161.555       &       2015-05-19      &       2.060   $\pm$   0.034   &       V16     \\
57162.639       &       2015-05-20      &       1.668   $\pm$   0.025   &       57162.606       &       2015-05-20      &       2.125   $\pm$   0.042   &       V16     \\
57163.722       &       2015-05-21      &       1.760   $\pm$   0.038   &       57163.553       &       2015-05-21      &       2.013   $\pm$   0.041   &       V16     \\
57164.639       &       2015-05-22      &       1.721   $\pm$   0.024   &       57164.553       &       2015-05-22      &       2.034   $\pm$   0.035   &       V16     \\
        &               &                               &       57165.658       &       2015-05-23      &       2.057   $\pm$   0.036   &       V16     \\
57167.083       &       2015-05-24      &       1.760   $\pm$   0.070   &       57167.083       &       2015-05-24      &       2.090   $\pm$   0.090   &       SW      \\
57167.359       &       2015-05-24      &       1.724   $\pm$   0.019   &       57167.359       &       2015-05-24      &       2.099   $\pm$   0.019   &       S         \\
        &               &                               &       57169.598       &       2015-05-27      &       2.018   $\pm$   0.031   &       V16     \\
57170.546       &       2015-05-28      &       1.770   $\pm$   0.047   &       57170.533       &       2015-05-28      &       2.108   $\pm$   0.014   &       BII, V16     \\
57171.364       &       2015-05-28      &       1.699   $\pm$   0.012   &       57171.364       &       2015-05-28      &       2.056   $\pm$   0.022   &       S         \\
57171.601       &       2015-05-29      &       1.575   $\pm$   0.028   &       57171.554       &       2015-05-29      &       2.028   $\pm$   0.019   &       BII, V16     \\
57173.816       &       2015-05-31      &       1.740   $\pm$   0.081   &       57173.816       &       2015-05-31      &       2.050   $\pm$   0.100   &       SW      \\
57174.628       &       2015-06-01      &       1.568   $\pm$   0.040   &       57174.586       &       2015-06-01      &       1.965   $\pm$   0.021   &       BII, V16     \\
57175.595       &       2015-06-02      &       1.548   $\pm$   0.032   &       57175.553       &       2015-06-02      &       2.047   $\pm$   0.019   &       BII, V16     \\
57176.595       &       2015-06-03      &       1.596   $\pm$   0.041   &       57176.558       &       2015-06-03      &       2.089   $\pm$   0.024   &       BII, V16     \\
        &               &                               &       57177.614       &       2015-06-04      &       1.967   $\pm$   0.014   &       V16     \\
57178.595       &       2015-06-05      &       1.716   $\pm$   0.034   &       57178.566       &       2015-06-05      &       2.020   $\pm$   0.019   &       BII, V16     \\
57179.595       &       2015-06-06      &       1.525   $\pm$   0.029   &       57179.552       &       2015-06-06      &       2.016   $\pm$   0.020   &       BII, V16     \\
57180.596       &       2015-06-07      &       1.530   $\pm$   0.030   &       57180.568       &       2015-06-07      &       2.029   $\pm$   0.021   &       BII, V16     \\
57181.000       &       2015-06-07      &       1.530   $\pm$   0.061   &       57181.000       &       2015-06-07      &       1.970   $\pm$   0.080   &       SW      \\
57181.596       &       2015-06-08      &       1.502   $\pm$   0.032   &       57181.552       &       2015-06-08      &       1.998   $\pm$   0.014   &       BII, V16     \\
57182.330       &       2015-06-08      &       1.497   $\pm$   0.010   &       57182.330       &       2015-06-08      &       1.950   $\pm$   0.015   &       S         \\
57182.600       &       2015-06-09      &       1.457   $\pm$   0.022   &       57182.553       &       2015-06-09      &       2.036   $\pm$   0.012   &       BII, V16     \\
57183.570       &       2015-06-10      &       1.527   $\pm$   0.027   &       57183.515       &       2015-06-10      &       1.978   $\pm$   0.014   &       BII, V16     \\
57184.569       &       2015-06-11      &       1.506   $\pm$   0.018   &       57184.515       &       2015-06-11      &       1.954   $\pm$   0.014   &       BII, V16     \\
57185.569       &       2015-06-12      &       1.576   $\pm$   0.035   &       57185.515       &       2015-06-12      &       2.053   $\pm$   0.012   &       BII, V16     \\
57186.570       &       2015-06-13      &       1.511   $\pm$   0.032   &       57186.515       &       2015-06-13      &       1.983   $\pm$   0.010   &       BII, V16     \\
57187.319       &       2015-06-13      &       1.445   $\pm$   0.018   &       57187.319       &       2015-06-13      &       1.941   $\pm$   0.011   &       S         \\
57187.570       &       2015-06-14      &       1.467   $\pm$   0.026   &       57187.516       &       2015-06-14      &       1.990   $\pm$   0.012   &       BII, V16     \\
57187.806       &       2015-06-14      &       1.430   $\pm$   0.061   &       57187.806       &       2015-06-14      &       2.010   $\pm$   0.090   &       SW      \\
        &               &                               &       57189.516       &       2015-06-16      &       1.998   $\pm$   0.019   &       V16     \\
57192.308       &       2015-06-18      &       1.600   $\pm$   0.011   &       57192.308       &       2015-06-18      &       2.023   $\pm$   0.019   &       S         \\
        &               &                               &       57192.518       &       2015-06-19      &       2.012   $\pm$   0.020   &       V16     \\
57195.424       &       2015-06-21      &       1.470   $\pm$   0.070   &       57195.424       &       2015-06-21      &       1.980   $\pm$   0.090   &       SW      \\
57195.510       &       2015-06-22      &       1.476   $\pm$   0.022   &       57195.497       &       2015-06-21      &       1.948   $\pm$   0.014   &       BII, V16     \\
57196.295       &       2015-06-22      &       1.608   $\pm$   0.013   &       57196.295       &       2015-06-22      &       2.020   $\pm$   0.017   &       S         \\
        &               &                               &       57198.497       &       2015-06-24      &       1.970   $\pm$   0.017   &       V16     \\
57201.271       &       2015-06-27      &       1.467   $\pm$   0.010   &       57201.271       &       2015-06-27      &       1.944   $\pm$   0.017   &       S         \\
57201.879       &       2015-06-28      &       1.490   $\pm$   0.070   &       57201.879       &       2015-06-28      &       1.950   $\pm$   0.100   &       SW      \\
57203.486       &       2015-06-29      &       1.617   $\pm$   0.032   &       57203.461       &       2015-06-29      &       2.005   $\pm$   0.013   &       BII, V16     \\
57206.265       &       2015-07-02      &       1.705   $\pm$   0.014   &       57206.265       &       2015-07-02      &       2.098   $\pm$   0.018   &       S         \\
57209.188       &       2015-07-05      &       1.660   $\pm$   0.061   &       57209.188       &       2015-07-05      &       2.060   $\pm$   0.090   &       SW      \\
57211.488       &       2015-07-07      &       1.613   $\pm$   0.023   &               &               &                               &       BII     \\
57212.488       &       2015-07-08      &       1.654   $\pm$   0.030   &               &               &                               &       BII     \\
57213.491       &       2015-07-09      &       1.625   $\pm$   0.038   &               &               &                               &       BII     \\
57215.489       &       2015-07-11      &       1.619   $\pm$   0.029   &               &               &                               &       BII     \\
57216.519       &       2015-07-13      &       1.624   $\pm$   0.038   &               &               &                               &       BII     \\
57217.227       &       2015-07-13      &       1.718   $\pm$   0.030   &       57217.227       &       2015-07-13      &       2.117   $\pm$   0.025   &       S         \\
57218.219       &       2015-07-14      &       1.590   $\pm$   0.070   &       57218.219       &       2015-07-14      &       1.880   $\pm$   0.090   &       SW      \\
57222.629       &       2015-07-19      &       1.510   $\pm$   0.061   &       57222.629       &       2015-07-19      &       1.980   $\pm$   0.090   &       SW      \\
57229.917       &       2015-07-26      &       1.560   $\pm$   0.070   &       57229.917       &       2015-07-26      &       2.030   $\pm$   0.090   &       SW      \\
57236.969       &       2015-08-02      &       1.580   $\pm$   0.070   &       57236.969       &       2015-08-02      &       2.090   $\pm$   0.090   &       SW      \\
57337.576       &       2015-11-11      &       1.950   $\pm$   0.081   &       57337.576       &       2015-11-11      &       2.130   $\pm$   0.100   &       SW      \\
57399.510       &       2016-01-12      &       1.913   $\pm$   0.017   &       57399.510       &       2016-01-12      &       2.240   $\pm$   0.020   &       S         \\
57421.351       &       2016-02-02      &       2.000   $\pm$   0.070   &       57421.351       &       2016-02-02      &       2.340   $\pm$   0.100   &       SW      \\
        &               &                               &       57473.593       &       2016-03-26      &       2.084   $\pm$   0.014   &       V16     \\
57474.588       &       2016-03-27      &       1.748   $\pm$   0.025   &       57474.570       &       2016-03-27      &       1.956   $\pm$   0.013   &       V16     \\
57475.558       &       2016-03-28      &       1.647   $\pm$   0.017   &       57475.540       &       2016-03-28      &       1.977   $\pm$   0.010   &       V16     \\
57477.599       &       2016-03-30      &       1.616   $\pm$   0.018   &       57477.581       &       2016-03-30      &       1.856   $\pm$   0.013   &       V16     \\
57478.581       &       2016-03-31      &       1.659   $\pm$   0.017   &       57478.563       &       2016-03-31      &       1.968   $\pm$   0.013   &       V16     \\
57479.555       &       2016-04-01      &       1.687   $\pm$   0.018   &       57479.537       &       2016-04-01      &       1.874   $\pm$   0.011   &       V16     \\
57480.747       &       2016-04-02      &       1.636   $\pm$   0.018   &       57480.729       &       2016-04-02      &       1.974   $\pm$   0.015   &       V16     \\
57481.686       &       2016-04-03      &       1.697   $\pm$   0.019   &       57481.668       &       2016-04-03      &       1.888   $\pm$   0.012   &       V16     \\
57482.737       &       2016-04-04      &       1.691   $\pm$   0.018   &       57482.676       &       2016-04-04      &       1.980   $\pm$   0.014   &       V16     \\
57484.606       &       2016-04-06      &       1.702   $\pm$   0.017   &       57484.587       &       2016-04-06      &       1.913   $\pm$   0.010   &       V16     \\
57485.552       &       2016-04-07      &       1.724   $\pm$   0.016   &       57485.534       &       2016-04-07      &       1.932   $\pm$   0.008   &       V16     \\
57486.667       &       2016-04-08      &       1.775   $\pm$   0.015   &       57486.649       &       2016-04-08      &       1.967   $\pm$   0.013   &       V16     \\
57487.577       &       2016-04-09      &       1.772   $\pm$   0.020   &       57487.558       &       2016-04-09      &       1.975   $\pm$   0.014   &       V16     \\
57488.553       &       2016-04-10      &       1.820   $\pm$   0.024   &       57488.535       &       2016-04-10      &       2.037   $\pm$   0.011   &       V16     \\
57492.493       &       2016-04-13      &       1.945   $\pm$   0.025   &       57492.475       &       2016-04-13      &       2.177   $\pm$   0.021   &       V16     \\
57493.515       &       2016-04-15      &       1.941   $\pm$   0.026   &       57493.497       &       2016-04-14      &       2.061   $\pm$   0.014   &       V16     \\
57494.495       &       2016-04-15      &       1.991   $\pm$   0.025   &       57494.477       &       2016-04-15      &       2.112   $\pm$   0.014   &       V16     \\
57499.489       &       2016-04-20      &       1.893   $\pm$   0.038   &       57499.471       &       2016-04-20      &       2.148   $\pm$   0.017   &       V16     \\
57500.488       &       2016-04-21      &       1.897   $\pm$   0.034   &       57500.470       &       2016-04-21      &       2.181   $\pm$   0.021   &       V16     \\
57504.252       &       2016-04-25      &       1.854   $\pm$   0.025   &       57504.253       &       2016-04-25      &       2.155   $\pm$   0.012   &       MS      \\
57505.513       &       2016-04-27      &       1.839   $\pm$   0.077   &       57505.514       &       2016-04-27      &       2.140   $\pm$   0.024   &       MS      \\
57507.599       &       2016-04-29      &       1.840   $\pm$   0.018   &       57507.581       &       2016-04-29      &       2.123   $\pm$   0.015   &       V16     \\
57508.497       &       2016-04-29      &       1.797   $\pm$   0.017   &       57508.478       &       2016-04-29      &       2.011   $\pm$   0.015   &       V16     \\
57509.484       &       2016-04-30      &       1.865   $\pm$   0.021   &       57509.466       &       2016-04-30      &       2.031   $\pm$   0.020   &       V16     \\
57510.565       &       2016-05-02      &       1.879   $\pm$   0.024   &       57510.547       &       2016-05-02      &       2.070   $\pm$   0.019   &       V16     \\
57511.399       &       2016-05-02      &       1.826   $\pm$   0.023   &               &               &                               &       MS      \\
57511.530       &       2016-05-03      &       1.920   $\pm$   0.017   &       57511.512       &       2016-05-03      &       2.114   $\pm$   0.010   &       V16     \\
57512.489       &       2016-05-03      &       1.895   $\pm$   0.018   &       57512.470       &       2016-05-03      &       2.082   $\pm$   0.014   &       V16     \\
57513.282       &       2016-05-04      &       2.018   $\pm$   0.016   &       57513.284       &       2016-05-04      &       2.323   $\pm$   0.021   &       MS      \\
57514.333       &       2016-05-05      &       2.017   $\pm$   0.015   &       57514.334       &       2016-05-05      &       2.365   $\pm$   0.020   &       MS      \\
57515.553       &       2016-05-07      &       2.032   $\pm$   0.023   &       57515.533       &       2016-05-07      &       2.369   $\pm$   0.014   &       V16     \\
57516.517       &       2016-05-08      &       2.132   $\pm$   0.018   &       57516.499       &       2016-05-07      &       2.329   $\pm$   0.021   &       V16     \\
57518.273       &       2016-05-09      &       1.873   $\pm$   0.025   &       57518.274       &       2016-05-09      &       2.174   $\pm$   0.011   &       MS      \\
57519.391       &       2016-05-10      &       1.944   $\pm$   0.005   &       57519.391       &       2016-05-10      &       2.311   $\pm$   0.014   &       S         \\
57520.353       &       2016-05-11      &       1.862   $\pm$   0.026   &       57520.354       &       2016-05-11      &       2.171   $\pm$   0.012   &       MS      \\
57521.629       &       2016-05-13      &       1.955   $\pm$   0.025   &       57521.647       &       2016-05-13      &       2.299   $\pm$   0.016   &       V16     \\
57522.461       &       2016-05-13      &       1.866   $\pm$   0.018   &       57522.479       &       2016-05-13      &       2.100   $\pm$   0.014   &       V16     \\
57524.471       &       2016-05-15      &       1.752   $\pm$   0.025   &       57524.490       &       2016-05-15      &       2.014   $\pm$   0.033   &       V16     \\
57526.462       &       2016-05-17      &       1.731   $\pm$   0.023   &               &               &                               &       V16     \\
57527.460       &       2016-05-18      &       1.558   $\pm$   0.025   &               &               &                               &       V16     \\
57529.463       &       2016-05-20      &       1.789   $\pm$   0.039   &               &               &                               &       V16     \\
57530.295       &       2016-05-21      &       1.740   $\pm$   0.256   &       57530.296       &       2016-05-21      &       2.137   $\pm$   0.114   &       MS      \\
57530.459       &       2016-05-21      &       1.728   $\pm$   0.022   &               &               &                               &       V16     \\
57531.639       &       2016-05-23      &       1.652   $\pm$   0.044   &               &               &                               &       V16     \\
57532.509       &       2016-05-24      &       1.636   $\pm$   0.018   &               &               &                               &       V16     \\
57533.458       &       2016-05-24      &       1.705   $\pm$   0.022   &               &               &                               &       V16     \\
57534.519       &       2016-05-26      &       1.839   $\pm$   0.024   &               &               &                               &       V16     \\
57535.293       &       2016-05-26      &       1.784   $\pm$   0.023   &       57535.294       &       2016-05-26      &       2.130   $\pm$   0.011   &       MS      \\
57536.314       &       2016-05-27      &       1.804   $\pm$   0.023   &       57536.315       &       2016-05-27      &       2.133   $\pm$   0.011   &       MS      \\
57536.458       &       2016-05-27      &       1.805   $\pm$   0.025   &               &               &                               &       V16     \\
57537.309       &       2016-05-28      &       1.915   $\pm$   0.013   &       57537.310       &       2016-05-28      &       2.240   $\pm$   0.013   &       MS      \\
57538.230       &       2016-05-29      &       1.927   $\pm$   0.014   &       57538.232       &       2016-05-29      &       2.262   $\pm$   0.013   &       MS      \\
57538.626       &       2016-05-30      &       1.900   $\pm$   0.046   &               &               &                               &       V16     \\
57539.487       &       2016-05-30      &       1.908   $\pm$   0.018   &               &               &                               &       V16     \\
57540.351       &       2016-05-31      &       1.888   $\pm$   0.008   &       57540.351       &       2016-05-31      &       2.196   $\pm$   0.018   &       S         \\
57540.599       &       2016-06-01      &       1.902   $\pm$   0.063   &               &               &                               &       V16     \\
57541.542       &       2016-06-02      &       1.843   $\pm$   0.032   &               &               &                               &       V16     \\
57542.536       &       2016-06-03      &       1.960   $\pm$   0.028   &               &               &                               &       V16     \\
57546.234       &       2016-06-06      &       1.822   $\pm$   0.023   &       57546.236       &       2016-06-06      &       2.173   $\pm$   0.011   &       MS      \\
57551.512       &       2016-06-12      &       2.086   $\pm$   0.035   &               &               &                               &       V16     \\
57555.463       &       2016-06-15      &       2.030   $\pm$   0.024   &               &               &                               &       V16     \\
57558.358       &       2016-06-18      &       1.692   $\pm$   0.250   &       57558.359       &       2016-06-18      &       2.145   $\pm$   0.098   &       MS      \\
57558.459       &       2016-06-18      &       1.860   $\pm$   0.022   &               &               &                               &       V16     \\
57560.459       &       2016-06-20      &       1.994   $\pm$   0.038   &               &               &                               &       V16     \\
57561.329       &       2016-06-21      &       1.897   $\pm$   0.103   &       57561.330       &       2016-06-21      &       2.215   $\pm$   0.031   &       MS      \\
57561.463       &       2016-06-21      &       1.969   $\pm$   0.018   &               &               &                               &       V16     \\
57562.459       &       2016-06-22      &       2.152   $\pm$   0.028   &               &               &                               &       V16     \\
57567.460       &       2016-06-27      &       2.000   $\pm$   0.027   &       57567.480       &       2016-06-27      &       2.277   $\pm$   0.022   &       V16     \\
57569.461       &       2016-06-29      &       1.881   $\pm$   0.023   &       57569.480       &       2016-06-29      &       2.165   $\pm$   0.021   &       V16     \\
57575.467       &       2016-07-05      &       2.105   $\pm$   0.031   &       57575.486       &       2016-07-05      &       2.451   $\pm$   0.016   &       V16     \\
57581.468       &       2016-07-11      &       2.080   $\pm$   0.023   &               &               &                               &       V16     \\
57582.469       &       2016-07-12      &       2.159   $\pm$   0.028   &               &               &                               &       V16     \\
57583.468       &       2016-07-13      &       2.214   $\pm$   0.025   &               &               &                               &       V16     \\
57584.464       &       2016-07-14      &       2.141   $\pm$   0.031   &               &               &                               &       V16     \\
57585.504       &       2016-07-16      &       2.059   $\pm$   0.032   &               &               &                               &       V16     \\
57889.381       &       2017-05-15      &       1.997   $\pm$   0.017   & 57889.381     & 2017-05-15    &  2.263    $\pm$ 0.016   &       S       \\ 
\label{optcontfluxlog}
\end{longtable}
%
%
%
\twocolumn
Table~\ref{optcontfluxlog}
 gives the derived B and V fluxes of HE\,1136-2304 based on
the \swift{} data, SALT spectra, photometric data obtained with the
 Cerro Armazones, and the MONET/North and South telescopes from 2014 to 2017.
All these photometric data have been  intercalibrated with respect to the
\swift{} data. In addition, we list the B and V values based on the ESO
spectrum taken in 1993  (Reimers et al.\citealt{reimers96}),
and those based on the 6dF spectrum taken in 2002
(Jones et al.\citealt{jones04}). 
%
%
%

\begin{figure}
\centering
\includegraphics[width=9.cm,angle=0]{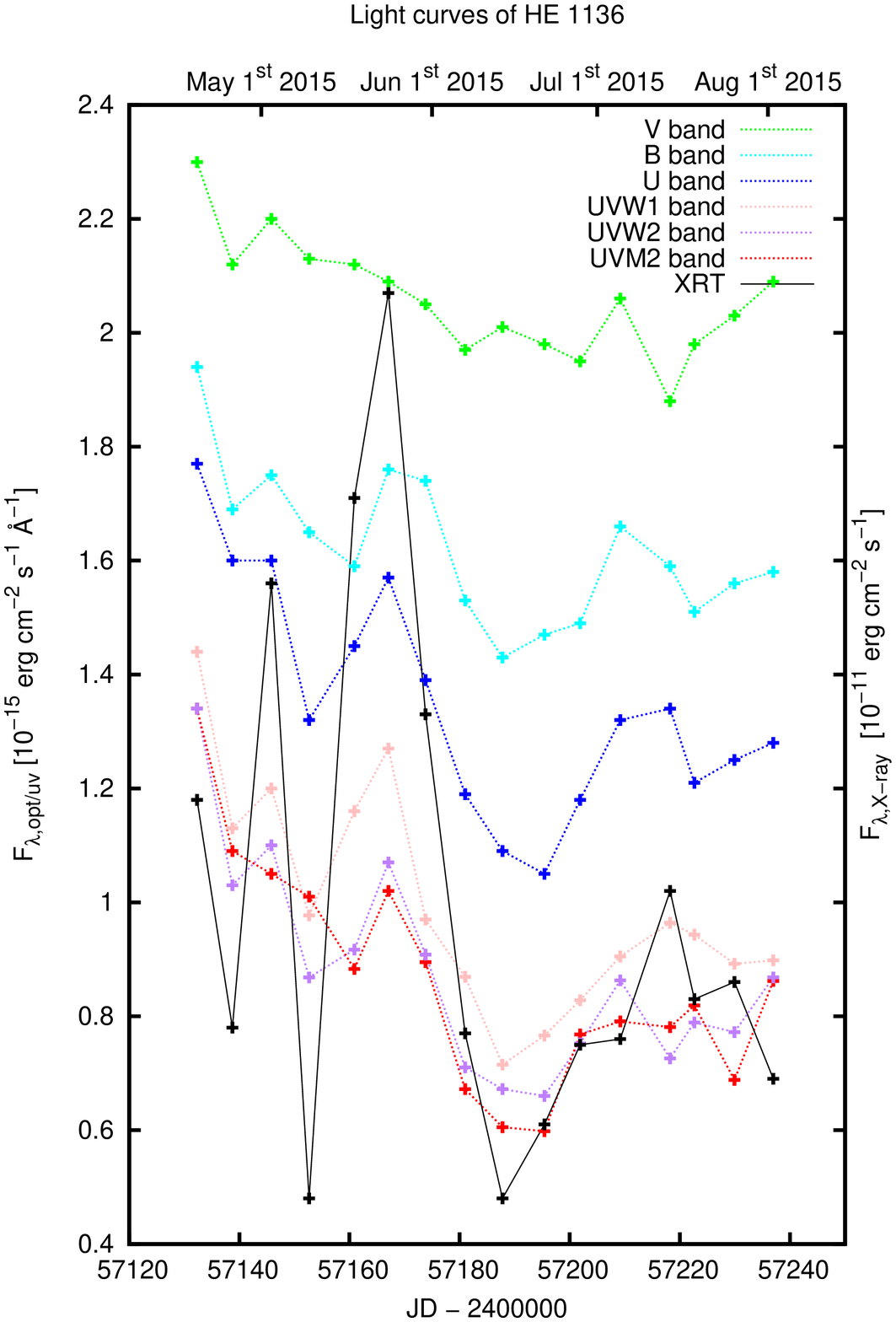}
\caption{Combined optical, UV, and X-ray light curves
taken with the \swift{} satellite for the dedicated campaign in 2015. 
}
\label{lc_2015.ps}
\end{figure}

Figure~\ref{lc_2015.ps} 
shows the X-ray, UV, and optical \swift{} light curves
for our detailed campaign in 2015 from April until August
in one plot to compare their amplitudes. 
During these months the source was observed
weekly.
Figure~\ref{lc_2015.ps} is a zoom-in of the middle part of
Figure~\ref{lc_swift_2014_2016}.
The UV and optical \swift{} bands closely follow the X-ray light curve. The
X-ray light curve exhibits the strongest variability amplitudes.
On the other hand, the \swift{}
V band only shows minor variations in contrast to the other bands. 

Table~\ref{swiftvarstatistics}
gives the variability statistics based on
the \swift{} continua (XRT, W2, M2, W1,
 U, B, V). 
We indicate the minimum and maximum fluxes F$_{\rm min}$ and F$_{\rm max}$,
peak-to-peak amplitudes R$_{\rm max}$ = F$_{\rm max}$/F$_{\rm min}$, the mean flux
over the period of observations $<$F$>$, the
standard deviation $\sigma_{\rm F}$,  and the fractional variation 
\begin{align*}
F_{\rm var} = \frac{\sqrt{{\sigma_F}^2 - \Delta^2}}{<F>}~, 
\end{align*}
as defined by Rodr\'\i{}guez-Pascual et al.\cite{rodriguez97}.
The quantity $\Delta^2$ is the mean square value of the uncertainties 
$\Delta_\text{i}$ associated with the fluxes $F_\text{i}$.
The $F_{\rm var}$ uncertainties are defined in Edelson et al.\cite{edelson02}.
The peak-to-peak amplitude and the fractional variation decrease as a function
of wavelength.
Additionally, we present the variability statistics
based on the combined B and V light curves including all optical 
ground-based telescopes 
 (MONET, Cerro Armazones, SALT) and \swift{} in units of
 10$^{-15}$\,erg\,s$^{-1}$\,cm$^{-2}$\,\AA$^{-1}$.
The results are similar to those based on the \swift{} data only.
Furthermore, we give the variability statistics
based solely on the dedicated variability campaign in 2015. 
In comparison to the complete data set, the peak-to-peak amplitudes and the fractional variations are
smaller because the optical high state
in 2014 is not included (see 
Figures~\ref{LC_B_all_ochm.ps} and \ref{LC_V_all_ochm.ps}).

We  compare our results with those from other spectroscopic
AGN variability campaigns. In comparison to photometric campaigns,
these spectroscopic variability campaigns
are typically based on small apertures.
 Therefore, we additionally calculated
the variability statistics based on our small aperture spectral
data taken with SALT
without intercalibration with respect to the large aperture
photometric data. 

Finally,  we present the variability statistics
 after subtracting the
nonvariable flux contribution of the host galaxy.
 This results in significantly higher variability
amplitudes in all individual wavebands
(Table~\ref{swiftvarstatistics}, Col. 2).
The derivation of the host galaxy flux contribution is described
 in the following two sections.

\subsection{Host galaxy contribution to the optical continuum flux}

\begin{figure}
\centering
\includegraphics[width=6.5cm,angle=0]{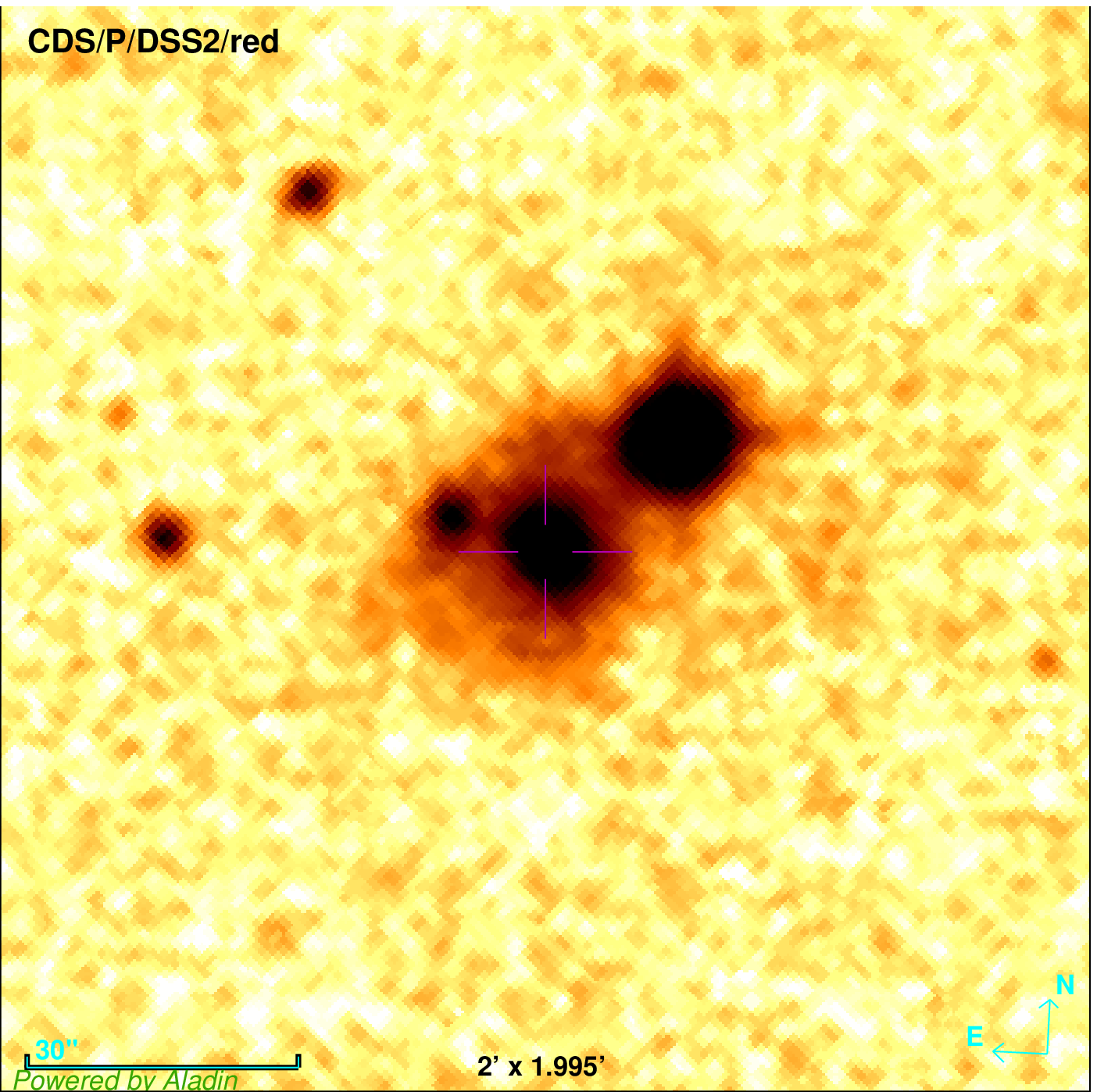}
\includegraphics[width=5.5cm,angle=0]{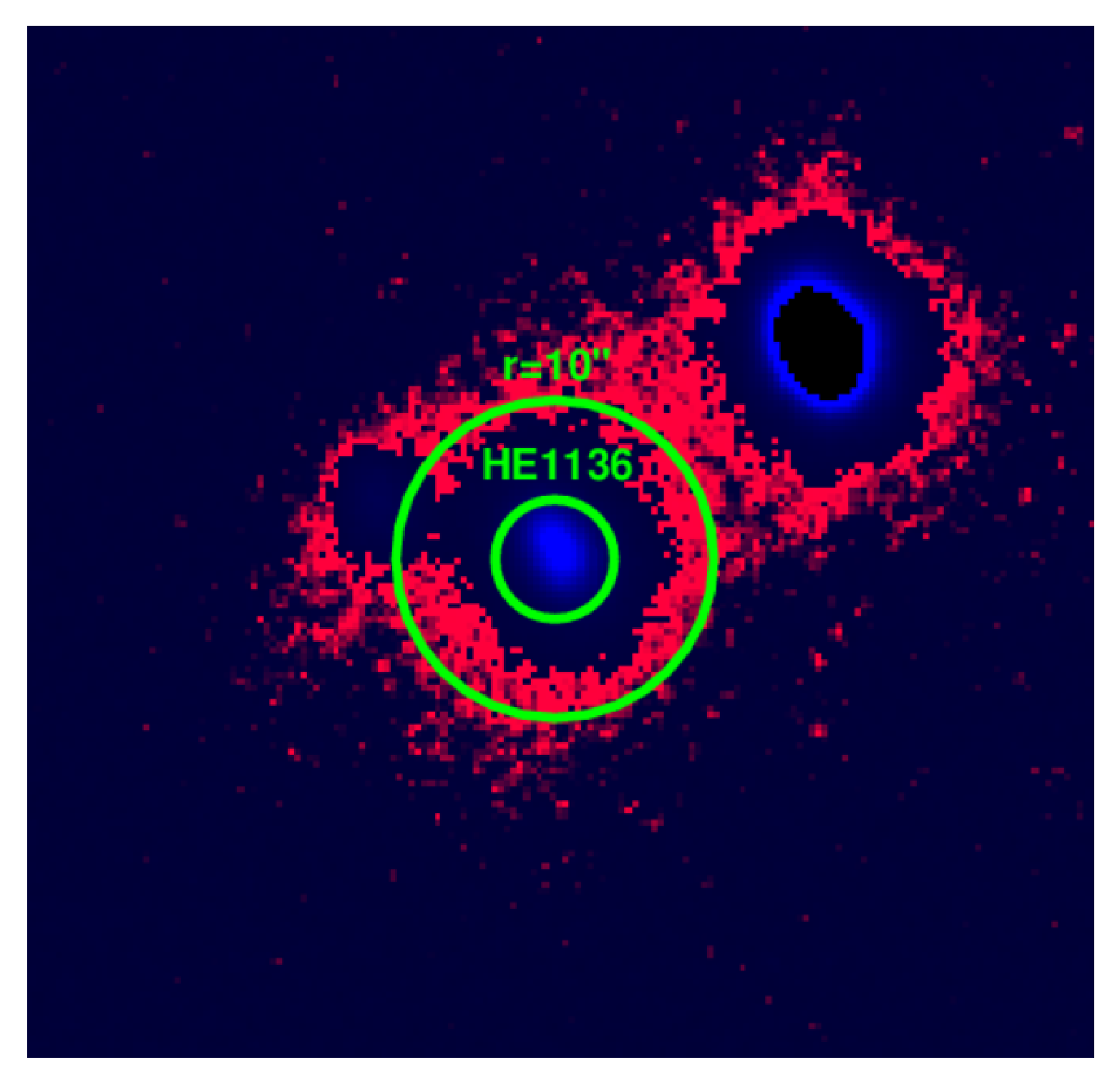}
\caption{Top: DSS1 image of HE\,1136-2304. Scale: 2 x 2 arcmin. North is
to the top, East to the left.
Bottom: Enlarged B-V two color image based on
    VYSOS\,16 data.
    The small green circle has $r = 3\farcs75$, indicating 
    that the aperture used for the OCA photometry is sufficiently small to be 
    not contaminated by the star located 
    about 20$\arcsec$ in the NW and the faint source in the E. 
    The large green circle with $r=10\arcsec$ as labeled indicates 
    the projected distance to the eastern source. 
  }
\label{DSSbildHE11_2arcmin.ps}
\end{figure}

Figure~\ref{DSSbildHE11_2arcmin.ps}
displays the DSS1 image of HE\,1136-2304
(Scale: 2 x 2 arcmin; pixel size 1.7 arcsec)
as well as a B-V two-color image (bottom) based on
VYSOS\,16 data.  The nucleus of HE\,1136-2304 is surrounded by
a spiral or S0 host galaxy;
the  radial profile of the surface brightness shows a central bulge structure and an extended disk structure in
the DSS1 image. Some asymmetry of the outer isophotes might be connected
with the object located to the east at a distance of 12 arcsec.

The observed flux of the variable AGN component is contaminated by the
flux contribution of the host galaxy. The relative contribution of the host
galaxy in the individual bands differs
since the central nonthermal component has
a different flux distribution from the stellar component of the host galaxy.
Furthermore, the flux contribution of the host galaxy depends on the size of
the aperture.
In addition, we compare the accuracy and  the results
 based on the photometric observations taken with \swift{}
 on the one hand
and spectroscopic observations taken with SALT on the other hand.
These photometric and spectroscopic observations were carried out
with different apertures.
All other photometric data were intercalibrated with respect
to the absolute fluxes of \swift{}. 
Finally, we compare our results with those of 
 Parker et al.\cite{parker16}.
Their results are based on only two spectra obtained in 
 1993 and 2014 and taken
with different instruments and apertures.

We estimate the relative contribution of the constant host galaxy flux
by means of the flux variation gradient (FVG) method (Choloniewski
\citealt{choloniewski81}, Winkler et al.\citealt{winkler92}, Haas et al.
\citealt{haas11},
Ramolla et al.\citealt{ramolla15}). This method
 disentangles the varying AGN flux in our aperture from the
constant host galaxy contribution.
We obtained B and V flux values of HE\,1136-2304 based on the 5" aperture 
of the  \swift{} UVOT. 
Furthermore, we derived B-, V-, and R-band fluxes for the SALT
spectra by convolving them with the B, V, and R filter curves
(IRAF task {\it sbands}).
We measured the fluxes at wavelengths close to the maxima of
the filter curves
(B-band filter: 4300 \AA{}; V-band filter: 5400 \AA{};
 R-band  filter: 6100 \AA{})
with widths of a few hundred \AA{}.
In this way we excluded the contribution of
emission lines in the spectra (see Fig.~\ref{filter_format.eps}).
These B, V, and R values 
are presented in Table~\ref{bvrsaltspecflux}.

\begin{figure}
\centering
\includegraphics[width=6.5cm,angle=-90]{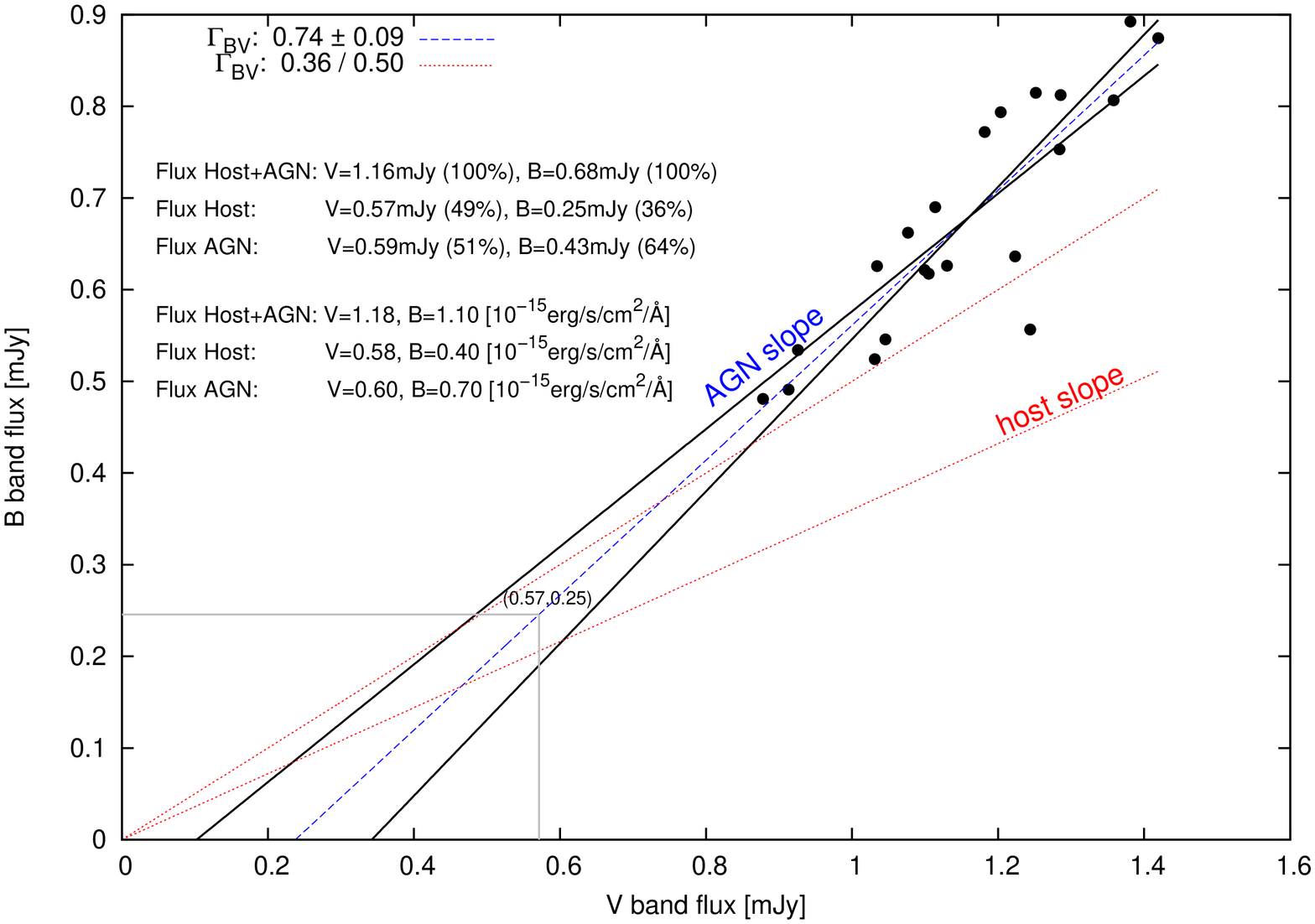}
\caption{Flux variations (B vs. V)  of HE\,1136-2304 based on the
SALT spectra. The blue dashed line is the best linear fit to the
B vs. V fluxes. The black solid lines cover the upper  
and lower standard deviations of the interpolated AGN slope. 
The red dashed lines give the range of host slopes as
determined by Sakata et al.\cite{sakata10}. 
The gray lines indicate the central B and V values of the host galaxy.
Listed are the B and V flux values (in units of mJy and
 10$^{-15}$\,erg\,s$^{-1}$\,cm$^{-2}$\,\AA$^{-1}$) for the combined
 mean host galaxy+AGN flux,
for the host galaxy flux, and for the mean AGN flux.}
\label{fvg_BvsV.ps}
\end{figure}

\begin{figure}
\centering
\includegraphics[width=6.5cm,angle=-90]{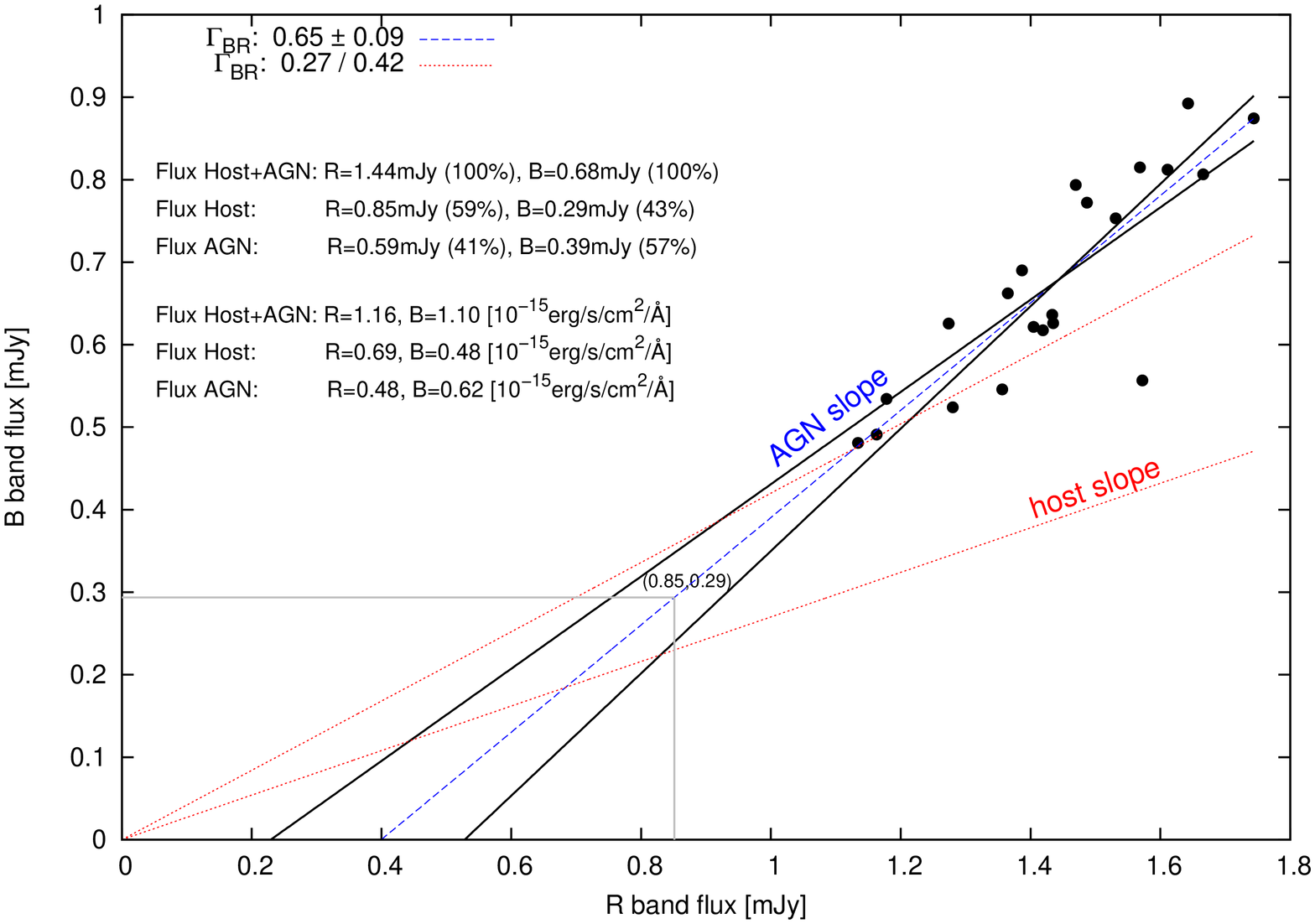}
\caption{Flux variations (B vs. R)  of HE\,1136-2304 based on the
SALT spectra. The blue dashed line presents the best linear fit to the
B vs. R fluxes. The black solid lines cover the upper  
and lower standard deviations of the interpolated AGN slope.
The red dashed lines show the range of host slopes.
The gray lines indicate the B and R values of the host galaxy.
The B and R flux values are listed as in Fig.~\ref{fvg_BvsV.ps}.
}
\label{fvg_BvsR.ps}
\end{figure}

Figures~\ref{fvg_BvsV.ps} and
\ref{fvg_BvsR.ps} show the B versus V and 
B versus R fluxes (black solid circles) of HE\,1136-2304 based on the
SALT spectra (aperture: 2 x 2 arcsec).  
The blue dashed line gives the best linear fit to the
B versus V and B versus R fluxes. The black solid lines cover the upper  
and lower standard deviations of the interpolated AGN slope. 
The red dashed lines give the range of host
slopes for nearby AGN as determined by Sakata et al.\cite{sakata10}.
The intersection point between the AGN
and host galaxy slopes gives the host galaxy fluxes in the B,
V, and R bands. The gray lines indicate these B, V, and R values
of the host galaxy.

\begin{figure}
\centering
\includegraphics[width=6.5cm,angle=-90]{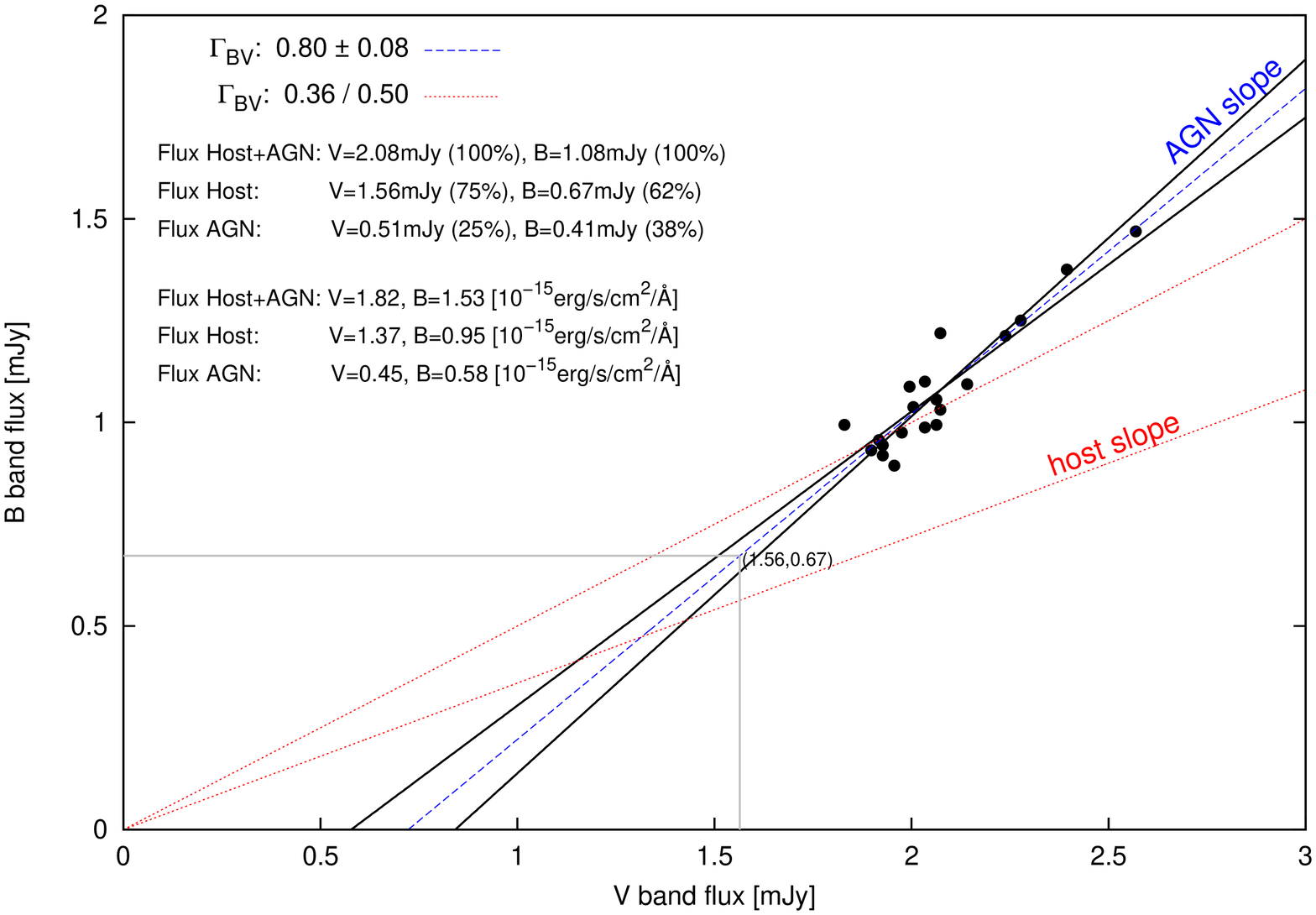}
\caption{Flux variations (B vs. V)  of HE\,1136-2304 based on the photometric
\swift{} data. The blue dashed line presents the best linear fit to the
B vs. V fluxes. The black solid lines cover the upper  
and lower standard deviations of the interpolated AGN slope.
The red dashed lines show the range of host slopes.
The gray lines indicate the central B and V values of the host galaxy.
The B and V flux values are listed as in Fig.~\ref{fvg_BvsV.ps}.
}
\label{fvg_BvsV_swift.ps}
\end{figure}

Based on the intersection in the two figures{\bf,} we derive a  B-band flux
of 0.27  mJy for the host galaxy contribution (mean of 0.25 and 0.29 mJy).
The corresponding values are 0.57 mJy for the V band and 0.85 mJy for
the R band.
Our derived flux values for the contribution of the host galaxy
in the B- and R-band spectra are  15\%\ 
higher than  those derived in Parker et al.\cite{parker16}.
The new values are of  higher confidence.
They are based on spectra taken under identical conditions
at 21 epochs. Parker et al.\cite{parker16} used
only two spectra taken with different apertures.
Figure~\ref{fvg_BvsV_swift.ps} shows the 
B versus V flux variations based on the \swift{} photometric data
taken with a 12 arcsec aperture.

Once  we know the integrated flux values of the host galaxy plus AGN as well as the host galaxy contribution, we can derive the AGN flux contribution in the individual bands. 
All these values are listed in Table~\ref{bvrhostflux}.
We present these values separately for the measurements based on the
SALT spectra (based on the smaller aperture) and for the \swift{} data
(based on the larger aperture).
Furthermore, we give all these flux values in units of mJy and in units of
10$^{-15}$\,erg\,s$^{-1}$\,cm$^{-2}$\,\AA$^{-1}$ with
the conversion formula
  \begin{equation}
    \label{eq:f2mJy}
    F_{\rm mJy,\lambda}=F\,\frac{\lambda^{2}}{29979245.8},
  \end{equation}
where F$_{\rm mJy, \lambda}$ is the flux in units of mJy, F the flux in
units of 10$^{-15}$\,erg\,s$^{-1}$\,cm$^{-2}$\,\AA$^{-1}$, and $\lambda$ the 
wavelength in $\AA$.

 The derived host galaxy fluxes in the B and V bands
(based on the \swift{} data) are  a factor two higher than those based on the SALT
spectra because the larger extraction area
of the \swift{} UVOT (10 arcsec diameter) collects more flux
of the extended host galaxy than the SALT spectra do
(2 x 2 arcsec only).
However, the mean AGN fluxes derived on the basis of the SALT spectra
are similar to those based on the \swift{} UVOT data (last three rows in
Table~\ref{bvrhostflux}).
The AGN contribution based on the SALT spectra
corresponds to 60\%, 51\%, and 41\%\ in the B, V, and R band, respectively.
The AGN contribution in the \swift\,UVOT B and V band
decreases to 38\%\ and 25\%,\  respectively, because of their larger aperture. 

\begin{table*}
\centering
\tabcolsep+1.5mm
\caption{Variability statistics based on the \swift{} continua
 (XRT, W2, M2, W1, U, B, V) and on the combined B and V light curves
 (\swift{}, SALT, MONET,
 Cerro Armazones)
in units of 
10$^{-15}$\,erg\,s$^{-1}$\,cm$^{-2}$\,\AA$^{-1}$ and $10^{-11}$ ergs s$^{-1}$ cm$^{-2}$ for the 0.3--10 keV X-ray data.
In addition, we give the statistics solely for the dedicated campaign in 2015.
Finally, we present the variability statistics based solely on the SALT spectra
with their small aperture. 
In the second column the variability statistics is given after subtraction
of the host galaxy flux.
}
\begin{tabular}{lcccccc|cccccc}
\hline 
\noalign{\smallskip}
Cont. & F$_{\rm max}$ & F$_{\rm min}$ & R$_{\rm max}$ & $<$F$>$ & $\sigma_{\rm F}$ & F$_{\rm var}$ & F$_{\rm max}$ & F$_{\rm min}$ & R$_{\rm max}$ & $<$F$>$ & $\sigma_F$ & F$_{\rm var}$ \\
&\multicolumn{6}{c}{with host}&\multicolumn{6}{c}{without host}\\
\noalign{\smallskip}
(1) & (2) & (3) & (4) & (5) & (6) & (7) & (8) & (9) & (10) & (11) & (12) & (13) \\ 
\noalign{\smallskip}
\hline 
\noalign{\smallskip}
Cont.~\swift{} XRT all     & 3.75 & 0.48 & 7.81 & 1.65 & 0.86 & 0.516 $\pm$ 0.016&\\
Cont.~\swift{} W2 all      & 1.83 & 0.66 & 2.77 & 1.00 & 0.31 & 0.308 $\pm$ 0.015& 1.58 & 0.41 & 3.88 & 0.74 & 0.31 & 0.413 $\pm$ 0.020\\  
Cont.~\swift{} M2 all      & 2.22 & 0.60 & 3.71 & 1.31 & 0.53 & 0.402 $\pm$ 0.009& 1.99 & 0.37 & 5.43 & 1.08 & 0.53 & 0.488 $\pm$ 0.011\\  
Cont.~\swift{} W1 all      & 1.92 & 0.71 & 2.69 & 1.14 & 0.34 & 0.288 $\pm$ 0.016& 1.64 & 0.44 & 3.74 & 0.86 & 0.34 & 0.380 $\pm$ 0.021\\  
Cont.~\swift{} U all       & 2.49 & 1.05 & 2.37 & 1.53 & 0.40 & 0.253 $\pm$ 0.012& 1.99 & 0.55 & 3.60 & 1.03 & 0.40 & 0.375 $\pm$ 0.018\\  
Cont.~\swift{} B all       & 2.35 & 1.43 & 1.64 & 1.72 & 0.25 & 0.138 $\pm$ 0.010& 1.40 & 0.48 & 2.93 & 0.77 & 0.25 & 0.309 $\pm$ 0.022\\ 
Cont.~\swift{} V all       & 2.64 & 1.88 & 1.40 & 2.13 & 0.19 & 0.073 $\pm$ 0.012& 1.27 & 0.51 & 2.49 & 0.76 & 0.19 & 0.206 $\pm$ 0.033\\ 
\noalign{\smallskip}    
\hline 
\noalign{\smallskip}    
Cont. B all & 2.35 & 1.43 & 1.64 & 1.76 & 0.19 & 0.103 $\pm$ 0.002 & 1.40 & 0.48 & 2.93 & 0.80 & 0.19 & 0.224 $\pm$ 0.004\\ 
Cont. V all & 2.64 & 1.86 & 1.42 & 2.09 & 0.13 & 0.058 $\pm$ 0.002 & 1.27 & 0.49 & 2.62 & 0.72 & 0.13 & 0.167 $\pm$ 0.005\\ 
\noalign{\smallskip}    
\hline
\hline
\noalign{\smallskip}    
Cont.~\swift{} XRT 2015    & 2.07 & 0.48 & 4.31 & 0.99 & 0.46 & 0.452 $\pm$ 0.025&\\
Cont.~\swift{} W2 2015     & 1.34 & 0.66 & 2.03 & 0.88 & 0.18 & 0.198 $\pm$ 0.017& 1.09 & 0.41 & 2.67 & 0.62 & 0.18 & 0.278 $\pm$ 0.024\\
Cont.~\swift{} M2 2015     & 1.34 & 0.60 & 2.24 & 0.87 & 0.20 & 0.217 $\pm$ 0.019& 1.11 & 0.37 & 3.03 & 0.64 & 0.20 & 0.296 $\pm$ 0.025\\
Cont.~\swift{} W1 2015     & 1.44 & 0.71 & 2.01 & 1.00 & 0.19 & 0.178 $\pm$ 0.021& 1.16 & 0.44 & 2.65 & 0.72 & 0.19 & 0.246 $\pm$ 0.029\\
Cont.~\swift{} U 2015      & 1.77 & 1.05 & 1.69 & 1.35 & 0.20 & 0.138 $\pm$ 0.015& 1.27 & 0.55 & 2.30 & 0.86 & 0.20 & 0.218 $\pm$ 0.023\\
Cont.~\swift{} B 2015      & 1.94 & 1.43 & 1.36 & 1.62 & 0.13 & 0.069 $\pm$ 0.012& 0.99 & 0.48 & 2.07 & 0.67 & 0.13 & 0.168 $\pm$ 0.029\\ 
Cont.~\swift{} V 2015      & 2.30 & 1.88 & 1.22 & 2.06 & 0.10 & 0.017 $\pm$ 0.026& 0.93 & 0.51 & 1.82 & 0.69 & 0.10 & 0.051 $\pm$ 0.078\\ 
\noalign{\smallskip}    
\hline
Cont. B 2015, Feb.--Aug. & 2.03 & 1.43 & 1.42 & 1.66 & 0.14 & 0.081 $\pm$ 0.003 & 1.07 & 0.48 & 2.25 & 0.70 & 0.14 & 0.192 $\pm$ 0.007\\ 
Cont. V 2015, Feb.--Aug. & 2.35 & 1.88 & 1.25 & 2.07 & 0.10 & 0.043 $\pm$ 0.003 & 0.98 & 0.51 & 1.93 & 0.70 & 0.10 & 0.128 $\pm$ 0.009\\ 
\noalign{\smallskip}          
\hline 
\hline
Cont. 4570 SALT 2015 & 1.26 & 0.78 & 1.61 & 0.99 & 0.14 & 0.141 $\pm$ 0.003 & 0.72 & 0.25 & 2.90 & 0.45 & 0.14 & 0.307 $\pm$ 0.007\\ 
Cont. 5360 SALT 2015 & 1.20 & 0.86 & 1.39 & 1.03 & 0.12 & 0.114 $\pm$ 0.006 & 0.64 & 0.31 & 2.08 & 0.47 & 0.12 & 0.248 $\pm$ 0.012\\ 
\noalign{\smallskip}          
\hline 
\hline
\end{tabular}
\label{swiftvarstatistics}
\end{table*}

\begin{table}
\tabcolsep+2.5mm
\caption{Julian date, UT date, and B, V, and R values (in units of mJy)
 determined by convolving the SALT spectra with the
 corresponding filter curves.
}
\centering
\begin{tabular}{lcccc}
\hline
\noalign{\smallskip}
Julian Date &               &   B band & V band  & R band \\
2\,400\,000+&  \rb{UT Date} &  \multicolumn{3}{c}{[mJy]}   \\
\hline 
\noalign{\smallskip}
56846.248       &       2014-07-07   &  0.87  &  1.42  &   1.74    \\
57016.559       &       2014-12-25   &  0.52  &  1.03  &   1.28    \\
57070.399       &       2015-02-16   &  0.63  &  1.03  &   1.27    \\
57082.362       &       2015-02-28   &  0.75  &  1.28  &   1.53    \\
57088.594       &       2015-03-07   &  0.81  &  1.36  &   1.67    \\
57100.539       &       2015-03-19   &  0.79  &  1.20  &   1.47    \\
57112.285       &       2015-03-30   &  0.66  &  1.08  &   1.37    \\
57121.256       &       2015-04-08   &  0.69  &  1.11  &   1.39    \\
57131.243       &       2015-04-18   &  0.77  &  1.18  &   1.49    \\
57167.359       &       2015-05-24   &  0.64  &  1.22  &   1.43    \\
57171.364       &       2015-05-28   &  0.62  &  1.10  &   1.40    \\
57182.330       &       2015-06-08   &  0.53  &  0.93  &   1.18    \\
57187.319       &       2015-06-13   &  0.49  &  0.91  &   1.16    \\
57192.308       &       2015-06-18   &  0.56  &  1.24  &   1.57    \\
57196.295       &       2015-06-22   &  0.55  &  1.05  &   1.36    \\
57201.271       &       2015-06-27   &  0.48  &  0.88  &   1.13    \\
57206.265       &       2015-07-02   &  0.63  &  1.13  &   1.43    \\
57217.227       &       2015-07-13   &  0.62  &  1.11  &   1.42    \\
57399.510       &       2016-01-12   &  0.81  &  1.25  &   1.57    \\
57519.391       &       2016-05-10   &  0.89  &  1.38  &   1.64    \\
57540.351       &       2016-05-31   &  0.81  &  1.29  &   1.61    \\
\hline 
\end{tabular}
\label{bvrsaltspecflux}
\end{table}
\begin{table}
\tabcolsep+2mm
\caption{B, V, and R values (in units of mJy and
 10$^{-15}$\,erg\,s$^{-1}$\,cm$^{-2}$\,\AA$^{-1}$) for the combined
 host galaxy+AGN fluxes
 as well as for the host galaxy and AGN fluxes alone.
These flux values are based on flux variation diagrams in combination 
with the SALT spectra and  with the photometric
 \swift{} data (see
 Figs.~\ref{fvg_BvsV.ps}, \ref{fvg_BvsR.ps}, and \ref{fvg_BvsV_swift.ps}).}
\centering
\begin{tabular}{lccc}
\hline
\noalign{\smallskip}
Diagram               & B band & V band  & R band \\
& \multicolumn{3}{c}{[mJy]}\\
(1)                   & (2)  & (3)   & (4) \\ 
\hline
Host+AGN (BvsR SALT)  &0.68  &       &1.44\\
Host+AGN (BvsV SALT)  &0.68  &1.16   &\\
Host+AGN (BvsV \swift{}) &1.08  &2.08   &\\
\hline
Host (BvsR SALT)      &0.29  &       &0.85\\
Host (BvsV SALT)      &0.25  &0.57   &\\
Host (BvsV \swift{})     &0.67  &1.56   &\\
\hline
AGN (BvsR SALT)       &0.39  &       &0.59\\
AGN (BvsV SALT)       &0.43  &0.59   &\\
AGN (BvsV \swift{})      &0.41  &0.51   &\\
\hline
\hline
& \multicolumn{3}{c}{[10$^{-15}$\,erg\,s$^{-1}$\,cm$^{-2}$\,\AA$^{-1}$]}\\
\hline
\hline
Host+AGN (BvsR SALT)  &1.10  &       &1.16\\
Host+AGN (BvsV SALT)  &1.10  &1.18   &\\
Host+AGN (BvsV \swift{}) &1.53  &1.82   & \\
\hline
Host (BvsR SALT)      &0.48  &       &0.69\\
Host (BvsV SALT)      &0.40  &0.58   &\\
Host (BvsV \swift{})     &0.95  &1.37   &\\
\hline
AGN (BvsR SALT)       &0.62  &       &0.48\\
AGN (BvsV SALT)       &0.70  &0.60   &\\
AGN (BvsV \swift{})      &0.58  &0.45   &\\
\hline
\end{tabular}
\label{bvrhostflux}
\end{table}


\subsection{\swift{} inter-band correlation analysis and host
 galaxy contribution in the UV/optical bands}

The \swift{} X-ray, UV, and optical light curves
based on the variability campaign in 2015 are shown 
 in Figure~\ref{lc_2015.ps}.
They all exhibit a similar variability pattern except for the V band, which exhibits no major variability amplitudes.

\begin{figure}
\centering
\includegraphics[width=6.5cm,angle=-90]{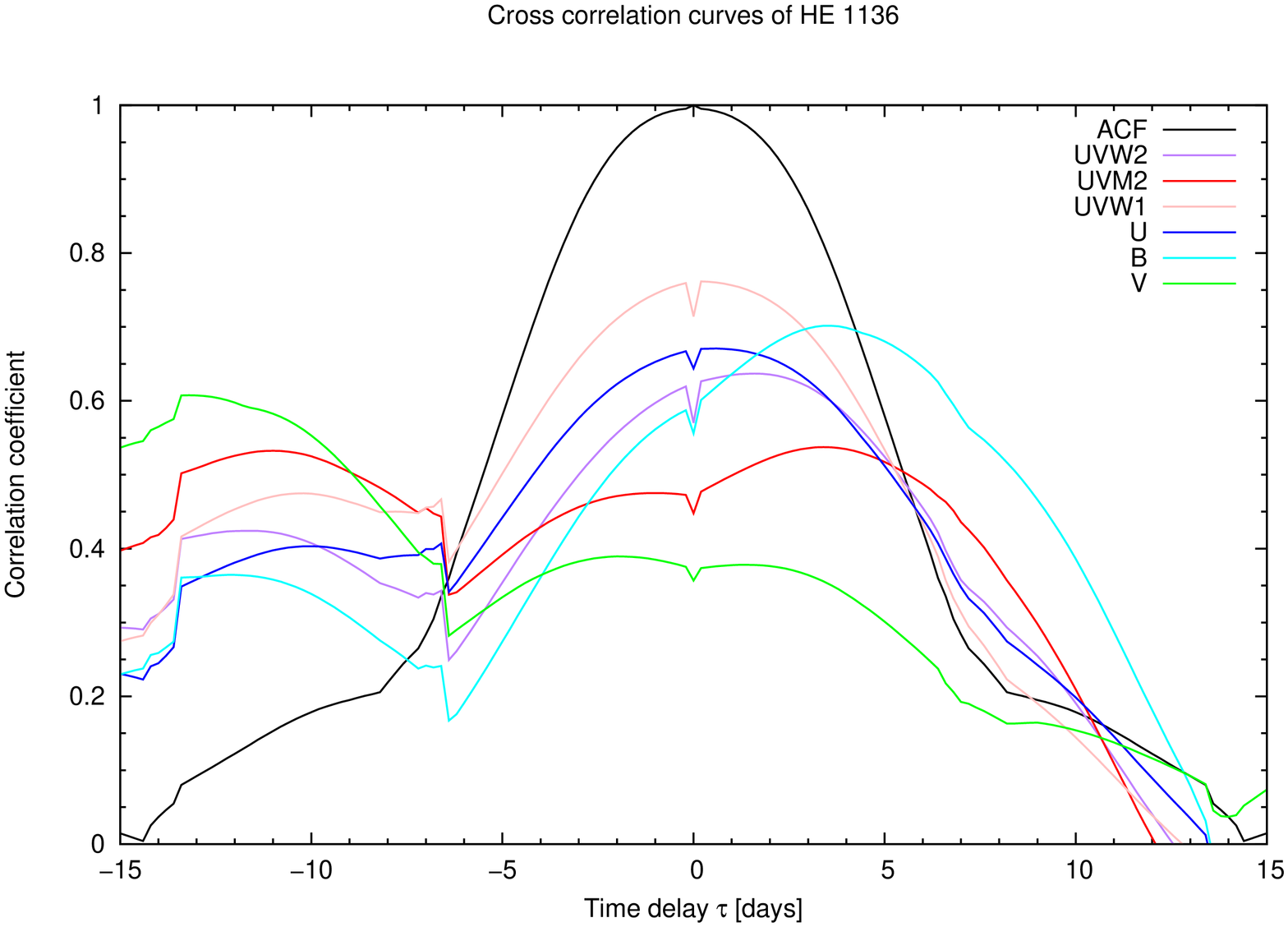}
\caption{Cross-correlation functions ICCF($\tau$) of the \swift{} bands UVW2, UVM2,
UVW1, U, B, and V with respect to the XRT light curve. Also shown is  an auto-correlation
function (ACF) of the XRT band. 
}
\label{ccf_xrt3_2015.ps}
\end{figure}

Based on these light curves 
we present the cross-correlation functions
ICCF($\tau$) of all the \swift{} UVOT bands 
with respect to the XRT light curve
in Figure~\ref{ccf_xrt3_2015.ps}.
In addition, we show the auto-correlation
function (ACF) of the XRT band.
We used the cross-correlation method as
described in, e.g., Dietrich \& Kollatschny\cite{dietrich95}  
and Kollatschny et al.\cite{kollatschny14}.
Table~\ref{optuvcontdelays}
lists the maximum correlation coefficient $r_{\rm max}$ of the individual
\swift{} bands with respect to the XRT band as well as the lags with respect
to the XRT band. We derive the centroids of these ICCF, $\tau_{\rm cent}$,
by using only the part of the CCF above 80\% of the peak value.
 It has been shown by Peterson et
al.\cite{peterson04} 
that a threshold value of 0.8 $r_{\rm max}$ is generally a good choice.
We determine the uncertainties of our cross-correlation results by
calculating the cross-correlation lags many times using
a model-independent Monte Carlo method known as
{flux redistribution/random subset selection} (FR/RSS).
This method was described
by Peterson et al.\cite{peterson98}.
The uncertainties correspond to 68\% confidence levels.

\begin{table}
\centering
\tabcolsep+9mm
\caption{\swift{} inter-band correlation coefficients \textbf{($r_{max}$)} and
  lags \textbf{($\tau$)}.}
\begin{tabular}{lcd}
\hline 
\noalign{\smallskip}
Band &$r_{max}$ &\multicolumn{1}{c}{$\tau$} \\
     & &\multicolumn{1}{c}{[days]}\\
(1)  &(2) &\multicolumn{1}{c}{(3)}\\
\noalign{\smallskip}
\hline
\noalign{\smallskip}
XRT (ACF) &    1.00 &       0\\
UVW2      &    0.64 &    1.3_{-3.5}^{+3.0} \\  
UVM2      &    0.54 &    2.6_{-4.0}^{+4.4} \\  
UVW1      &    0.76 &    0.3_{-4.0}^{+2.9} \\ 
U         &    0.67 &    0.5_{-3.0}^{+3.9} \\ 
B         &    0.70 &    3.5_{-2.6}^{+6.6} \\ 
V         &    0.39 &    - \\
\noalign{\smallskip}
\hline 
\end{tabular}
\label{optuvcontdelays}
\end{table}

The V-band light curve does not show any significant correlation
 with respect to the X-ray light curve
 (see Figure~\ref{ccf_xrt3_2015.ps}).
This might be caused by the nonthermal AGN contribution
in the V band being  less than 25\%\   (see Table~\ref{bvrhostflux}).
Furthermore, the light distribution of the host galaxy is not exactly point-like,
as seen in Figure~\ref{DSSbildHE11_2arcmin.ps}. Therefore, measurements
made with a large aperture in the V band are more
sensitive to small-scale deviations from an exact centering.
By contrast, the X-ray and UV bands are dominated by the central nonthermal
point source. Additionally, the V band is contaminated by the variable
H$\beta$ line (see Figure~\ref{filter_format.eps}).

\begin{figure}
\centering
\includegraphics[width=6.5cm,angle=-90]{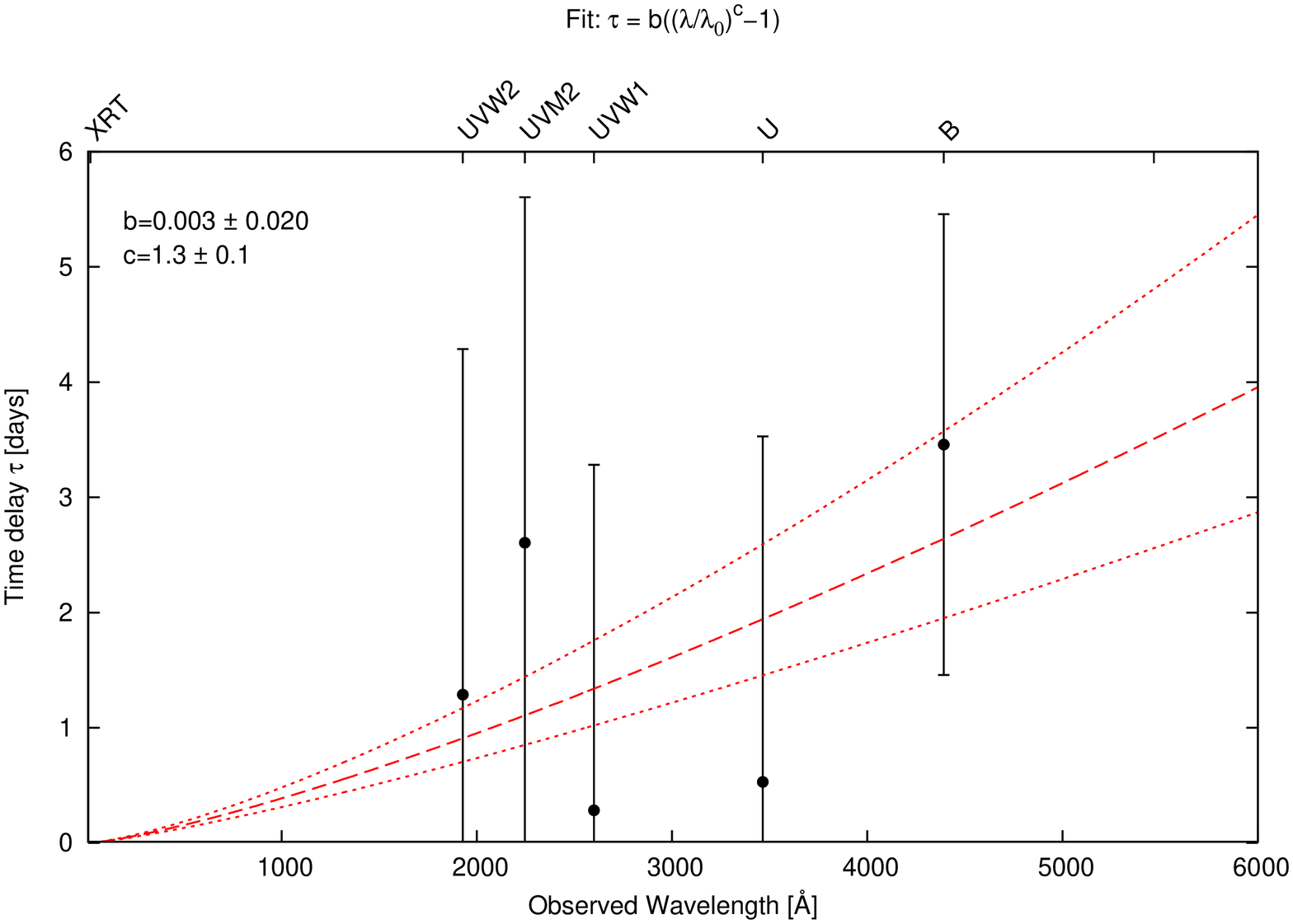}
\caption{Time delay of the \swift{} UV and optical bands with respect to the
\swift{} XRT light curve as a function of the wavelength of the \swift{} bands. 
The V band has been excluded as it showed  a very low correlation
coefficient. The dashed line shows the best fit to the data. The dotted
lines give the error of the exponent c.}
\label{delay_vs_wavelength4.ps}
\end{figure}

Figure~\ref{delay_vs_wavelength4.ps} shows the time delay of the \swift{} UV
and optical bands with respect to the
\swift{} X-ray light curve as a function of wavelength. 
The V band has been excluded here as it showed no correlation.
The dashed line shows the most general fit to the data: 
\begin{align*}
 \tau = b((\lambda/\lambda_{0})^{c}-1)  
\end{align*}
with $\lambda_{0}=25\,\AA$.
The b-value and the power-law index  
c have been allowed to vary. First we determined the
 fit parameter b = 0.003$\pm$0.020 light-days giving a
hint on the size of the X-ray emitting region at $\lambda_{0}=25\,\AA$
(corresponding to $\lambda_{\rm pivot}$ of XRT filter).
Afterwards we kept b fixed and calculated the exponent c.
The best fit to the data gives c = 1.3$\pm$0.1.
This value is consistent with a theoretically expected value c = 1.33 = 4/3
for an irradiated accretion disk (see discussion section).
 
\begin{figure}
\centering
\includegraphics[width=6.5cm,angle=-90]{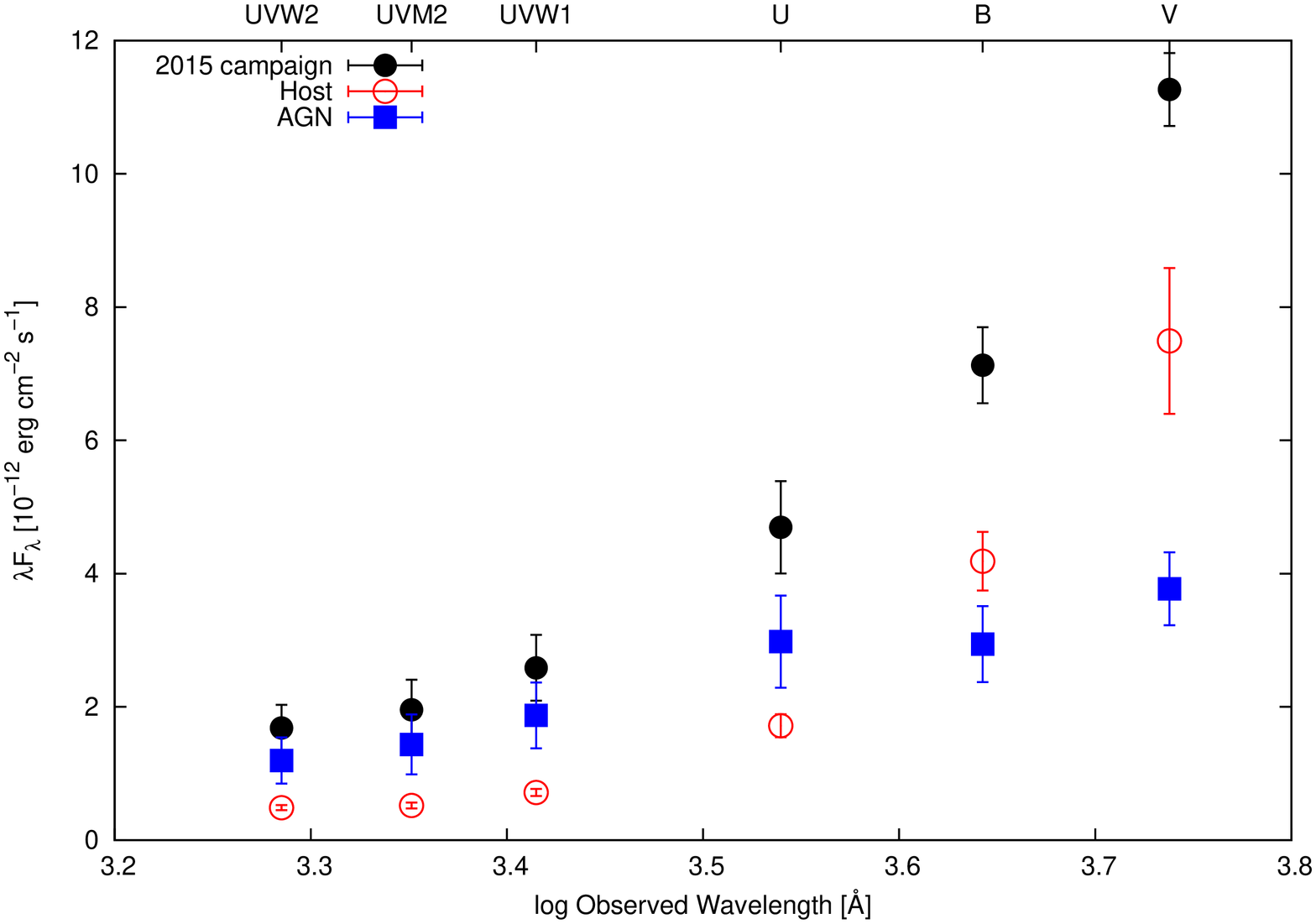}
\caption{Mean UV and optical spectral energy distribution of HE\,1136-2304
based on the \swift{} data taken in 2015 (black filled circles). 
The red open circles and the blue squares indicate the
contributions of the host galaxy and the AGN, respectively.  
  }
\label{lambdaF_lambda_vs_loglambda_2015.ps}
\end{figure}

\begin{figure}
\centering
\includegraphics[width=6.5cm,angle=-90]{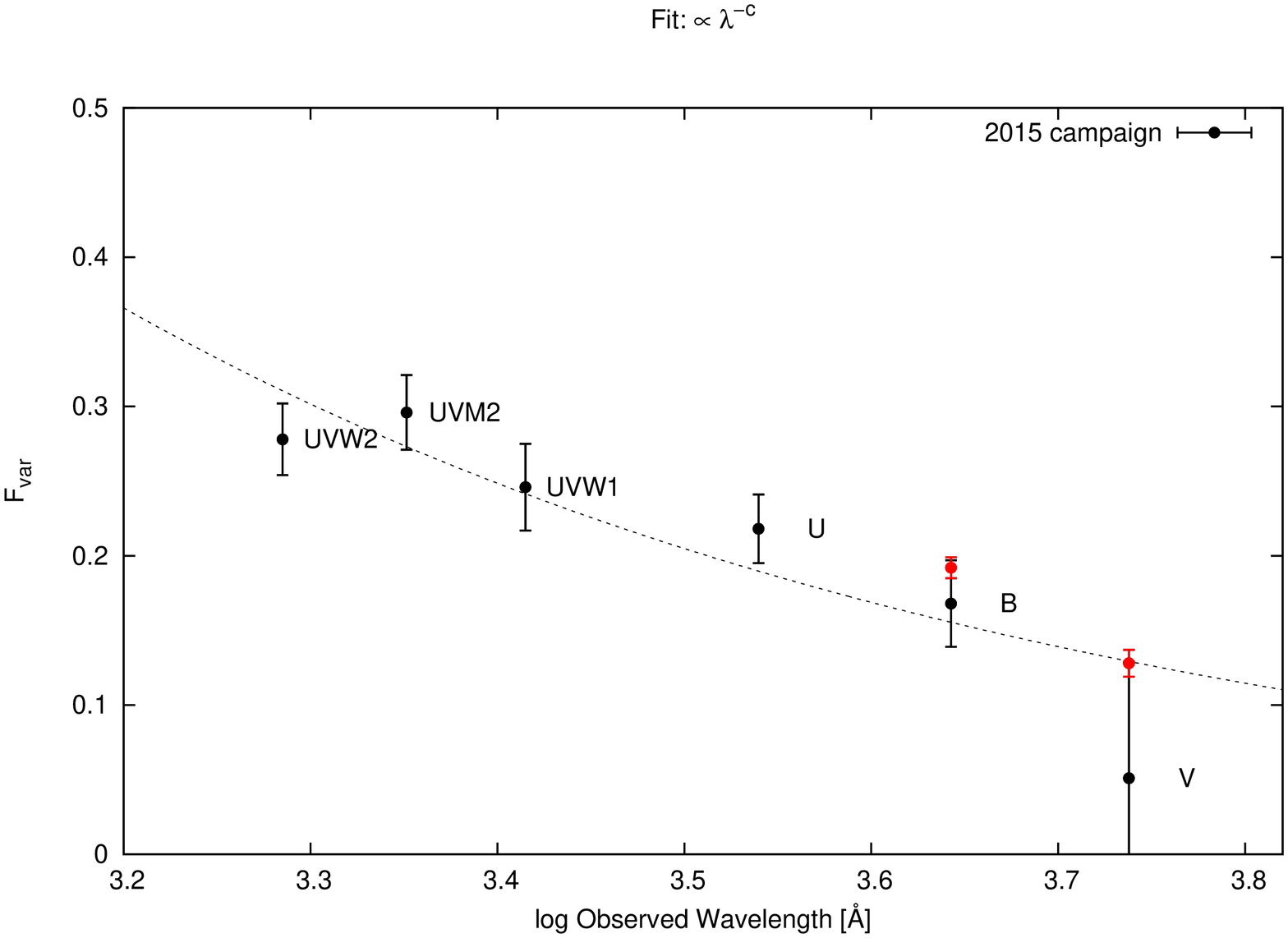}
\caption{Fractional variations of the UV and optical continuum bands derived from
the \swift{} data in 2015 as a function of wavelength.
 Furthermore, the B and V band
measurements  based on the photometric data taken in 2015 have been added.
The contribution of the host
galaxy has been subtracted in all cases. The dashed line shows a general
fit with an exponent c = 0.84.
}
\label{fvar_vs_loglambda_2015_nox.ps}
\end{figure}

The UV and optical
spectral energy distribution of HE\,1136-2304
based on our \swift{} data taken in 2015
is presented in Figure~\ref{lambdaF_lambda_vs_loglambda_2015.ps}
with black symbols.
The red open circles show the
contribution of the host galaxy in the individual bands.
The host contribution in the B and V bands
is based on the flux variation gradient analysis (section 3.2).
We calculated the contribution of the host galaxy in the  UV bands
by scaling an Sb spectrum 
(Kinney et al.\citealt{kinney96}) with respect to the B and V 
fluxes of the host galaxy.
The AGN flux in the individual bands has been determined  
by subtracting the flux of the host galaxy from the observed flux.
The blue filled squares in Figure~\ref{lambdaF_lambda_vs_loglambda_2015.ps} 
give the AGN flux contribution in the
individual \swift{} bands.
Knowing the AGN flux contribution in the individual \swift{} bands, we can
derive the pure fractional variations in those bands.
We present the fractional variations
of the UV and optical continuum bands recorded
with \swift{} in 2015 as a function of wavelength in
Figure~\ref{fvar_vs_loglambda_2015_nox.ps}. 
The contribution of the host
galaxy flux has been subtracted from the individual filter bands.
We then add (in red) the fractional variations in the B and V bands
on basis of  our measurements with the different telescopes in 2015.
There is a clear trend of increasing fractional variation of the AGN towards the UV. 
The dashed line shows a general
fit to the data with a 
value c = 0.84.

\begin{figure}
\centering
\includegraphics[width=6.5cm,angle=-90]{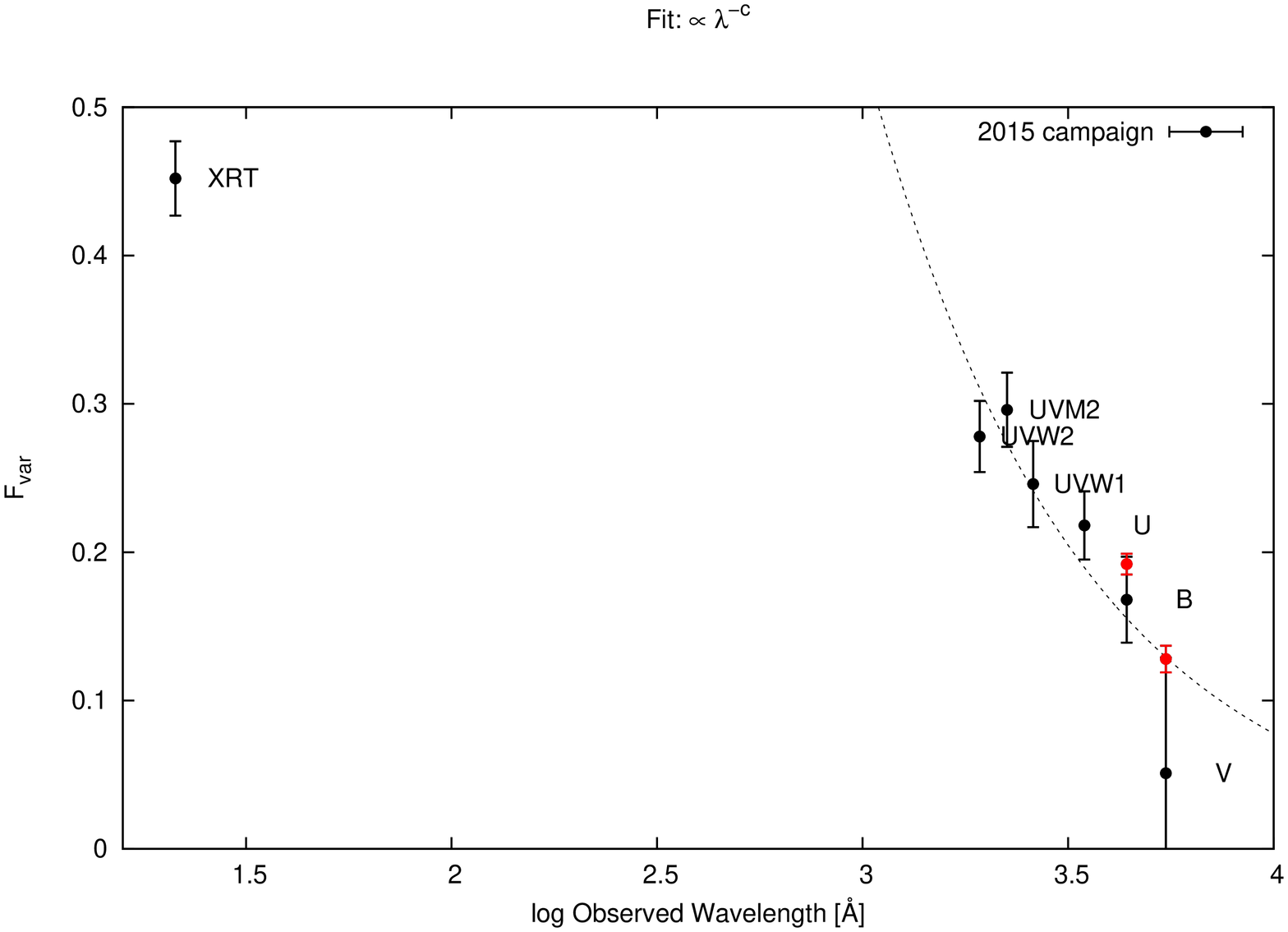}
\caption{Fractional variations of the X-ray, UV, and optical continuum bands
measured from the
 \swift{} data  in 2015 as a function of wavelength.
}
\label{fvar_vs_loglambda_2015.ps}
\end{figure}

We present the fractional variation of the X-ray band together with the fractional variations of the UV/optical bands in Figure~\ref{fvar_vs_loglambda_2015.ps}.
The fractional variations in X-rays are the strongest
(as seen in Table~\ref{swiftvarstatistics}). However, the
fractional variations in X-rays do not follow the same trend as
seen for the fractional variations in the UV and optical bands.
An extrapolation of the fit in the UV and optical bands
does not line up with the X-ray observations.
This is an indication that the origin of the X-ray continuum emission is
not connected in a simple way with the origin of the UV/optical emission
(see Section 4.2).

\subsection{Spectral type changes and
long-term variability of HE\,1136-2304 }

The first spectrum of HE\,1136-2304 was taken in 1993
(Reimers et al.\citealt{reimers96}).
At that time no broad H$\beta$ emission line component was present
in the spectrum. Only a faint broad H$\alpha$ component was visible.  
Therefore, this galaxy was of nearly Seyfert 2 (1.95) type in 1993.
Another spectrum of HE\,1136-2304 was obtained on 2002 May 16 as part
of the 6dF Galaxy Survey  (Jones et al.\citealt{jones04}). At this time 
HE\,1136-2304 was of the same spectral AGN type as it was
nine years earlier.
The AGN type had changed to Seyfert 1.5
when we took an optical spectrum in 2014 July (Parker et al.\citealt{parker16}).
This happened together with an
increase in the optical continuum flux and with a strong increase
in X-ray flux.

\begin{figure*}
\centering
\includegraphics[width=11.7cm,angle=-90]{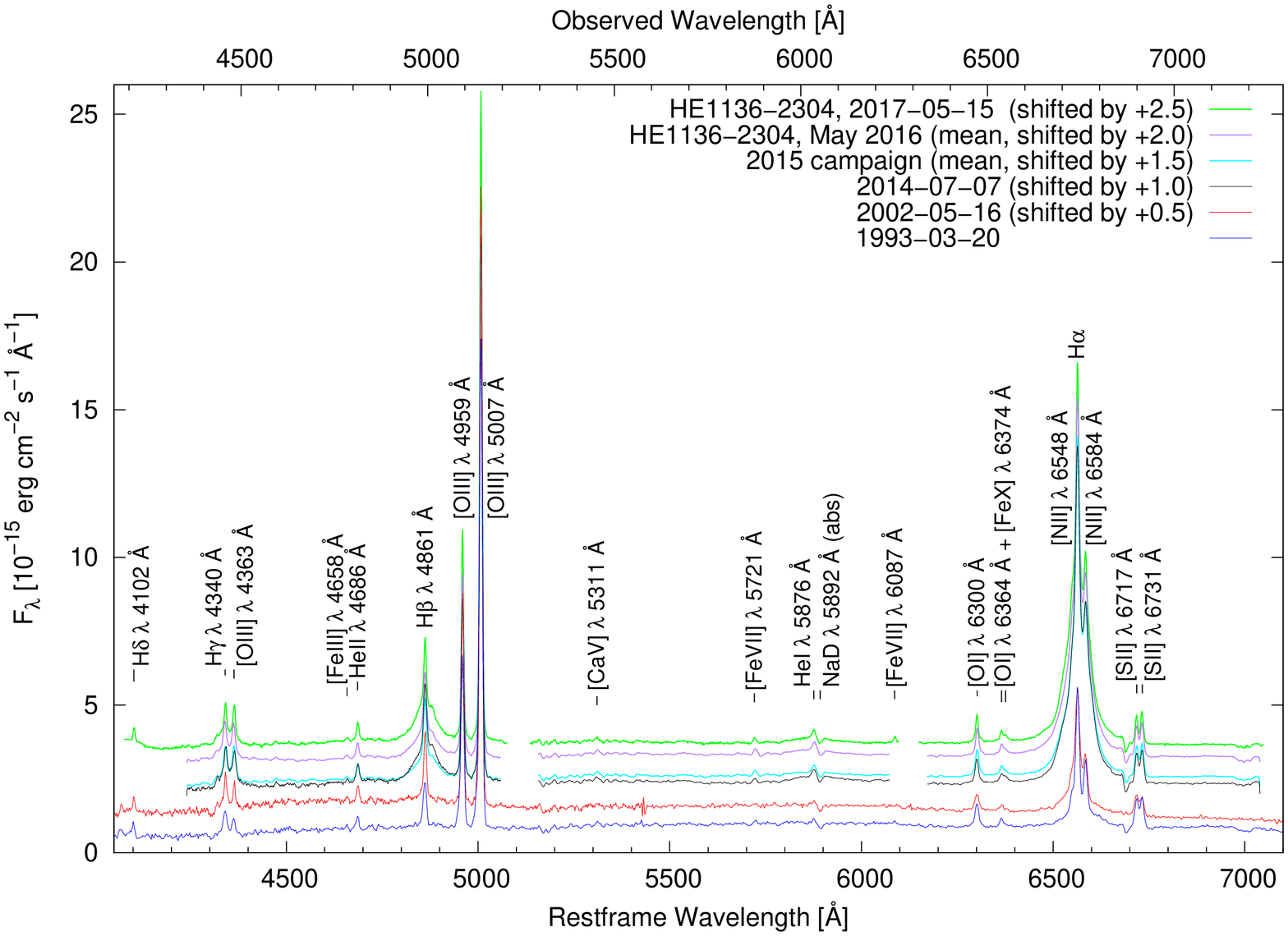}
\caption{Optical spectra of HE\,1136-2304 for the epochs 1993, 2002, 2014, 2017
as well as mean spectra for 2015 and 2016.}
\label{spec_1993_2002_2014_2015_2016May_2017.ps}
\end{figure*}

\begin{figure}
\centering
\includegraphics[width=6.2cm,angle=0]{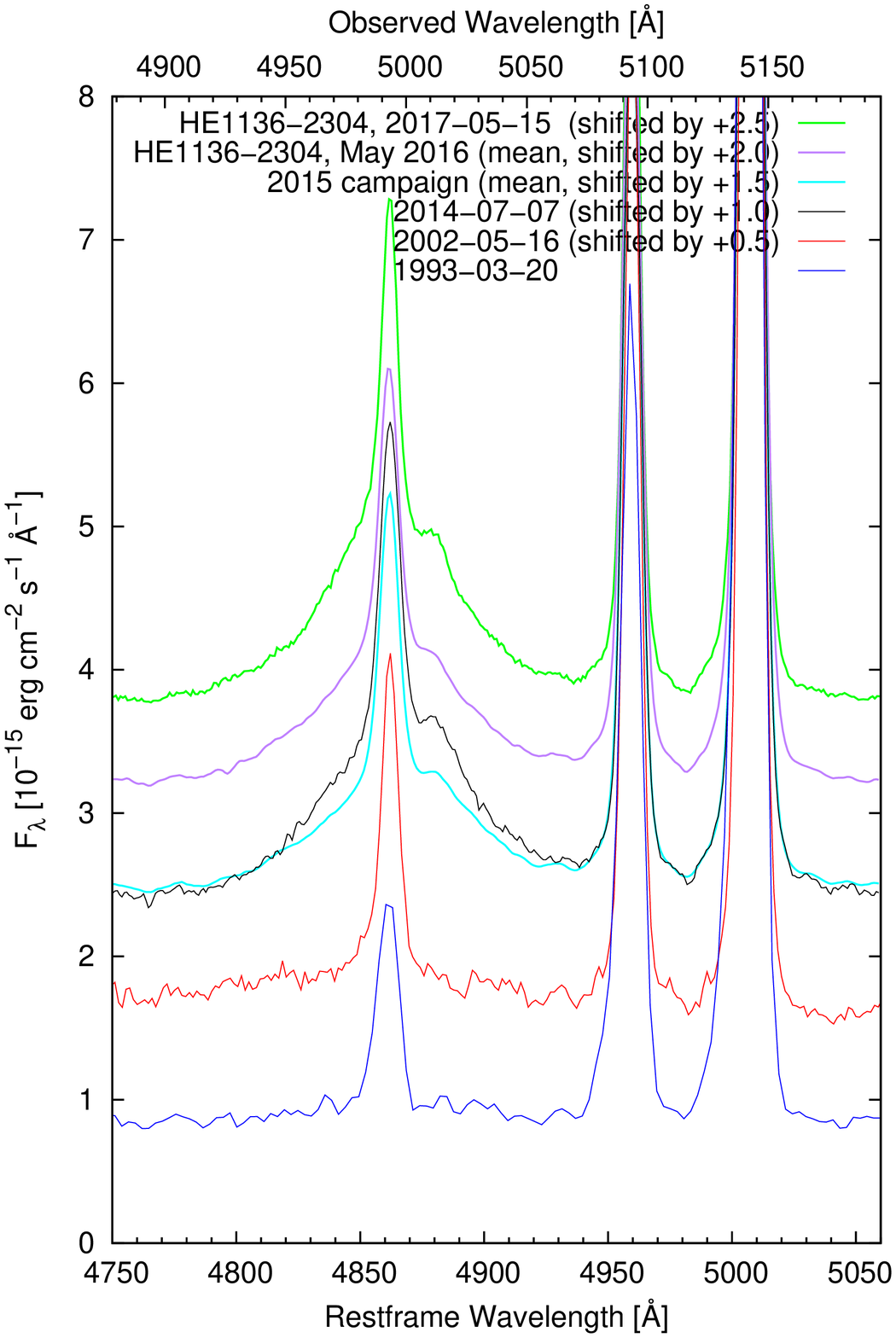}
\caption{Optical spectra of HE\,1136-2304,  as in Fig. 
\ref{spec_1993_2002_2014_2015_2016May_2017.ps}, but showing 
the H$\beta$ profiles in more detail.
}
\label{spec_6times_hb.ps}
\end{figure}

\begin{figure}
\centering
\includegraphics[width=6.2cm,angle=0]{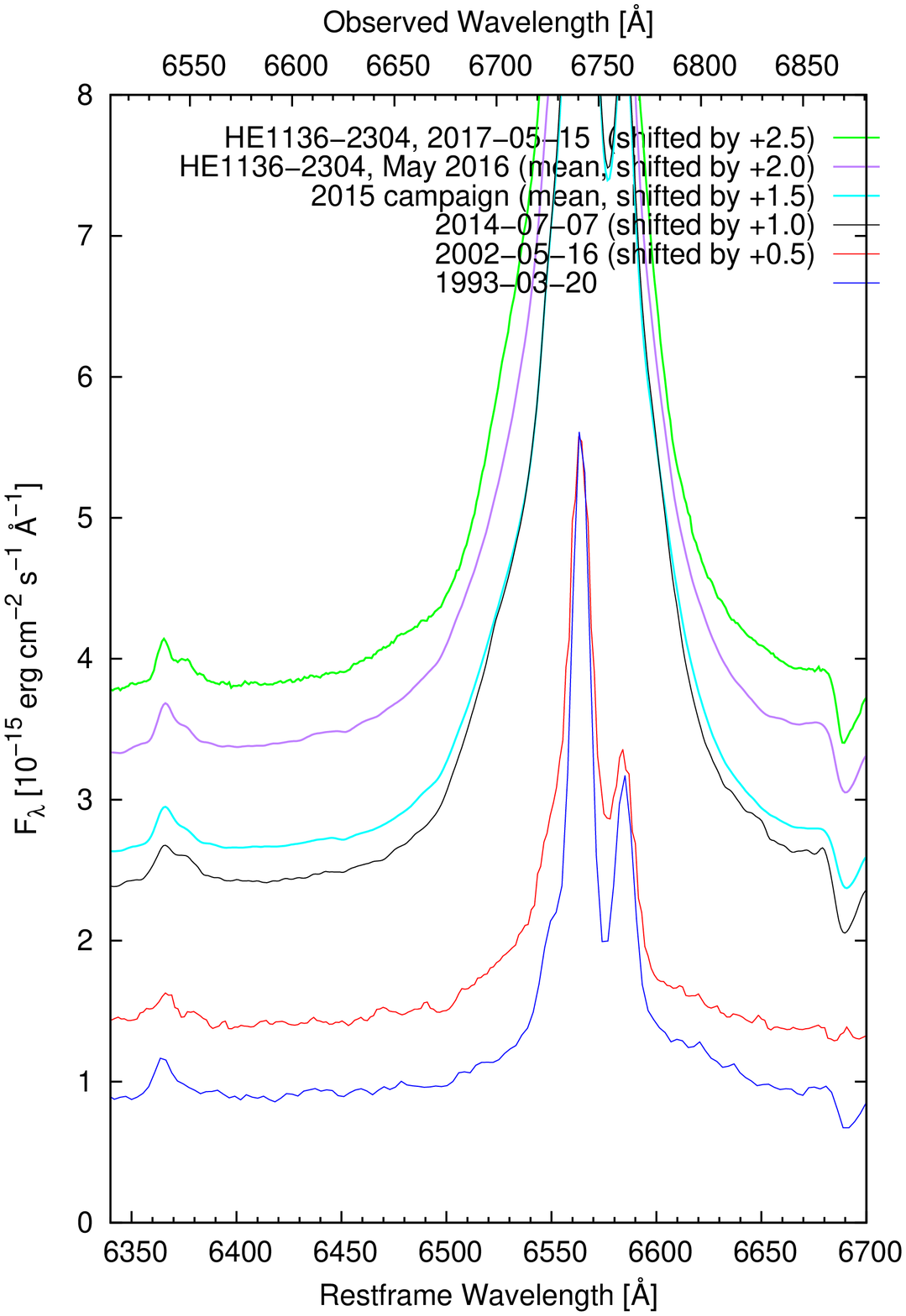}
\caption{Optical spectra of HE\,1136-2304,  as in Fig. 
\ref{spec_1993_2002_2014_2015_2016May_2017.ps}, but showing  
the H$\alpha$ profiles in more detail.
}
\label{spec_6times_ha.ps}
\end{figure}

In Figure~\ref{spec_1993_2002_2014_2015_2016May_2017.ps}
the spectra of
HE\,1136-2304 taken in 1993, 2002, 2014, 2015, 2016,
and 2017 are shown to present line profile changes.
The spectra are shifted by a constant with respect to each other.
Figures~\ref{spec_6times_hb.ps} and
\ref{spec_6times_ha.ps} show the H$\beta$ and H$\alpha$ spectral regions
in more detail.
%
%
The mean spectrum for  2015 is based on the variability
campaign carried out in 2015.
We will present details of this campaign
in a separate publication.
The spectrum shown for 2016 is the mean of two spectra
taken in May 2016.
The strong broad component in the H$\beta$ line profile 
that appeared in  2014 remained there for the subsequent
years until 2017. No major profile changes occurred. 
HE\,1136-2304 remained  a Seyfert 1.5 type.

\begin{figure*}
\centering
\includegraphics[width=9.cm,angle=-90]{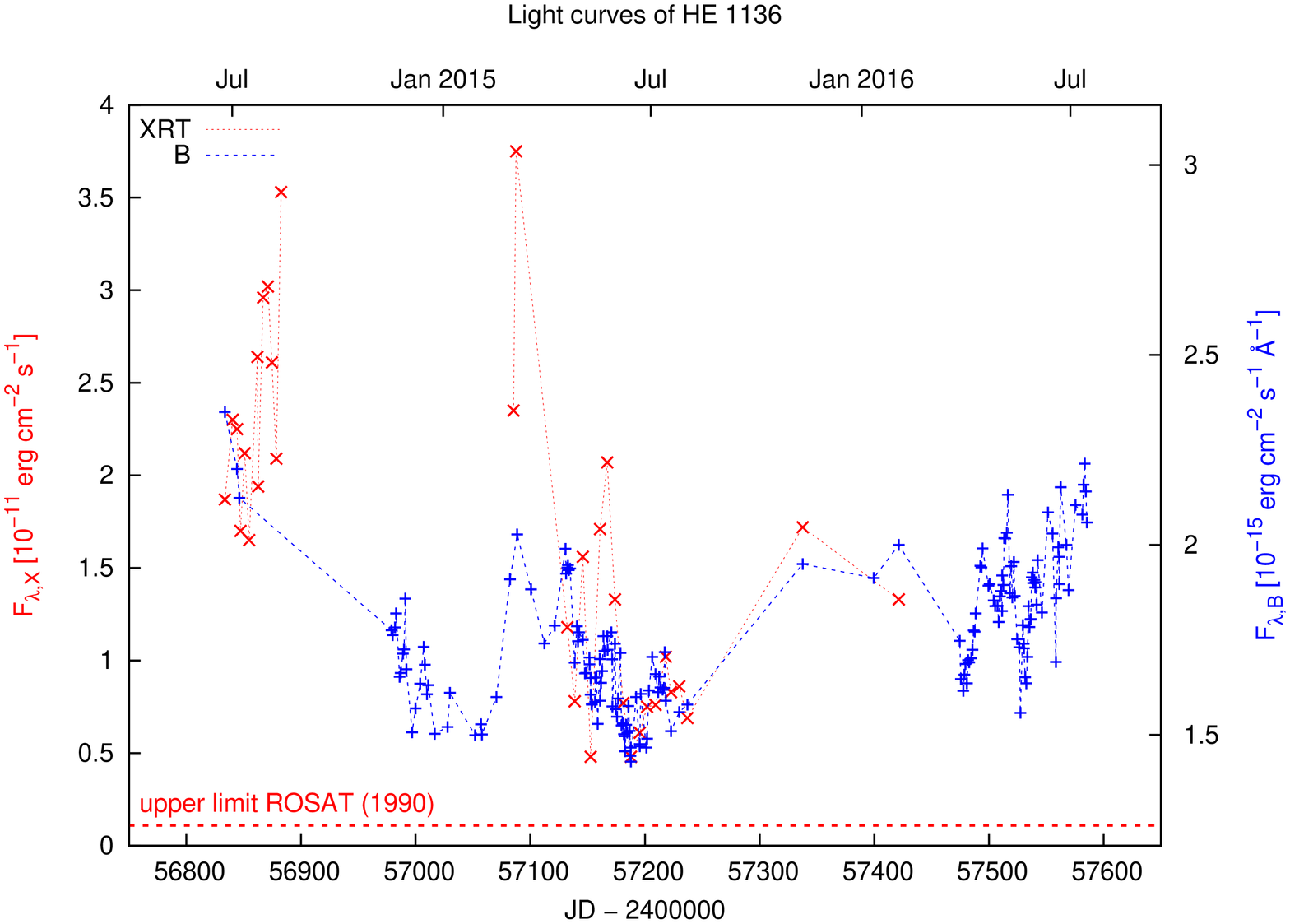}
\caption{Optical (blue) and X-ray (red) light curve from 2014 to 2016. 
The upper limit of the X-ray flux in 1990 (ROSAT) is shown by a horizontal
dashed line.}
\label{lc_xrt_plus_b_short.ps}
\end{figure*}

\begin{figure*}
\centering
\includegraphics[width=9.cm,angle=-90]{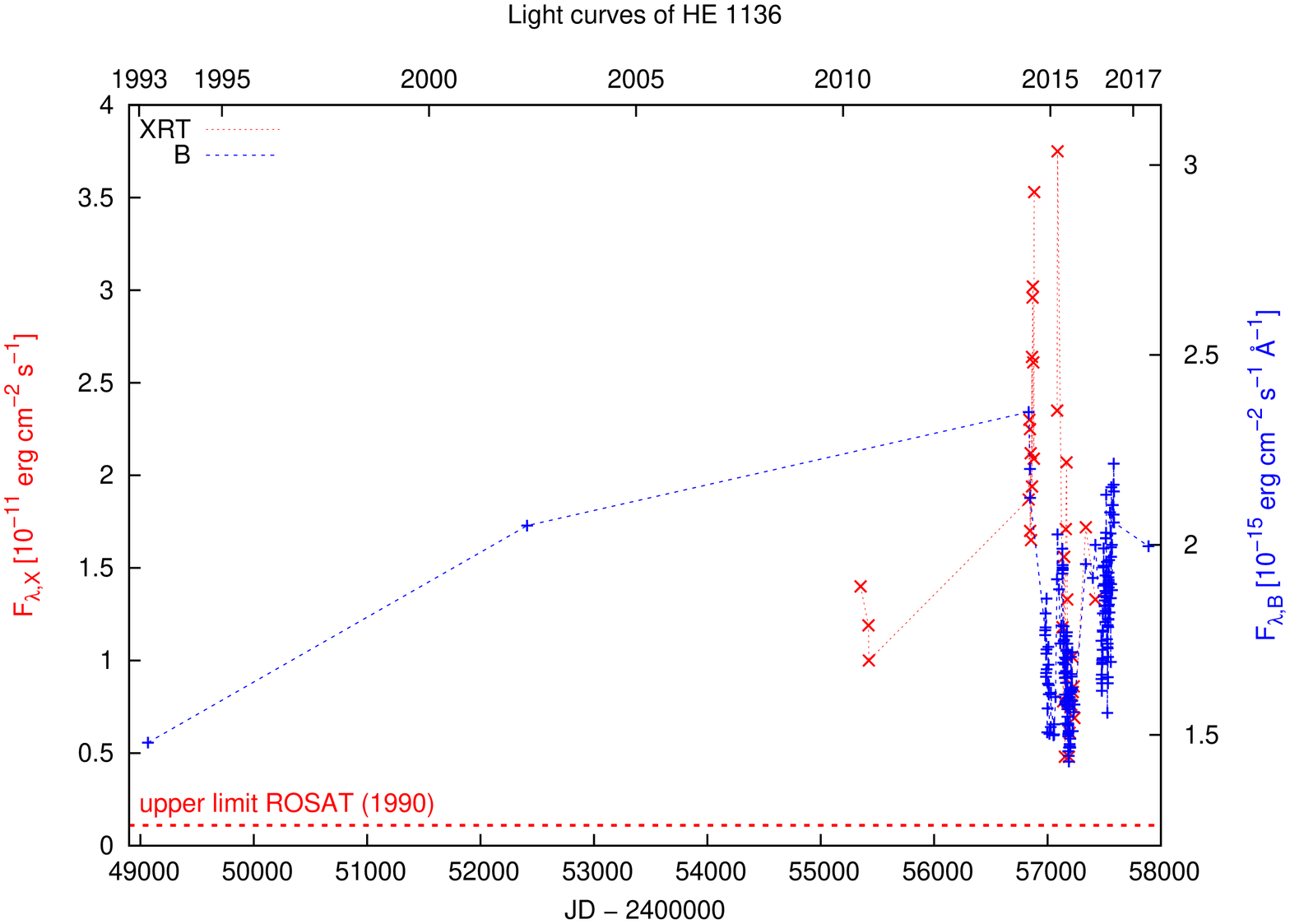}
\caption{Long-term optical (blue) and X-ray (red) light curves 
of HE\,1136-2304 from 1993--2017. The \swift{} X-ray data and 
the upper X-ray limit in 1990 based on ROSAT is presented in red
(left axis label).
The B-band data are scaled with respect to the X-ray data, and  are shown in blue
(right axis label).}

\label{lc_xrt_plus_b_long.ps}
\end{figure*}

We compared the spectral variations of the data from  
1993 to 2017 with the variability behavior
in the optical and X-ray continuum. 
Figure~\ref{lc_xrt_plus_b_short.ps} shows
the X-ray
and optical B-band continuum variations from 2014 to 2016.
The long-term trends for 1993--2017 are presented in Figure~\ref{lc_xrt_plus_b_long.ps}.
The \swift{} X-ray data and the ROSAT upper limit for 
1990 are presented in red (axis label on the left side).
The optical continuum variations are given in blue.

We scaled the amplitude of the optical light curve with respect
to the X-ray light curve. The 
scaling has been carried out with regard to nearly simultaneous
observations in the optical and X-rays
in 2014 July and for the combined variability 
campaign in 2015. The axis label for the blue continuum
is given on the right side.
Dramatic continuum
variations in X-rays and in the optical occurred between 2014 and 2016.
The optical continuum closely follows the X-ray
flux.
HE\,1136-2304 was in a low state in X-rays and in the optical before 2000.

\section{Discussion}

One of the main goals of  our optical, UV, and X-ray variability campaign was to investigate the variability behavior of the changing-look AGN HE\,1136-2304 subsequent to the outburst in 2014 July.
Our campaigns in X-rays and UV,  and at optical wavelengths 
lasted for two and three years, respectively.

A strong and sudden outburst in AGN can in principle be caused by three
different events: gravitational lensing, a tidal disruption event (TDE),
or major changes in the accretion process.
Light curves caused by a lensing effect should exhibit a characteristic
smooth, single-peaked shape (e.g., Bruce et al.
\citealt{bruce17}).
A tidal disruption event is characterized
 by a sudden dramatic rise in luminosity
and a steady decline to quiescence on timescales of months to years
(Rees\citealt{rees88}). Some candidates for TDEs 
  in X-rays, UV, and optical bands have been presented by,
e.g., Komossa \& Bade\cite{komossa99}, Gezari et al.\cite{gezari08}, and 
Holoien et al.\cite{holoien16}.
However, the variability pattern of HE\,1136-2304
 following the outburst
in 2014 shows various outbursts on timescales of days to months 
typical for ``ordinary'' AGN variability (see Figures~\ref{LC_B_all_ochm.ps}, \ref{LC_V_all_ochm.ps}, and 
\ref{lc_swift_2014_2016}). Therefore,
we can rule out a micro lensing
or tidal disruption event as the cause of the observed variability pattern
seen in HE\,1136-2304.

\subsection{Optical continuum variability in
 HE\,1136-2304}

HE\,1136-2304 shows no systematic long-term trends in the continuum light curves
(see Figure~\ref{LC_B_all_ochm.ps})
since the start of our  variability campaign in 2014 July.
After two years the light curve reaches approximately
 the same flux level as in 2014 July, while showing unsystematic flux variations down to about 50\%\ in between.

 One way to measure the strength of the  variability in AGN
is to determine their
fractional variation F$_{\rm var}$. 
The fractional variation depends on the duration
of the monitoring campaign, 
on the examined wavelength, and on the (accurate) decomposition
of the host galaxy contribution. 
A typical value for the fractional variation F$_{\rm var}$ 
of the continuum at  5100$\,\AA$ is 0.05 to 0.15
for variability periods of 6--12 months:
e.g., NGC\,5548 (Peterson et al.\citealt{peterson02}, Fausnaugh
et al.\citealt{fausnaugh16}),
3C\,120 (Kollatschny et al.\citealt{kollatschny14}).
For variability campaigns  over longer periods, 
typical  F$_{\rm var}$  continuum values at  5100$\,\AA$ are 
to 0.1 to 0.25: e.g., NGC\,5548 (Peterson et al.\citealt{peterson02}),
Ark\,564  (Shapovalova et al.\citealt{shapovalova12}),
Mrk\,110  (Kollatschny et al.\citealt{kollatschny01}),
or 3C\,120 (Kollatschny et al.\citealt{kollatschny14}),
 and a collection of
many AGN in Kollatschny et al.\citealt{kollatschny06}.
This higher F$_{\rm var}$  value is caused by the irregular variations of AGN
on longer timescales. There
is a higher probability for observing stronger variability amplitudes 
when monitoring over longer periods of time.  
We determined optical F$_{\rm var}$ values
of 0.11 (5360$\,\AA$) and 0.14 (4570$\,\AA$) 
for our campaign in 2015 (based on the SALT spectra). 
Our value of 0.11 for F$_{\rm var}$ indicates
that the continuum variations of HE\,1136-2304 
were equal to or even stronger than other AGN, in particular after correcting for the flux contribution
of the host galaxy.

For a more detailed inspection of the AGN variability, the contribution of the host galaxy starlight should be subtracted
 before comparing the amplitudes of different AGN.
The relative contribution of the host galaxy flux
is quite different
in spectroscopic and photometric data
(see  Table~\ref{bvrhostflux}).
The typical contribution of the host galaxy is larger if it is based on broadband photometry  because the typical aperture for broadband photometry is larger
than that for spectral photometry.
One way to estimate the contribution of the host galaxy is to create
nucleus-free images of the AGN based on HST images 
(e.g., Bentz et al.\citealt{bentz09}) or by decomposition 
of the observed AGN spectra (e.g., Barth et al.\citealt{barth15}). 
Typical values for the relative host galaxy flux contribution
are on the order of 20\%\ to 60\%\  
in optical AGN spectra (see Figure 4 in  Barth et al.\citealt{barth15}).
The flux variation gradient (FVG) method (Choloniewski
\citealt{choloniewski81},  Winkler et al.\citealt{winkler92}, Haas et al.
\citealt{haas11}, Ramolla et al.\citealt{ramolla15}) is
another way to estimate the relative contribution of the host galaxy
flux to the variable continuum flux. 
A typical value for the relative  contributions of the host galaxy flux
is on the order of 50\%
(e.g., Haas et al.\citealt{haas11}) for broadband photometry. 
The contribution of the host galaxy flux in HE\,1136-2304 
amounts to 50\% (for spectrum photometry) and to 75\%
(for broadband photometry)
in the V band. Therefore, the variability amplitude in HE\,1136-2304
remains quite high in comparison to other AGN after subtraction of the host galaxy flux.

\subsection{Comparison of X-ray variations against UV/optical
variations}

The time delays of the individual \swift{} UV/optical light curves with respect to the
\swift{} X-ray light curve
 are presented in Figure~\ref{delay_vs_wavelength4.ps}. 
There is a trend that the UV/optical light curves
at higher frequencies show shorter  delays.
A general fit to the data in Figure~\ref{delay_vs_wavelength4.ps}
resulted in a value of 0.003$\pm$0.020 light-days
for the fit parameter b in 
$\tau = b((\lambda/\lambda_{0})^{c}-1)$, with $\lambda_{0}=25\AA$.\\ 
This functional form of $\tau$ has been discussed before by
Edelson et al.\cite{edelson15} and Fausnaugh et al.\cite{fausnaugh16}
in the context of
the \swift{} and HST reverberation mapping campaign on NGC\,5548.
The value of b gives an
estimate of the size of the X-ray emitting region.
A value of 0.020 light-days (based on the error of the b value)
corresponds to $5.1 \times 10^{11}$ m.
We can compare this size with the Schwarzschild radius for a central
black hole mass of $M = 4 \times 10^{7} M_{\odot}$ (Kollatschny et al.,
in prep.): 
$1.2 \times 10^{11}$ m. This indicates that
the X-ray emission originates at a distance of a few 
Schwarzschild radii from the center which is consistent
with the last stable orbit of a Schwarzschild black hole.

The second parameter we derived from the general fit shown in
Figure~\ref{delay_vs_wavelength4.ps} is the parameter c = 1.3$\pm$0.1.
This value is close to a theoretically expected value c = 1.33 = 4/3
for an irradiated accretion disk where the geometrically thin, optically
thick accretion disk is hotter in the inner radii and cooler in the outer radii
 (e.g., Cackett et al.\citealt{cackett07},
Edelson et al.\citealt{edelson15}). The optical continuum
is delayed by about three light-days with respect to the X-ray
variations.  
Similar delays of approximately 3 to 4 light-days
 of the optical continuum bands with respect to the UV/X-ray bands
 have also been seen in \swift{} variability campaigns
of NGC\,2617 (Shappee
et al.\citealt{shappee14}) and NGC\,5548
 (Fausnaugh et al.\citealt{fausnaugh16}).
A further indication that the optical continuum in HE\,1136-2304 is delayed with
respect to the ionizing continuum will be presented
in a future paper (Kollatschny et al., in prep.) where we show that
the outer line wings, for example  in  H$\beta$, respond faster than the adjacent
optical continuum.

As shown in Figure~\ref{fvar_vs_loglambda_2015_nox.ps}, the fractional variability F$_{\rm var}$ of the continuum bands
is a function of their
wavelength.
The variations are stronger at shorter wavelength bands.
This means that the strength of the fractional variability can be considered
as a proxy for the distance of the
continuum emitting region with respect to the center.
Similar to the time delay of the individual continuum bands, we can test whether
a power-law model 
$F_{\rm var}=a\cdot\lambda^{-c}$  is consistent
with the observations. 
The optimal c value c\,=\,0.84 we found is close to a simple
power law with c\,=\,1: $F_{\rm var}=a\cdot\lambda^{-1}$. 
Furthermore, as shown in 
Figure~\ref{fvar_vs_loglambda_2015.ps}, the  
fractional variations in the X-ray band are  50\% stronger than those in the UV bands.
However, the
magnitude of the fractional variation in  X-rays
is not simply a continuation of the general trend 
seen in the UV and optical bands.
This indicates that the observed X-ray emission does not 
exactly follow the same trend as the UV/optical emission. 
The UV/optical continuum emission is generally associated with 
blackbody emission from the accretion disk
(e.g., Hubeny et al. \citealt{hubeny01}).

\begin{figure}
\centering
\includegraphics[width=6.5cm,angle=-90]{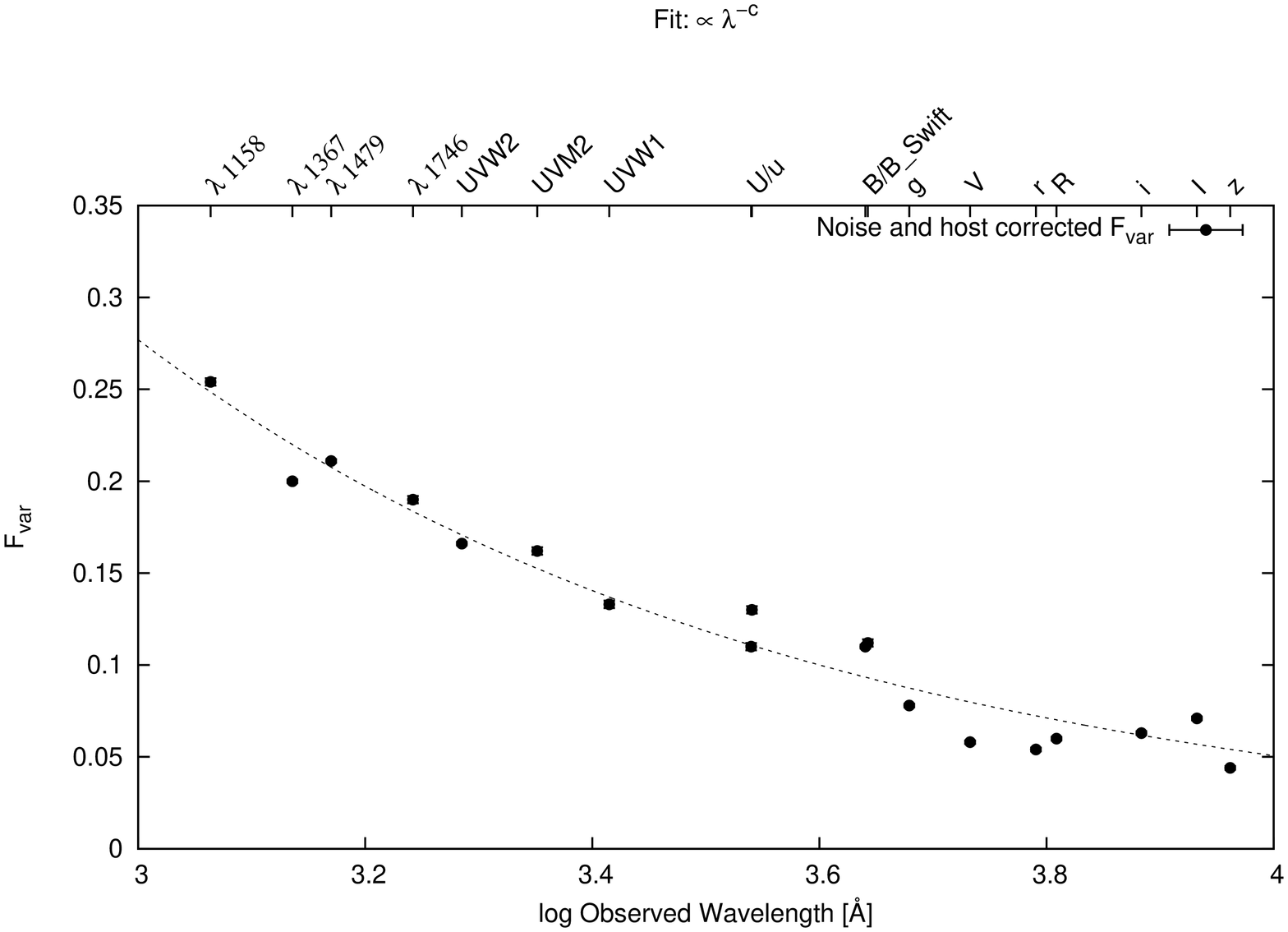}
\caption{Fit to the fractional variations in the UV and optical bands of an
 extensive variability campaign of NGC\,5548 (with c=0.74).
 The data are from
Fausnaugh et al.\cite{fausnaugh16}.
}
\label{Fausnaugh16_fig3_agnonly.ps}
\end{figure}

We tested whether the observed trend of the fractional variability
in the UV/optical bands of HE\,1136-2304 is present in other galaxies
as well, for example in NGC\,5548.
An extensive variability campaign of NGC\,5548 has been carried out in 2014
(Edelson et al.\citealt{edelson15}, Fausnaugh et al.\citealt{fausnaugh16}).
Their fractional variability data
of NGC\,5548 are shown in 
Figure~\ref{Fausnaugh16_fig3_agnonly.ps}.
A fit with $F_{\rm var}=a\cdot\lambda^{-c}$ and c\,=\,0.74 perfectly matches the
observations of NGC\,5548. This c value is close to  
the optimal c value of 0.84 we found for HE\,1136-2304. 
A comparison of the fractional variations of NGC\,5548
(Figure~\ref{Fausnaugh16_fig3_agnonly.ps}) with those of HE\,1136-2304
(Figure~\ref{fvar_vs_loglambda_2015_nox.ps}) shows that the
variations in the UV/X-ray bands of HE\,1136-2304 are stronger
by a factor of 2.3.

However, there are different trends when comparing the variably 
pattern in X-rays and in the UV/optical observed in HE\,1136-2304 with
those seen in  NGC\,5548
(Edelson et al.\citealt{edelson15}).
The UV/optical light curves of HE\,1136-2304 show the same pattern as the
X-ray light curve, while  this is not the case in NGC\,5548
(Edelson et al.\citealt{edelson15}). Furthermore,  in HE\,1136-2304 the 
strongest variability is observed in X-rays, while 
 this is not the case in NGC\,5548. 
Edelson et al.\cite{edelson15}  
suspected that the X-ray flux may not drive
the UV/optical light curves in NGC\,5548 because of his findings.
Such a statement cannot be made for HE\,1136-2304 based on its
light curves.

\subsection{Comparison of optical spectral changes  with continuum
variations in HE\,1136-2304 }

Early optical spectra of HE\,1136-2304 were taken  
in 1993 and 2002. At that time, it was of nearly Seyfert 2 (1.95) type. 
We took a spectrum in 2014 July. The spectral type
of HE\,1136-2304 had changed to be of Seyfert 1.5. 
Since then the spectral type remained the same
(see Figures~\ref{spec_1993_2002_2014_2015_2016May_2017.ps}
to \ref{spec_6times_ha.ps}).
There are no major variations present in the Balmer line profiles
for the period from 2014 to the present.
However, the optical and X-ray continua varied a lot
at the same time (see Figure~\ref{lc_xrt_plus_b_short.ps}).

It has been discussed by Parker et al.\cite{parker16}
 whether the outburst
seen in 2014 was caused by a flare due to a stellar disruption event
(e.g., Komossa \& Bade\citealt{komossa99}). In that case we would
have expected
a general decline in the continuum flux of HE\,1136-2304 over
time. However, the observed long-term behavior with repeated phases
of decreasing and increasing continuum flux 
between 2014 and 2017 contradicts this scenario.

On the other hand,
Elitzur et al.\cite{elitzur14} present a model in which the
broad-line spectral evolution
is connected with the AGN luminosity.
This might be controlled by the accretion rate onto the central black hole.
The long-term variations of HE\,1136-2304 from 1993 to 2014 
support this model:
 the X-ray flux increased by a factor of more than ten. 
The B-band continuum flux (without correction for the host galaxy contribution) 
increased by more than 60\%,
and the spectral type changed from a nearly Seyfert 2 type to a 
Seyfert 1.5 type.
A similar scenario,  but with decreasing continuum flux,  has been found
for Fairall 9.
The continuum flux dropped  to 20\%\ of its original flux
in 1978, and its spectral type changed
 from a quasar/Seyfert 1 type to a Seyfert 1.95 type
within  six years (Kollatschny \& Fricke,\citealt{kollatschny85}).
Spectral variations of this kind generally occur on timescales of years.

However, the spectral type
variations seem not to follow
the continuum intensity variations on shorter timescales (weeks to months). 
In 2015, HE\,1136-2304 varied in the X-rays by a factor of eight
 within  two months
(see Figure~\ref{lc_xrt_plus_b_short.ps}), whereas
the broad-line profiles varied only marginally over the period
from 2014 to 2017
(see Figure~\ref{spec_1993_2002_2014_2015_2016May_2017.ps}).
A detailed discussion of the spectral variability campaign
carried out in 2015 will be presented in a separate paper
 (Kollatschny et al., in prep.).

\section{Summary}

We present results of an optical, UV, and X-ray monitoring campaign of the
changing-look AGN HE\,1136-2304 carried out from 2014 to 2017.
This campaign took place after a continuum outburst in the
optical and X-rays in 2014 July connected to a spectral change from a Seyfert 1.95
to a Seyfert 1.5 type.
 Our findings can be summarized as follows: 

\begin{enumerate}[(1)]
\item The optical, UV, and X-ray continuum light curves show the same
 variability pattern. The amplitude decreases with increasing
wavelength. It 
varies by a factor of eight in X-rays, by a factor of four
in the UV, and by a factor of two in the optical continuum between   
2014 and 2016. The amplitude in the optical increases by a factor of three
after correction for the host galaxy contribution.
\item No general trend was visible in the
variability pattern. This rules out that the outburst in 2014 was caused by
gravitational lensing or by a tidal disruption event. 
 In these cases we would have
expected a general decrease in the emitted continuum flux.

\item The optical B-band continuum light curve is delayed by about three days
with respect to the X-ray light curve.

\item The spectral type of HE\,1136-2304
 remained as Seyfert 1.5 between 2014 and 2017
despite its strong continuum variations at the same time.
\end{enumerate}

\begin{acknowledgements}
We are thankful to the late Neil Gehrels for approving our ToO requests.
 We also thank the \swift{} team for performing the ToO observations. 
This research has made use of the
  XRT Data Analysis Software (XRTDAS) developed under the responsibility
  of the ASI Science Data Center (ASDC), Italy.
  This research has made use of the NASA/IPAC Extragalactic
Database (NED) which is operated by the Jet Propulsion Laboratory,
Caltech, under contract with the National Aeronautics and Space
Administration. 
This work has been supported by the DFG grants Ha 3555/12-1, Ko 857/32-2 and
Ko 857/33-1.
\end{acknowledgements}

\clearpage

\begin{appendix}
\section{\bf Additional tables}
\begin{table*}
\caption{\label{swiftlog} 
 XRT and UVOT monitoring observation log: 
Julian date, UT Date, and XRT and UVOT exposure times in seconds.
}
\begin{tabular*}{\textwidth}{@{\extracolsep{\fill} }ccrrrrrrr}
\hline 
\noalign{\smallskip}
Julian Date &  \\
2\,400\,000+&  \rb{UT Date}  &  \rb{XRT} & \rb{V} & \rb{B} & \rb{U} & \rb{UV W1} & \rb{UV M2} & \rb{UVW2}   \\
\hline 
55350.7604    &       2010-06-03 06:15 & 3828 & --- & --- & --- & --- & 3805 & ---  \\
55420.4688    &       2010-08-11 23:15 & 1643 & --- & --- & --- & 1639 & --- & ---  \\
55424.4167    &       2010-08-15 23:00 & 3685 & --- & --- & --- & 2665 & --- & ---  \\
56833.6090    &       2014-06-25 02:45 & 1573 & 130 & 130 & 130 & 260 & 380 & 520  \\
56840.3437    &       2014-07-01 20:15 & 1021 & --- & --- & --- & --- & --- & 1018  \\
56844.1424    &       2014-07-05 15:25 & 1019 &  83 &  83 &  83 & 166 & 246 & 333  \\
56847.2729    &       2014-07-08 18:33 &  979 & --- & --- & --- & --- & 979 & ---  \\
56850.9375    &       2014-07-12 10:30 & 1039 & --- & --- & --- & --- & 1025 & ---  \\
56854.8646    &       2014-07-16 08:10 &  817 & --- & --- & --- & --- &  820 & ---  \\
56861.9479    &       2014-07-23 10:43 & 1051 & --- & --- & --- & --- & 1060 & ---  \\   
56862.6007    &       2014-07-24 02:26 &  724 & --- & --- & --- & --- &  735 & --- \\
56866.9965    &       2014-07-28 11:53 &  949 & --- & --- & --- & --- &  936 & --- \\
56871.0729    &       2014-08-01 13:46 &  998 & --- & --- & --- & --- &  756 & --- \\
56874.1007    &       2014-08-05 02:25 &  948 & --- & --- & --- & --- &  946 & --- \\
56878.5347    &       2014-08-09 00:49 &  599 & --- & --- & --- & --- &  592 & --- \\
56882.7340    &       2014-08-13 05:38 & 1017 & --- & --- & --- & --- & 1008 & ---  \\
57085.3507    &       2015-03-03 20:23 &  957 & --- & --- & 952 & --- & --- & ---  \\
57087.5208    &       2015-03-06 00:31 &  826 & --- & --- & --- & 828 & --- & ---  \\
57132.3125    &       2015-04-19 19:35 & 1186 &  95 &  95 &  95 & 190 & 267 & 381  \\
57138.7500    &       2015-04-26 06:00 & 1892 & 153 & 153 & 153 & 306 & 451 & 613  \\
57145.8368    &       2015-05-03 08:05 & 1326 & 109 & 109 & 106 & 216 & 297 & 438  \\
57152.6875    &       2015-05-10 04:10 & 2253 & 182 & 182 & 182 & 365 & 100 & 730  \\
57160.9028    &       2015-05-18 09:39 &  587 &  52 &  52 &  52 & 103 & 114 & 207  \\
57167.0833    &       2015-05-24 14:00 & 2208 & 182 & 182 & 182 & 365 & 494 & 734  \\
57173.8160    &       2015-05-31 07:25 & 1436 & 116 & 116 & 116 & 233 & 331 & 466  \\
57181.0000    &       2015-06-07 12:00 & 2662 & 216 & 216 & 216 & 432 & 623 & 866  \\
57187.8056    &       2015-06-14 07:20 & 2003 & 164 & 164 & 164 & 326 & 465 & 654  \\
57194.9236    &       2015-06-21 22:30 & 1888 & 154 & 154 & 154 & 310 & 430 & 618  \\
57201.8785    &       2015-06-28 09:03 & 1521 & 126 & 126 & 126 & 251 & 366 & 503  \\
57209.3986    &       2015-07-05 21:34 & 2597 & 205 & 205 & 205 & 414 & 616 & 827  \\
57218.2188    &       2015-07-14 17:15 & 2138 & 175 & 175 & 175 & 350 & 481 & 701  \\
57222.6285    &       2015-07-19 03:05 & 2165 & 172 & 172 & 172 & 346 & 515 & 692  \\
57229.9167    &       2015-07-26 09:55 & 1963 & 159 & 159 & 159 & 317 & 472 & 636  \\
57236.9687    &       2015-08-02 11:17 & 1955 & 158 & 158 & 158 & 316 & 480 & 632  \\
57337.5763    &       2015-11-11 01:50 & 1983 & 161 & 161 & 161 & 321 & 482 & 642  \\
57421.3507    &       2016-02-02 20:25 & 1963 & 157 & 157 & 157 & 314 & 472 & 630  \\
\hline 
\end{tabular*}
\label{swiftlog}
\end{table*}

\begin{table*}
\tabcolsep+5mm
\caption{Log of photometric observations with MONET/North (N) and MONET/South
  (S): Julian date, UT date, used filters, and telescope.
}
\centering
\vspace{-0.5mm}
\begin{tabular}{cccc}
\hline 
\noalign{\smallskip}
Julian Date & UT Date & Filter& Telescope \\
2\,400\,000+&         &      \\
\hline 
56979.010       &       2014-11-17      &       B, V, R &       N       \\
56979.968       &       2014-11-18      &       B, V, R &       N       \\
56981.980       &       2014-11-20      &       B, V, R &       N       \\
56982.971       &       2014-11-21      &       B, V, R &       N       \\
56985.947       &       2014-11-24      &       B, V, R &       N       \\
56986.936       &       2014-11-25      &       B, V, R &       N       \\
56988.945       &       2014-11-27      &       B, V, R &       N       \\
56989.936       &       2014-11-28      &       B, V, R &       N       \\
56991.034       &       2014-11-29      &       B, V, R &       N       \\
56991.987       &       2014-11-30      &       B, V, R &       N       \\
56996.924       &       2014-12-05      &       B, V, R &       N       \\
56997.921       &       2014-12-06      &       B, V, R &       N       \\
56999.890       &       2014-12-08      &       B, V, R &       N       \\
57004.007       &       2014-12-12      &       B, V, R &       N       \\
57006.928       &       2014-12-15      &       B, V, R &       N       \\
57008.023       &       2014-12-16      &       B, V, R &       N       \\
57009.928       &       2014-12-18      &       B, V, R &       N       \\
57011.018       &       2014-12-19      &       B, V, R &       N       \\
57027.909       &       2015-01-05      &       B, V, R &       N       \\
57029.911       &       2015-01-07      &       B, V, R &       N       \\
57051.844       &       2015-01-29      &       B, V, R &       N       \\
57056.946       &       2015-02-03      &       B, V, R &       N       \\
57057.827       &       2015-02-04      &       B, V, R &       N       \\
57504.252       &       2016-04-25      &       B, V, R &       S       \\
57505.513       &       2016-04-26      &       B, V, R &       S       \\
57511.399       &       2016-05-02      &       B, R    &       S       \\
57513.282       &       2016-05-04      &       B, V    &       S       \\
57514.333       &       2016-05-05      &       B, V    &       S       \\
57518.273       &       2016-05-09      &       B, V, R &       S       \\
57520.353       &       2016-05-11      &       B, V, R &       S       \\
57530.295       &       2016-05-21      &       B, V, R &       S       \\
57535.293       &       2016-05-26      &       B, V, R &       S       \\
57536.314       &       2016-05-27      &       B, V, R &       S       \\
57537.309       &       2016-05-28      &       B, V, R &       S       \\
57538.230       &       2016-05-29      &       B, V, R &       S       \\
57546.234       &       2016-06-06      &       B, V, R &       S       \\
57558.358       &       2016-06-18      &       B, V, R &       S       \\
57561.329       &       2016-06-21      &       B, V, R &       S       \\

\hline 
\end{tabular}
\label{monetlog}
\end{table*}

\tabcolsep+4mm
\begin{longtable}{lll|lll}
\caption{Log of photometric observations with VYSOS 16 (V 16) 
and BEST II (B II): Julian date, used filter, and telescope.
}\\
  \hline 
 Julian Date  & Filter& Telescope & Julian Date  & Filter& Telescope \\
2\,400\,000+&         &           &2\,400\,000+&         &      \\
 \hline 
  \endfirsthead
\caption{continued.}\\
  \hline 
 Julian Date  & Filter& Telescope & Julian Date  & Filter& Telescope \\
2\,400\,000+&         &           &2\,400\,000+&         &      \\
 \hline 
\endhead
\hline
\endfoot
57130.705       &       B, V, NB$_{670}$        &       V 16    &      57203.486       &       B       &       B II    \\                                  
57132.680       &       B, V, NB$_{670}$        &       V 16    &      57211.467       &       NB$_{670}$      &       V 16    \\                          
57133.627       &       B, V, NB$_{670}$        &       V 16    &      57211.488       &       B       &       B II    \\                                  
57134.649       &       B, V, NB$_{670}$        &       V 16    &      57212.468       &       NB$_{670}$      &       V 16    \\                          
57140.622       &       B, V, NB$_{670}$        &       V 16    &      57212.488       &       B       &       B II    \\                                  
57141.572       &       B, V, NB$_{670}$        &       V 16    &      57213.491       &       B       &       B II    \\                                  
57142.571       &       B, V, NB$_{670}$        &       V 16    &      57215.465       &       NB$_{670}$      &       V 16    \\                          
57143.696       &       NB$_{670}$      &       V 16    &              57215.489       &       B       &       B II    \\                                  
57144.670       &       NB$_{670}$      &       V 16    &              57216.488       &       NB$_{670}$      &       V 16    \\                          
57145.621       &       NB$_{670}$      &       V 16    &              57216.519       &       B       &       B II    \\                                  
57145.670       &       V       &       V 16    &                      57219.465       &       NB$_{670}$      &       V 16    \\                          
57146.695       &       NB$_{670}$      &       V 16    &              57220.465       &       NB$_{670}$      &       V 16    \\                          
57147.620       &       NB$_{670}$      &       V 16    &              57221.465       &       NB$_{670}$      &       V 16    \\                          
57147.651       &       B       &       V 16    &                      57229.468       &       NB$_{670}$      &       V 16    \\                          
57148.656       &       B, V, NB$_{670}$        &       V 16    &      57230.468       &       NB$_{670}$      &       V 16    \\                          
57150.644       &       V       &       V 16    &                      57231.468       &       NB$_{670}$      &       V 16    \\                         
57151.581       &       B, V, NB$_{670}$        &       V 16    &      57474.588       &       V       &       V16     \\                                 
57151.663       &       B       &       B II    &                      57475.558       &       B, V, NB$_{670}$        &       V16     \\                 
57152.727       &       B       &       B II    &                      57477.599       &       B, V, NB$_{670}$        &       V16     \\                 
57153.585       &       B, V, NB$_{670}$        &       V 16    &      57478.581       &       B, V, NB$_{670}$        &       V16     \\                 
57153.636       &       B       &       B II    &                      57479.555       &       B, V, NB$_{670}$        &       V16     \\                 
57156.555       &       B, V, NB$_{670}$        &       V 16    &      57480.747       &       B, V, NB$_{670}$        &       V16     \\                 
57156.667       &       B       &       B II    &                      57481.686       &       B, V, NB$_{670}$        &       V16     \\                 
57157.556       &       V, NB$_{670}$   &       V 16    &              57482.737       &       B, V, NB$_{670}$        &       V16     \\                 
57157.663       &       B       &       B II    &                      57484.606       &       B, V, NB$_{670}$        &       V16     \\                 
57158.653       &       V, NB$_{670}$   &       V 16    &              57485.552       &       B, V, NB$_{670}$        &       V16     \\                 
57158.735       &       B       &       B II    &                      57486.667       &       B, V, NB$_{670}$        &       V16     \\                 
57160.600       &       NB$_{670}$      &       V 16    &              57492.493       &       B, V, NB$_{670}$        &       V16     \\                 
57160.681       &       B       &       B II    &                      57493.515       &       B, V, NB$_{670}$        &       V16     \\                 
57161.555       &       V, NB$_{670}$   &       V 16    &              57494.495       &       B, V, NB$_{670}$        &       V16     \\                 
57161.637       &       B       &       B II    &                      57499.489       &       B, V, NB$_{670}$        &       V16     \\                 
57162.606       &       V, NB$_{670}$   &       V 16    &              57500.488       &       B, V, NB$_{670}$        &       V16     \\                 
57162.639       &       B       &       B II    &                      57507.599       &       B, V, NB$_{670}$        &       V16     \\                 
57163.553       &       V, NB$_{670}$   &       V 16    &              57508.497       &       B, V, NB$_{670}$        &       V16     \\                 
57163.722       &       B       &       B II    &                      57509.484       &       B, V, NB$_{670}$        &       V16     \\                 
57164.553       &       V, NB$_{670}$   &       V 16    &              57510.565       &       B, V, NB$_{670}$        &       V16     \\                 
57164.639       &       B       &       B II    &                      57511.530       &       B, V, NB$_{670}$        &       V16     \\                 
57165.658       &       V       &       V 16    &                      57512.489       &       B, V, NB$_{670}$        &       V16     \\                 
57169.598       &       V, NB$_{670}$   &       V 16    &              57515.553       &       B, V, NB$_{670}$        &       V16     \\                 
57170.533       &       V, NB$_{670}$   &       V 16    &              57516.517       &       B, V, NB$_{670}$        &       V16     \\                 
57170.546       &       B       &       B II    &                      57521.629       &       B, V, NB$_{670}$        &       V16     \\                 
57171.554       &       V, NB$_{670}$   &       V 16    &              57522.461       &       B, V, NB$_{670}$        &       V16     \\                 
57171.601       &       B       &       B II    &                      57524.471       &       B, V, NB$_{670}$        &       V16     \\                 
57174.586       &       V, NB$_{670}$   &       V 16    &              57526.462       &       B, V, NB$_{670}$        &       V16     \\                 
57174.628       &       B       &       B II    &                      57527.460       &       B, NB$_{670}$   &       V16     \\                         
57175.553       &       V, NB$_{670}$   &       V 16    &              57529.463       &       B, NB$_{670}$   &       V16     \\                         
57175.595       &       B       &       B II    &                      57530.459       &       B       &       V16     \\                                 
57176.558       &       V, NB$_{670}$   &       V 16    &              57531.639       &       B, NB$_{670}$   &       V16     \\                         
57176.595       &       B       &       B II    &                      57532.509       &       B, NB$_{670}$   &       V16     \\                         
57177.614       &       V       &       V 16    &                      57533.458       &       B, NB$_{670}$   &       V16     \\                         
57178.566       &       V, NB$_{670}$   &       V 16    &              57534.519       &       B, NB$_{670}$   &       V16     \\                         
57178.595       &       B       &       B II    &                      57536.458       &       B, NB$_{670}$   &       V16     \\                         
57179.552       &       V, NB$_{670}$   &       V 16    &              57538.626       &       B, NB$_{670}$   &       V16     \\                         
57179.595       &       B       &       B II    &                      57539.487       &       B       &       V16     \\                                 
57180.568       &       V, NB$_{670}$   &       V 16    &              57540.599       &       B, NB$_{670}$   &       V16     \\                         
57180.596       &       B       &       B II    &                      57541.542       &       B, NB$_{670}$   &       V16     \\                         
57181.552       &       V, NB$_{670}$   &       V 16    &              57542.536       &       B, NB$_{670}$   &       V16     \\                         
57181.596       &       B       &       B II    &                      57551.512       &       B, NB$_{670}$   &       V16     \\                         
57182.553       &       V, NB$_{670}$   &       V 16    &              57555.463       &       B, NB$_{670}$   &       V16     \\                         
57182.600       &       B       &       B II    &                      57558.459       &       B, NB$_{670}$   &       V16     \\                         
57183.515       &       V, NB$_{670}$   &       V 16    &              57560.459       &       B, NB$_{670}$   &       V16     \\                         
57183.570       &       B       &       B II    &                      57561.463       &       B, NB$_{670}$   &       V16     \\                         
57184.515       &       V, NB$_{670}$   &       V 16    &              57562.459       &       B, NB$_{670}$   &       V16     \\                         
57184.569       &       B       &       B II    &                      57567.460       &       B, NB$_{670}$   &       V16     \\                         
57185.515       &       V, NB$_{670}$   &       V 16    &              57569.461       &       B, V, NB$_{670}$        &       V16     \\                 
57185.569       &       B       &       B II    &                      57575.467       &       B, V, NB$_{670}$        &       V16     \\                 
57186.515       &       V, NB$_{670}$   &       V 16    &              57581.468       &       B, V, NB$_{670}$        &       V16     \\                 
57186.570       &       B       &       B II    &                      57582.469       &       B, NB$_{670}$   &       V16     \\                         
57187.516       &       V, NB$_{670}$   &       V 16    &              57583.468       &       B, NB$_{670}$   &       V16     \\                         
57187.570       &       B       &       B II    &                      57584.464       &       B, NB$_{670}$   &       V16     \\                         
57189.516       &       V, NB$_{670}$   &       V 16    &              57585.504       &       B, NB$_{670}$   &       V16     \\                         
57192.518       &       V, NB$_{670}$   &       V 16    &              57473.593       &       B       &       V16     \\                                 
57195.497       &       V, NB$_{670}$   &       V 16    &              57474.570       &       NB$_{670}$      &       V16     \\                         
57195.510       &       B       &       B II    &                      57475.540       &       NB$_{670}$      &       V16     \\                         
57198.497       &       V, NB$_{670}$   &       V 16    &              57477.581       &       NB$_{670}$      &       V16     \\                         
57203.461       &       V, NB$_{670}$   &       V 16    &              57478.563       &       NB$_{670}$      &       V16     \\                         
\label{bochumlog}
\end{longtable}
\twocolumn
%
%
\twocolumn
\end{appendix}

\end{document}